\documentclass[conference]{IEEEtran}
\IEEEoverridecommandlockouts
\usepackage{cite}
\usepackage{amsmath,amssymb,amsfonts}
\usepackage{algorithmic}
\usepackage{graphicx}
\usepackage{textcomp}
\usepackage[dvipsnames]{xcolor}

\usepackage{contour}
\contournumber{80}

\usepackage[a4paper,left=1.32cm,right=1.32cm,top=1.9cm,bottom=4.3cm]{geometry} 
\usepackage[font=footnotesize]{caption}
\usepackage[font=footnotesize]{subcaption}      
\usepackage{booktabs}
\usepackage{tcolorbox}
\usepackage{bm}			
\usepackage[per-mode=symbol]{siunitx}    		
\usepackage{cite} 
\DeclareSIUnit{\dBm}{dBm}	
\DeclareSIUnit{\eq}{eq}	    
\DeclareSIUnit{\sqrtW}{\ensuremath{\sqrt{\text{W}}}}
\usepackage{layouts}
\usepackage{overpic}		
\definecolor{RDlightgreen}{RGB}{141 192 69}
\definecolor{RDgreen}{RGB}{93 109 68}
\definecolor{mygray}{gray}{0.6}
\usepackage[    
                backgroundcolor=RDlightgreen!70,
                textcolor=black,
                bordercolor=RDgreen,
                textsize=tiny]{todonotes}
\makeatletter
\renewcommand{\todo}[2][]{\tikzexternaldisable\@todo[#1]{#2}\tikzexternalenable}
\makeatother
                
\usepackage{enumerate}  

\usepackage[xindy]{glossaries}
\makenoidxglossaries%
\newacronym{2d}{2D}{two-dimensional}
\newacronym{3d}{3D}{three-dimensional}
\newacronym{aoa}{AOA}{angle-of-arrival}
\newacronym{aod}{AOD}{angle-of-departure}
\newacronym{asic}{ASIC}{application specific integrated circuit}
\newacronym{awgn}{AWGN}{additive white Gaussian noise}
\newacronym{bf}{BF}{beamformer}
\newacronym{ce}{CE}{channel estimation}
\newacronym{ced}{CED}{cumulative energy density}
\newacronym{cdf}{CDF}{cumulative distribution function}
\newacronym{csi}{CSI}{channel state information}
\newacronym{crlb}{CRLB}{Cram\'er-Rao lower bound}
\newacronym{dc}{DC}{direct current}
\newacronym{dm}{DM}{diffuse multipath}
\newacronym{dmc}{DMC}{diffuse multipath component}
\newacronym{eh}{EH}{energy harvesting}
\newacronym{eirp}{EIRP}{equivalent isotropically radiated power}
\newacronym{em}{EM}{electromagnetic}
\newacronym{en}{EN}{energy neutral}
\newacronym{e2e}{E2E}{End-to-End}
\newacronym{epu}{EPU}{edge processing unit}
\newacronym{fa}{FA}{federation anchor}
\newacronym{ic}{IC}{integrated circuit}
\newacronym{iot}{IoT}{Internet of things}
\newacronym{id}{ID}{information decoding}
\newacronym{ism}{ISM}{industrial, scientific and medical}
\newacronym{kpi}{KPI}{key performance indicator}
\newacronym{lca}{LCA}{life cycle assessment}
\newacronym{lis}{LIS}{large intelligent surface}
\newacronym{liion}{Li-ion}{lithium-ion}
\newacronym{los}{LoS}{line-of-sight}
\newacronym{lmo}{LMO}{lithium ion manganese oxide}
\newacronym{mc}{MC}{Monte Carlo}
\newacronym{mf}{MF}{matched filter}
\newacronym{mmse}{MMSE}{minimum mean square error}
\newacronym{mmwave}{mmWave}{millimeter wave}
\newacronym{mimo}{MIMO}{multiple-input multiple-output}
\newacronym{miso}{MISO}{multiple-input single-output}
\newacronym{xlmimo}{XL-MIMO}{extremely large-scale \acrshort{mimo}}
\newacronym{mrc}{MRC}{maximum ratio combining}
\newacronym{mrt}{MRT}{maximum ratio transmission}
\newacronym{ofdm}{OFDM}{orthogonal frequency-division multiplexing}
\newacronym{ota}{OTA}{over-the-air}
\newacronym{nb}{NB}{narrowband}
\newacronym{nr}{NR}{New Radio}
\newacronym{pa}{PA}{power amplifier}
\newacronym{pdp}{PDP}{power delay profile}
\newacronym{per}{PER}{packet error rate}
\newacronym{pl}{PL}{path loss}
\newacronym{pg}{PG}{path gain}
\newacronym{pc}{PC}{pilot count}
\newacronym{pw}{PW}{planar wavefront}
\newacronym{rcs}{RCS}{radar cross-section}
\newacronym{re}{RE}{radio element}
\newacronym{rf}{RF}{radio frequency}
\newacronym{rfid}{RFID}{\acrshort{rf} identification}
\newacronym{ris}{RIS}{reflective intelligent surface}
\newacronym{rms}{RMS}{root mean square}
\newacronym{rss}{RSS}{received signal strength}
\newacronym{rssi}{RSSI}{received signal strength indicator}
\newacronym{rv}{RV}{random variable}
\newacronym{rw}{RW}{RadioWeave}
\newacronym{sa}{SA}{synthetic aperture}
\newacronym{sdg}{SDG}{Sustainable Development Goal}
\newacronym{sdn}{SDN}{software-defined network}
\newacronym{sinr}{SINR}{signal-to-interference-plus-noise ratio}
\newacronym{siso}{SISO}{single-input single-output}
\newacronym{slam}{SLAM}{simultaneous localization and mapping}
\newacronym{smc}{SMC}{specular multipath component}
\newacronym{snr}{SNR}{signal-to-noise ratio}
\newacronym{s-parameter}{S-parameter}{scattering parameter}
\newacronym{svd}{SVD}{singular value decomposition}
\newacronym{sw}{SW}{spherical wavefront}
\newacronym{swipt}{SWIPT}{simultaneous wireless information and power transfer}
\newacronym{tdd}{TDD}{time division duplexing}
\newacronym{tdoa}{TDOA}{time-difference-of-arrival}
\newacronym{toa}{TOA}{time-of-arrival}
\newacronym{tosm}{TOSM}{through–open–short–match}
\newacronym{ue}{UE}{user equipment}
\newacronym{uhf}{UHF}{ultra high frequency}
\newacronym{ula}{ULA}{uniform linear array}
\newacronym{upa}{UPA}{uniform planar array}
\newacronym{ura}{URA}{uniform rectangular array}
\newacronym{uwb}{UWB}{ultrawideband}
\newacronym{vna}{VNA}{vector network analyzer}
\newacronym{wb}{WB}{wideband}
\newacronym{wpt}{WPT}{wireless power transfer}
\newacronym{wsn}{WSN}{wireless sensor network}
\newacronym{xets}{XETS}{cross exponentially tapered slot}
\newacronym{zf}{ZF}{zero forcing}
\newacronym{csp}{CSP}{contact service point}
\newacronym{ecsp}{ECSP}{edge computing service point}
\newacronym{ap}{AP}{access point}

\usepackage{balance}
\usepackage{flushend}
\usepackage{multirow}

\tcbset{shield externalize} 

\newlength\figureheight
\newlength\figurewidth 
\newlength\hspaceing
\newlength\plotHeight
\newlength\plotWidth

\usepackage{ifthen}
\newcommand{\externalizeFigures}{false}
\ifthenelse{\equal{\externalizeFigures}{true}}
{
  \usepackage{pgfplots}
  \pgfplotsset{compat=newest}
  \usepgfplotslibrary{external}
  \tikzexternalize[prefix=externalized/]

  \usepackage{shellesc}
}
{}

\newcommand{\mrm}[1]{ \mathrm{#1} }
\DeclareMathOperator*{\argmax}{arg\,max}

\DeclareMathOperator*{\SNR}{SNR}
\DeclareMathOperator*{\Var}{Var}

\newcommand{\herm}{\mathsf{H}}
\newcommand{\trp}{\mathsf{T}}
\DeclareMathOperator{\diag}{diag}

\newcommand{\realset}[2]{ \mathbb{R}^{#1 \times #2}  }
\newcommand{\complexset}[2]{ \mathbb{C}^{#1 \times #2}  }

\newcommand{\vis}{\ensuremath{\chi}} 
\newcommand{\refcoeff}{\ensuremath{\gamma}} 
\newcommand{\house}{\ensuremath{\mathcal{H}}} 

\newcommand{\ticked}{$\text{\rlap{$\checkmark$}}\square$}
\newcommand{\unticked}{{$\square$}}
\newcommand{\tick}[1]{\ifthenelse{#1=1}{\ticked}{\unticked}}

  \usepackage{pgfplots}
  \pgfplotsset{compat=newest}
  \usetikzlibrary{plotmarks}
  \usetikzlibrary{arrows.meta}
  \usepgfplotslibrary{patchplots}
  \usetikzlibrary{calc}
  \usepackage{tikzscale}		
  
\def\BibTeX{{\rm B\kern-.05em{\sc i\kern-.025em b}\kern-.08em
    T\kern-.1667em\lower.7ex\hbox{E}\kern-.125emX}}
\begin{document}

\title{XL-MIMO Channel Modeling and Prediction for Wireless Power Transfer}

\author{\IEEEauthorblockN{
Benjamin J. B. Deutschmann\IEEEauthorrefmark{4},   
Thomas Wilding\IEEEauthorrefmark{4},   
Maximilian Graber\IEEEauthorrefmark{4},   
Klaus Witrisal\IEEEauthorrefmark{4}     
}                                     
\thanks{The project has received funding from the European Union’s Horizon 2020 research and innovation program under grant agreement No 101013425.}
\IEEEauthorblockA{\IEEEauthorrefmark{4}
Graz University of Technology, Austria}
\IEEEauthorblockA{ \emph{benjamin.deutschmann@tugraz.at} }
}

\maketitle

\begin{abstract}
Massive antenna arrays form physically large apertures with a beam-focusing capability, leading to outstanding \gls{wpt} efficiency paired with low radiation levels outside the focusing region. 
However, leveraging these features requires accurate knowledge of the multipath propagation channel and overcoming the (Rayleigh) fading channel present in typical application scenarios.
For that, reciprocity-based beamforming is an optimal solution that estimates the actual channel gains from pilot transmissions on the uplink. 
But this solution is unsuitable for passive backscatter nodes that are not capable of sending any pilots in the initial access phase.

Using measured channel data from an \gls{xlmimo} testbed, we compare geometry-based planar wavefront and spherical wavefront beamformers with a reciprocity-based beamformer, to address this initial access problem. 
We also show that we can predict \glspl{smc} based only on geometric environment information. 
We demonstrate that a transmit power of 1\,W is sufficient to transfer more than 1\,mW of power to a device located at a distance of 12.3\,m when using a 40\,$\times$\,25 array at 3.8\,GHz.
The geometry-based beamformer exploiting predicted SMCs suffers a loss of only 2\,dB compared with perfect channel state information. 
\end{abstract}

\glsresetall            


\section{Introduction}

\Gls{mimo} systems have proved to bear promising potential for \gls{wpt} in indoor scenarios exploiting array gain and diversity gain~\cite{ArnitzPC2016}.
The advent of distributed or \gls{xlmimo} has brought even more diversity to be leveraged but also challenges to overcome. 
Some of these challenges have been studied in~\cite{Zhang22NFwpt} where high frequencies in the mmWave range are promoted based on the findings that they 
result in smaller beam widths and a higher \gls{wpt} efficiency when considering a constant aperture array~\cite{Smith17NFwpt}. 
The power received at the device side is typically in the microwatt region. 
Contrary to the advantage of forming very narrow beams at mmWave frequencies, we propose using sub-10\,\SI{}{\giga\hertz} frequencies for \gls{wpt} because much larger apertures are realized with the same number of array elements.
Low power densities and range-dependent gain patterns facilitate higher efficiencies while meeting regulatory compliance constraints~\cite{D4_1}.
%
%
%
Possible devices operated through \gls{rf} \gls{wpt} are \gls{rfid} tags, which are batteryless \gls{en} devices communicating through backscatter communication.
The concurrent optimization goals of \gls{swipt}~\cite{PereraCST2018} have been subject to extensive research.
Particularly the nonlinear input impedance of the device frontend has attracted significant attention, which led to the development of optimal waveform designs~\cite{ClerckxTSP2016}.

Another challenge with batteryless \gls{en} devices is the initial access procedure, where no measured \gls{csi} is available and the device has to be woken up for the first time. 
Our approach to solving this task is applying geometry-based \glspl{bf} that rely on predicted rather than measured \gls{csi}.
We intentionally consider frequency-domain radio channels between the antennas of a \gls{miso} system but neglect frontend behavior and communication aspects, as these may be synthesized on top of our findings and exceed the scope of this work. 

\setlength{\figurewidth}{0.96\linewidth}
\begin{figure}[tb]\centering
\input{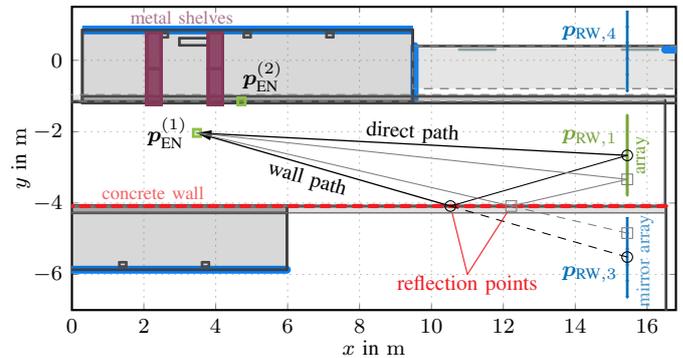}
\vspace{-4mm}
\caption{The measurement scenario: A long hallway with concrete walls. A large RW is indicated on the right side (positive $x$-direction) along with its virtual mirror sources. The two measurement positions $\bm{p}_\mrm{EN}^{(1)}$ and $\bm{p}_\mrm{EN}^{(2)}$ of hypothetical \gls{en} devices are indicated on the left side.}
\label{fig:scenario}
\vspace{-5mm}
\end{figure}	

We investigate the performance of geometry-based \glspl{bf} that work with predicted channels to solve the initial access problem. 
Analyses of the expected performance of reciprocity-based \glspl{bf} with imperfect \gls{csi} exist in the literature both in a communication and \gls{wpt}~\cite{Mishra2019} context.
In this work, we derive its expected efficiency and compare its performance to our geometry-based \glspl{bf} evaluated on real channel measurement data.
We further describe the typical asymptotic efficiency regimes w.r.t. the \gls{snr}.
%
We verify that \gls{xlmimo} systems operating at sub-10\,\SI{}{\giga\hertz} achieve the gains forecasted in~\cite{D4_1} and lift the receivable power at the device side from the microwatt to the milliwatt range while transmitting over distances greater than \SI{10}{\metre}.


\section{Problem Statement}\label{sec:problem}

We consider a \gls{miso} system, where we aim to transmit power wirelessly from one physically large array, denoted a \gls{rw}, to a single \gls{en} device.
We assume a narrowband frequency-flat fading channel $\bm{h}\in \complexset{L}{1}$ that models the transmission coefficients, i.e., \glspl{s-parameter}, from each transmit antenna $\ell \in \left\{1, \dots, L\right\}$ to the receiving antenna of the \gls{en} device. 
The \gls{rw} transmits an overall power $P_\mrm{t}$ distributed via the weights $\bm{w}\in \complexset{L}{1}$ to its antennas, where $\lVert \bm{w}\rVert = 1$.
The \gls{en} device receives a complex baseband amplitude, i.e., a phasor,
\begin{align}\label{eq:y_rx}
    y = \bm{h}^\trp \bm{w} \sqrt{P_\mrm{t}} + n 
\end{align}
where we assume that the power of the noise $n$ (i.e., for ambient \acrlong{eh}) is negligible when compared to intentional \gls{wpt}.
Consequently, the \gls{rf}-\gls{rf} transmission efficiency is
\begin{align}\label{eq:PG}
    \frac{P_\mrm{r}}{P_\mrm{t}} = \frac{|y|^2}{P_\mrm{t}} \approx \left| \bm{h}^\trp  \bm{w} \right|^2 = PG \,
\end{align}
where $P_\mrm{r}= |y|^2$ is the power received by the \gls{en} device and $PG$ denotes the 
path gain including 
antenna gains, polarization gain, and array gain as defined in~\cite{D4_1}.

A beamforming method aims 
to choose the optimum weights that maximize the path gain in~\eqref{eq:PG}.
These power-optimal weights are well-known to be computed through \gls{mrt} as
\begin{align}\label{eq:mrt}
    \bm{w} = \frac{\bm{h}^\ast}{\lVert \bm{h} \rVert}
\end{align}
which demands knowledge of the channel vector $\bm{h}$.
For the initial access to an \gls{en} device, i.e., before the initial wake-up and reception of a backscattered signal, this \gls{csi} is unknown and the weights $\bm{w}$ cannot be computed through~\eqref{eq:mrt}.

The challenge of the initial access procedure is to deliver power to an \gls{en} device that exceeds its sensitivity limit $P_\mrm{r,min}$, i.e., its minimum wake-up power, given that neither the channel vector $\bm{h}$, nor the device position $\bm{p}$ are known.
Having a calibrated\footnote{We consider the linear error network before the antennas as characterized and compensated for. I.e., we can coherently control the phasors at the phase centers of the antennas.} array 
with known antenna positions $\bm{p}_\mrm{RW}^{(\ell)}$, beamforming weights can be computed based on a geometry-based channel model $\widetilde{\bm{h}}(\bm{p})$.
To compensate for the a priori unknown \gls{en} device position $\bm{p}$, a common strategy is to perform  
an exhaustive search,
also known as beam sweeping.
Upon the first signal transmitted from the \gls{en} device, \gls{csi} can be estimated at the \gls{rw} and the channel estimate $\hat{\bm{h}}$ can be used to conduct reciprocity-based \gls{mrt}.

\section{Channel Model}\label{sec:channel}
In~\cite{Deutschmann22BeamDiversity}, we introduced a geometry-based channel model for 
\gls{wpt} with isotropic antennas. 
This initial model is based on an image source model~\cite{Leitinger2015}, where \glspl{smc} originating from extended, flat surfaces, e.g., walls, are modeled by mirroring the antennas $\ell$ of a transmitting \gls{rw} across these surfaces. 
We model $k \in \left\{1, \dots, K\right\}$ \glspl{smc} including the \gls{los}, where each reflection is associated with a reflection coefficient $\refcoeff_k$.
We extended the model in~\cite{D4_1} to accommodate antennas with arbitrary gain patterns $G(\theta,\varphi)$, where $\theta$ and $\varphi$ are elevation and azimuth angles within the local coordinate system of each antenna $\ell$.
Furthermore, possible polarization losses $g_{k,\ell}^{\mrm{pol}}$ due to a polarization mismatch between transmit and receive antennas as well as polarization shifts due to reflections are accounted for. 
Our most complete channel model is defined in~\cite{D1_2}, where we model the visibility of an \gls{smc} through a visibility vector $\bm{\vis}_{k}$.
This becomes of particular significance 
with distributed \glspl{rw} and physically large arrays where specific \glspl{smc} are only visible over a part of the antennas.
To consider these effects, we define the radio channel $\bm{h}$ as
\begin{align}\label{eq:SW-SMC-channel-model}
\bm{h} &= \sum\limits_{k=1}^{K} \bm{h}_k(\bm{p}) + \bm{h}_\mrm{DM}
\end{align}
consisting of $K$ \glspl{smc} related to the environment geometry, and \gls{dm} denoted as $\bm{h}_\mathrm{DM}$. 

\paragraph{Specular Multipath}\label{sec:deterministic}
Each \gls{smc} channel vector $\bm{h}_k(\bm{p})$ is modeled 
as 
\begin{align}\label{eq:SW-SMC-channel-k}
\bm{h}_k(\bm{p}) &= \bm{A}_k(\bm{p})\bm{G}_k(\bm{p})\bm{b}_k(\bm{p}) 
\end{align}
with the matrices $\bm{A}_k(\bm{p}) = \diag([\vis_{k,1}(\bm{p})\, \dots \, \vis_{k,L}(\bm{p})])$ accounting for the \gls{smc} visibility%
\ifdefined\reduceSize  
    \,
\else
    \footnote{Please refer to Appendix~\ref{app:visibility} for a description on how to compute the visibility vector $\bm{\vis}_{k}$.}
\fi
per array element, $\bm{G}_k(\bm{p}) = \diag([g_{k,1}(\bm{p}) \, \dots \, g_{k,L}(\bm{p})])$ containing the gain pattern values of each transmit antenna $\ell$ and the receive antenna as well as the \gls{siso} path loss, i.e.,
\begin{align}
    g_{k,\ell}(\bm{p}) = \sqrt{G_{\mrm{t},\ell}} \sqrt{G_{\mrm{r}}} \frac{\lambda}{4\pi d_{k,\ell}} \refcoeff_k g^\mrm{pol}_{k,\ell} \, ,
\end{align}
where the dependency on the look angles is omitted for notational brevity and $d_{k,\ell}$ is the distance of each antenna $\ell$ from each mirror source $k$ to the \gls{en} device.
The vector $\bm{b}_k(\bm{p}) = [{e}^{-jk_0 d_{k,1}(\bm{p})}\, \dots \, {e}^{-jk_0 d_{k,L}(\bm{p})}]^\mathsf{T}$ contains the phase shifts due to the distances $d_{k,\ell}$ traveled from the \gls{rw} to the \gls{en} device, with $k_0 = \frac{2\pi}{\lambda}$ denoting the spatial angular frequency.

\paragraph{Diffuse Multipath}\label{sec:stochastic}
The \gls{dm} contained in $\bm{h}_\mathrm{DM}$ is used to represent an often large number of components that are either not resolvable due to the measurement aperture or cannot be related to the environment without involved modeling, e.g., ray-tracing. A number of statistical models are popular in the literature. 
In \cite{KulmerPIMRC2018}, surface roughness was used to model non-specular reflections, allowing a smooth transition from specular to diffuse reflection. In \cite{SalmiEUCAP2010} an autoregressive process was used, whereas \cite{RichterPhD2005,WitrisalICC12} model the \gls{dm} as the result of the convolution of the transmit signal with a random process. 
A geometry-related approach was taken 
in~\cite{FlordelisTWC2020,SchubertTAP2013,D1_2}, 
where randomly distributed point source scatterers are used to model diffuse paths.
While important when aiming at accurate modeling of the received signal in terms of power, this work will focus on exploiting specular components and thus does not rely on a specific model for the \gls{dmc}.
Note that in contrast to geometry-based \glspl{bf}, a reciprocity-based \gls{bf} using measured \gls{csi} will inherently exploit the full radio channel, independent of an underlying model.

\section{Beamforming Strategies}\label{sec:beamforming-strategies}%
We introduce geometry-based \glspl{bf}, which compute the beamforming weights $\bm{w}_\textsc{pw}$ for a \gls{pw}, and $\bm{w}_\textsc{sw}$ for a \gls{sw}.
Both work with the vectorial distances $\bm{r}_\ell = \bm{p} - \bm{p}_\mrm{RW}^{(\ell)}$ from each transmit antenna $\ell$ of the \gls{rw} to an arbitrary point
$\bm{p}$ in space. 
When the position of an \gls{en} device $\bm{p}_\mrm{EN}$ is known, these weights can be used to conduct beamforming without measured \gls{csi} available, i.e., by computing the weights per \gls{smc} from the known geometry.
In the initial access procedure, however, both measured \gls{csi} and the positions of \gls{en} devices are unknown.
Thus, the two-dimensional or three-dimensional parameter spaces of the respective \gls{bf} have to be searched through beam sweeping.

In the following, we derive the beamforming weights $\bm{w}$ through \gls{mrt} on a modeled 
geometry-based channel $\widetilde{\bm{h}}$ as
\begin{align}
    \bm{w} = \frac{{\widetilde{\bm{h}}}^\ast}{\lVert \widetilde{\bm{h}} \rVert} \, \in \complexset{L}{1} \, .
\end{align}
%
Deviations from the ``true" channel manifest as power losses, 
such that geometry-based \glspl{bf} based on more detailed channel models perform more efficiently. 
The same accounts for a reciprocity-based \gls{bf}, which has an efficiency that is dependent on the quality of the estimated \gls{csi}. 
We derive its expected efficiency, i.e., its path gain, as a function of the quality of \gls{csi} analytically in Appendix~\ref{app:CSI-loss} and we compare its performance with our geometry-based \glspl{bf} in Section~\ref{sec:validation}.

\subsection{Planar Wavefront LoS Beamformer}\label{sec:plane-wave-beamformer}
The \gls{pw} \gls{bf} computes weights $\bm{w}_\textsc{pw}$ that cause the \gls{rw} to transmit planar wavefronts towards the targeted angular direction $[\theta \ \varphi]^ \trp$, where $\theta$ and $\varphi$ are the look angles in local spherical coordinates of the \gls{rw} pointing from its center of gravity $\bm{p}_\mrm{RW}$ to the 
position $\bm{p}$.
The spherical coordinates translate to the wave vector defined as~\cite{Dudgeon93ASP} 
\begin{align}\label{eq:wave-vector}
    \bm{k} =
    \begin{bmatrix}
        k_x \\ k_y \\ k_z
    \end{bmatrix}
    =
    k_0  \, 
    \begin{bmatrix}
        \sin{\theta }\cos{\varphi } \\ \sin{\theta }\sin{\varphi } \\ \cos{\theta } 
    \end{bmatrix} \in \realset{3}{1} \,.
\end{align}
With the array layout captured 
%
%
%
%
%
\ifdefined\reduceSize   
    in $\bm{P}_\mrm{RW} = \left[\bm{p}_\mrm{RW}^{(1)} \ \hdots \  \bm{p}_\mrm{RW}^{(L)}\right] \in \realset{3}{L}$, 
    the \gls{pw} beamforming weights can be computed as 
    \begin{align}\label{eq_wp}
        {\bm{w}}_\textsc{pw} = \frac{{\widetilde{\bm{h}}_\textsc{pw}}^\ast}{\lVert \widetilde{\bm{h}}_\textsc{pw} \rVert} ~ \text{with} ~ \left[\widetilde{\bm{h}}_\textsc{pw}\right]_\ell = e^{j \bm{p}_\mrm{RW}^{(\ell)\trp} \bm{k} } \, .
    \end{align}
    
    \subsection{Spherical Wavefront LoS Beamformer}\label{sec:sw-los-beamformer}
    An \gls{sw} \gls{bf} computes weights $\bm{w}_\textsc{sw}$ that cause the \gls{rw} to transmit spherical wavefronts towards the three-dimensional position $\bm{p}$. 
    It 
    projects the distances $\bm{r}_\mrm{\ell}$ 
    onto wave vectors $\bm{k}_\ell$ defined w.r.t. the phase center of each antenna $\ell$. 
    Consequently, the \gls{sw} \gls{los} \gls{bf} weights are given by
    \begin{align}\label{eq_sw_los}
        {\bm{w}}_\textsc{sw} &= \frac{{\widetilde{\bm{h}}_\textsc{sw}}^\ast}{\lVert \widetilde{\bm{h}}_\textsc{sw} \rVert} ~~ 
        \text{with} ~~  \left[\widetilde{\bm{h}}_\textsc{sw}\right]_\ell = e^{-j \bm{r}_{\ell}^\trp \bm{k}_\ell } = e^{-j k_0 \, \lVert\bm{r}_{\ell}\rVert } \, 
    \end{align}
    essentially modeling the phase shifts due to the distances $\lVert\bm{r}_{\ell}\rVert$ traveled from each antenna $\ell$ to the \gls{en} device.
\else
    in $\bm{P}_\mrm{RW} = \left[\bm{p}_\mrm{RW}^{(1)} \ \hdots \  \bm{p}_\mrm{RW}^{(L)}\right] \in \realset{3}{L}$ 
    as well as the position 
    $\bm{p}$,
    the \gls{pw} beamforming weights can be computed\footnote{An $(L\times1)$ vector of the exponents in~\eqref{eq_wp_tilde} can be expressed notationally compact as $-j \left(\bm{P} - \bm{P}_\mrm{RW} \right)^\trp \bm{k}$, where $\bm{P} = \bm{p} \, \bm{1}_{1\times L} \in \realset{3}{L}$. Likewise, a vector of the exponents in~\eqref{eq_wp} can be expressed as $j\bm{P}_\mrm{RW}^{\trp}\bm{k}$.} as 
    \begin{align}\label{eq_wp_tilde}
        \breve{\bm{w}}_\textsc{pw} = \frac{{\breve{\bm{h}}_\textsc{pw}}^\ast}{\lVert \breve{\bm{h}}_\textsc{pw} \rVert} ~ \text{with} ~ \left[\breve{\bm{h}}_\textsc{pw}\right]_\ell= e^{-j \left(\bm{p} - \bm{p}_\mrm{RW}^{(\ell)} \right)^\trp \bm{k} } \, .
    \end{align}
    It should be noted, however, that the position $\bm{p}$ in~\eqref{eq_wp_tilde} can be omitted for the purpose of beamforming for \gls{wpt}, since the alternative weights
    \begin{align}\label{eq_wp}
        {\bm{w}}_\textsc{pw} = \frac{{\widetilde{\bm{h}}_\textsc{pw}}^\ast}{\lVert \widetilde{\bm{h}}_\textsc{pw} \rVert} ~ \text{with} ~ \left[\widetilde{\bm{h}}_\textsc{pw}\right]_\ell = e^{j \bm{p}_\mrm{RW}^{(\ell)\trp} \bm{k} } \, .
    \end{align}
    only differ from $\breve{\bm{w}}_\textsc{pw}$ in a constant phase shift $\Delta \varphi = k_0 \lVert \bm{r} \rVert$ (the projection of $\bm{r}=\bm{p} - \bm{p}_\mrm{RW}$ on $\bm{k}$) present at all $L$ weights, i.e., $\breve{\bm{w}}_\textsc{pw}={\bm{w}}_\textsc{pw} {e}^{j\Delta \varphi}$.
    
    \subsection{Spherical Wavefront LoS Beamformer}\label{sec:sw-los-beamformer}
    A \gls{sw} \gls{bf} computes weights $\bm{w}_\textsc{sw}$ that cause the \gls{rw} to transmit spherical wavefronts towards the three-dimensional position $\bm{p}$. 
    It is strongly related to the \gls{pw} \gls{bf} in~\eqref{eq_wp_tilde} but rather than projecting the distances $\bm{r}_\mrm{\ell}$ onto a single look direction, i.e., onto $\bm{k}$, defined w.r.t. the center of gravity of the array, it projects them onto wave vectors $\bm{k}_\ell$ defined w.r.t. the phase center of each antenna $\ell$. 
    Consequently, the \gls{sw} \gls{los} \gls{bf} weights are given by
    \begin{align}\label{eq_sw_los}
        {\bm{w}}_\textsc{sw} &= \frac{{\widetilde{\bm{h}}_\textsc{sw}}^\ast}{\lVert \widetilde{\bm{h}}_\textsc{sw} \rVert} ~~ 
        \text{with} ~~  \left[\widetilde{\bm{h}}_\textsc{sw}\right]_\ell = e^{-j \bm{r}_{\ell}^\trp \bm{k}_\ell } = e^{-j k_0 \, \lVert\bm{r}_{\ell}\rVert } \, 
    \end{align}
    essentially modelling the phase shifts due to the distances $\lVert\bm{r}_{\ell}\rVert$ traveled from each antenna $\ell$ to the \gls{en} device.
\fi
%
%
%
%
%
%
%
%
%
\setlength{\figurewidth}{0.6\linewidth}
\setlength{\figureheight}{0.2\textheight}

\subsection{Spherical Wavefront SMC Beamformer}\label{sec:sw-smc-beamformer}
We introduce a \gls{sw} \gls{bf} using the geometry-based channel model introduced in Section~\ref{sec:channel}.
It leverages our most complete channel model and takes gain patterns of the used antennas into account, as well as image sources up to the \num{1}\textsuperscript{st} order. 
It also considers the \gls{smc} visibility per array element. 
Consequently, it computes beamforming weights from
\begin{align}\label{eq_sw_smc}
    \widetilde{\bm{h}}_\textsc{smc} 
    = \sum\limits_{k=1}^{K} {\bm{h}}_k(\bm{p})
\end{align}
with ${\bm{h}}_k(\bm{p})$ from the model in~\eqref{eq:SW-SMC-channel-k}.
%
Superimposing channel vectors for several spatially separated positions, \eqref{eq_sw_smc} can be used to extend the \gls{smc} beamformer for the multi-receiver case.
Weighting the individual channels gives control over the power directed to each device. 
Superimposing channel vectors for multiple \glspl{smc} of a single \gls{en} device,~\eqref{eq_sw_smc} results in multiple \gls{smc} beams that optimally interfere constructively at its location.
This simultaneous multibeam transmission exploits reflections from the walls and the floor.
Therefore, it needs a geometric environment model to compute the $K-1$ virtual mirror sources of the \gls{rw}.
Uncertainties in the environment model affect the locations of \gls{smc} beams and their phases at the intended focal point position $\bm{p}$. 
The impact of the latter may be quite severe, since unaligned phases of \gls{smc} beams may even interfere destructively at $\bm{p}$. 
Possibly 
unknown phase shifts incurring due to specular reflections may also cause this effect. 
During the initial access phase, we propose using a scheme of varying \gls{smc} beam phases to compensate for these possibly unknown phases, either randomly as proposed in~\cite{Deutschmann22BeamDiversity}, or by iterating through a predefined codebook.
The objective is to find optimal phase shifts $\widetilde{\varphi}_k$ for alternative weights
\begin{align}\label{eq:smc-weights}
    \widetilde{\bm{w}}_{\textsc{smc}} = \frac{\sum_{k=1}^K {\bm{w}}_k \, e^{j \, \widetilde{\varphi}_k}}{\lVert \sum_{k=1}^K {\bm{w}}_k \, e^{j \, \widetilde{\varphi}_k} \rVert}
    \hspace{0.4cm} \text{with} \hspace{0.4cm} 
    {\bm{w}}_k=\frac{{\bm{h}}_k^*}{\lVert {\bm{h}}_k \rVert}
\end{align}
attempting to solve the optimization problem 
\begin{align}\label{eq:phi-optimization-problem}
    \widetilde{\bm{\varphi}} = \begin{bmatrix}
        \widetilde{\varphi}_1 \\ \vdots \\ \widetilde{\varphi}_K 
    \end{bmatrix}= \argmax_{\varphi_2 \, \dots \, \varphi_K} 
    \left| \sum_{k=1}^K \bm{h}^\trp  \, \bm{w}_k \, e^{j \, \varphi_k}\right|^2 \, ,
\end{align}
which optimizes the path gain and thus the received power at the \gls{en} device.
Due to the fact that the ``true" channel $\bm{h}$ is generally unknown and measured \gls{csi} $\hat{\bm{h}}$ is unavailable in the initial access phase,~\eqref{eq:phi-optimization-problem} can only be solved approximately, 
e.g., either through a random search or a grid search.
After the initial wake-up of the \gls{en} device, feedback about the power $P_\mrm{r}$ received by the device may become available and~\eqref{eq:phi-optimization-problem} can be solved efficiently.
Note that the number of beam phases to be optimized is $K-1$, i.e., the phase of one beam can be kept constant (e.g., the \gls{los} beam, s.t. $\widetilde{\varphi}_1 = 0$) and all other beam phases are optimized.

Furthermore, a geometric environment model may not provide information on the reflection coefficients of surfaces dependent on their electromagnetic properties.
A similar optimization approach can be made to compensate for possibly unknown reflection coefficients $\refcoeff_k$.
The objective is to find optimal reflection coefficients $\widetilde{\refcoeff}_k$ that optimize
\begin{align}\label{eq:gamma-optimization-problem}
    \widetilde{\bm{\refcoeff}} = \begin{bmatrix}
        \widetilde{\refcoeff}_1 \\ \vdots \\ \widetilde{\refcoeff}_K 
    \end{bmatrix}= \argmax_{\refcoeff_2 \, \dots \, \refcoeff_K} 
    \left| 
    \frac{\sum_{k=1}^K {\bm{h}}^\herm_k(\refcoeff_k)}{\lVert \sum_{k=1}^K \bm{h}_k^\herm (\refcoeff_k) \rVert} \, \bm{h}\right|^2 \, ,
\end{align}
where the optimal phase shifts $e^{j \, \widetilde{\varphi}_k}$ in~\eqref{eq:smc-weights} can be applied to the vectors $\bm{h}^\herm_k$ before computing~\eqref{eq:gamma-optimization-problem}.
Although less crucial than the optimization of \gls{smc} beam phases, this optimization allows the \gls{bf} to vary the power directed into each beam and thus pronounce strong links and attenuate weak links.

\subsection{Reciprocity-based Beamformer}\label{sec:reciprocity-based-bf}
Only after an \gls{en} device has been woken up successfully for the first time, \gls{csi} can be obtained from a backscattered signal.
We assume to acquire a noisy channel estimate
\begin{align}\label{eq:channel-estimate}
    \hat{\bm{h}} = \bm{h} + \bm{n}_h
\end{align}
with the i.i.d. circular Gaussian noise samples  $\left[\bm{n}_h\right]_\ell \sim \mathcal{C}\mathcal{N}(0,\sigma_h^2)$.
The efficiency of a reciprocity-based \gls{bf} depends on the quality of the acquired \gls{csi}~\cite{Mishra2019}.
To express the quality of our channel estimate $\hat{\bm{h}}$, we define the channel \gls{snr} as 
\begin{align}\label{eq:channel-SNR}
    \SNR = \frac{P_\mrm{ch}}{P_n} \,,
    ~~ \text{where} ~~
    P_\mrm{ch} = \frac{1}{L} \lVert \bm{h} \rVert^2 \approx PG_\textsc{siso}
\end{align}
is the channel power (which is actually an efficiency) and $P_n = \sigma_h^2$ is the channel noise variance.
The efficiency of the reciprocity-based \gls{bf} versus channel \gls{snr} is derived in Appendix~\ref{app:CSI-loss} and performance assessments of all introduced \glspl{bf} are conducted in Section~\ref{sec:validation}.

\begin{table}[tb]
\caption{List of Measurement Parameters}
\vspace{-3mm}
\begin{center}
\begin{tabular}{ l c c c } \toprule[0.75pt] 
\textbf{Variable} & \textbf{Symbol} & \textbf{Unit} & \textbf{Value} \\[0.75pt] \hline \addlinespace[2pt] 
Carrier frequency & $f$ & \SI{}{\giga\hertz} & \SI{3.8}{}   \\
\gls{rw} (width $\times$ height) & $l_y\times l_z$ & \SI{}{\square\meter} & $(2.24 \times 1.38)$   \\
\gls{rw} position  & $\bm{p}_\mrm{RW}$ & \SI{}{\meter} & $\left[\SI{15.4}{}, \SI{-2.6}{}, \SI{3.6}{}\right]^ \trp$   \\
\gls{en} device position 1 & $\bm{p}_\mrm{EN}^{(1)}$ & \SI{}{\meter} & $\left[\SI{3.5}{}, \SI{-2}{}, \SI{1}{}\right]^ \trp$   \\
\gls{en} device position 2 & $\bm{p}_\mrm{EN}^{(2)}$ & \SI{}{\meter} & $\left[\SI{4.7}{}, \SI{-1.1}{}, \SI{1.1}{}\right]^ \trp$   \\
No. of TX antennas & $L_y \times L_z$ & - & $\num{40}{} \times \num{25}$ \\
\gls{rw} antenna spacing & $\Delta_y$,$\Delta_z$ & \SI{}{\meter} & $0.97\frac{3}{4}\lambda$ \\
No. of receiving pos. & $L_x^{\mrm{EN}} \times L_y^{\mrm{EN}}$ & - & $\num{8}{} \times \num{8}$ \\
RX SA ant. spacing & $\Delta_x^{\mrm{EN}}$,$\Delta_y^{\mrm{EN}}$ & \SI{}{\meter} & $\frac{3}{8}\lambda$ \\
\bottomrule[0.75pt]
\end{tabular}
\label{tab:meas_parameters}
\end{center}
\vspace{-7mm}
\end{table}

\section{Channel Measurements}\label{sec:measurements}

We use a Rohde \& Schwarz ZVA24 \gls{vna} in a two-port configuration to measure the transmission coefficient $S_{21,\ell}\triangleq \left[ \bm{h}\right]_\ell$ between a transmit antenna $\ell$ connected to Port\,$1$ and a receiving antenna connected to Port\,$2$. 
The transmit antenna is attached to a large mechanical positioner which forms a \gls{sa}, i.e., we subsequently measure the channel entries $\left[ \bm{h}\right]_\ell$ at desired positions $\bm{p}_\mrm{RW}^{(\ell)}$ parallel to the $yz$-plane (cf., the \gls{los} source $k=1$ in Fig.\,\ref{fig:scenario}).
Linear systematic measurement errors introduced by cables and connectors are removed by a \gls{tosm} calibration which effectively shifts the measurement reference planes to the antenna ports. 
We employ two \gls{xets} antennas~\cite{Costa2009XETS} which are \gls{uwb} antennas but we only evaluate channels at a frequency $f = \SI{3.8}{\giga\hertz}$ in this work. 
The antennas were characterized in an anechoic chamber, such that we have knowledge of their gain patterns $G(\theta,\varphi)$.
They are linearly polarized and their polarization vectors $\bm{\rho}$ are oriented parallel to the $z$-axis.
Our channel measurements still include conduction losses, dielectric losses, and matching losses introduced by the antennas.
The matching losses, i.e., $(1-|S_{ii}|^2)$ for the respective ports $i \in \{1,2\}$, are known and thus accounted for in our modeled channel vectors $\widetilde{\bm{h}}$, and conduction and dielectric losses are accounted for by the antenna gain patterns.
Inherently to using an \gls{sa}, parasitic coupling between adjacent antennas within the \gls{rw} is neglected, however, \gls{xets} antennas are generally suitable to be used in arrays~\cite{Costa2009XETS} and the spacing between our antennas $\Delta_y = \Delta_z = 0.97 \frac{3}{4}\lambda$ is reasonably large.
The measurement scenario\footnote{A more comprehensive documentation of the measurement scenario and the measurement testbed can be found in~\cite{D1_2}.} depicted in Fig.\,\ref{fig:scenario} is a long hallway where the synthetic \gls{rw} is mounted on a bridge between two concrete walls. 
Measurement and scenario-specific parameters are summarized in Table\,\ref{tab:meas_parameters}.
On the backside of the transmit antenna, there is a pyramidal absorber mounted which effectively removes the backlobe of the transmit antenna facing towards the positive $x$-direction and strongly attenuates any reflections caused by the bridge. 
Our synthetic \gls{rw} has a total of $L_y \times L_z = (40 \times 25)$ antennas and forms a physical aperture of $l_y\times l_z = (2.24 \times 1.38)\SI{}{\metre\squared}$. 
Lidar measurements of the hallway served as ground truth to generate the geometric model in Fig.\,\ref{fig:scenario}. 
We choose the two sidewalls and the floor as representative \glspl{smc} and use the corresponding $K=4$ mirror sources including the \gls{los} for predicting channels with the \gls{sw} \gls{smc} \gls{bf} from Section~\ref{sec:plane-wave-beamformer}.
\ifdefined\reduceSize  
\else
    The process of mirroring of the \gls{rw} across walls is described in Appendix~\ref{app:householder}.
\fi
The receiving antenna is moved by another mechanical positioner in a $\frac{3}{8}\lambda$-spaced grid parallel to the $xy$-plane to capture the spatial distribution of power around one antenna position chosen as hypothetical \gls{en} device. 
The measurements are conducted at two positions $\bm{p}_\mrm{EN}^{(1)}$ and $\bm{p}_\mrm{EN}^{(2)}$ in the hallway, where the latter is less favorable for beamforming due a metal shelf behind it, which is filled with water bottles and has a reflective metal grid mounted on its backside facing the negative $x$-direction.
The measurement positions are located at distances 
$\lVert\bm{r}^{(1)}\rVert \approx \SI{12.3}{\metre}$ and $\lVert\bm{r}^{(2)}\rVert \approx \SI{11.1}{\metre}$ from the center of gravity of the \gls{rw}.

\subsection{Analysis of \glspl{smc}}
We assume that our measured channels $[\bm{h}]_\ell$ are the ``true"\footnote{We make this assumption knowing that the ``true" value of a measurand is generally unknown and our corrected measurement result is merely its best estimate~\cite{GUM}.} channels given that we have a reasonably high measurement \gls{snr}.
We compute separate weights $\bm{w}_k$ as defined in~\eqref{eq:smc-weights} to analyze the path gain $PG_k$ for each of the $K$ \gls{smc} beams.
We apply these weights on the measured channel vector $\bm{h}$ and compute the achievable path gain for every \gls{smc} beam through \gls{mrt}.
The individual power budgets $PG_k = |\bm{h}^\trp\bm{w}_k|^2$ 
for each of these beams at measurement position $\bm{p}_\mrm{EN}^{(1)}$ are indicated in Table\,\ref{tab:SMC-parameters}. 
The \gls{smc} $k=2$ corresponds to the image source from the floor, $k=3$ to the reflection in the negative $y$-direction, and $k=4$ to the reflection in the positive $y$-direction.
With $PG_1\approx \SI{-33.4}{\dB}$, the \gls{los} is the strongest of the analyzed components and closely followed by $PG_3\approx \SI{-36.6}{\dB}$ which is a quite strong component, given that it suffers reflection losses at the wall, longer distances to the \gls{en} device, and lower antenna gains. 
Component $k=4$ is comparably weak, because of the limited visibility of the wall: only $141$ out of the $L=1000$ \gls{rw} antennas of \gls{smc} $4$ are visible from the \gls{en} device due to the limited extent of the respective wall, while all $L$ antennas of \gls{smc} $k=3$ are visible.
Furthermore, the optimization problems in~\eqref{eq:phi-optimization-problem} and~\eqref{eq:gamma-optimization-problem} have been solved numerically to compute the optimum phase shifts $\widetilde{\varphi}_k$ and reflection coefficients $\widetilde{\refcoeff}_k$.
After the optimization, we get a modeled reflection coefficient of $\widetilde{\refcoeff}_3 = \SI{-2.28}{\dB}$.
\begin{table}[tb!]
\caption{\gls{smc} Power Budgets and Optimized Parameters}
\vspace{-3mm}
\begin{center}
\begin{tabular}{ l c c | c c c c} \toprule[0.75pt]
\textbf{Variable} & \textbf{Symbol} & \textbf{Unit} & \multicolumn{4}{c}{\textbf{Values}}  \\[0.75pt] \hline \addlinespace[2pt] 
No.\,\gls{smc} & $k$ & - & \SI{1}{} & \SI{2}{}& \SI{3}{}& \SI{4}{}   \\
Path gain & $PG_k$ & \SI{}{\dB} & \SI{-33.4}{} & \SI{-52.7}{}& \SI{-36.6}{}& \SI{-49.0}{}   \\
Refl. coeff.& $\widetilde{\refcoeff}_{k}$ & \SI{}{\dB} & \SI{0}{} & \SI{-43.32}{}& \SI{-2.28}{}& \SI{-4.8}{}   \\
    Phase shifts & $\widetilde{\varphi}_k$ & \SI{}{\degree} & \SI{0}{} & \SI{118}{}& \SI{182}{}& \SI{212}{} \\
\bottomrule[0.75pt]
\end{tabular}
\label{tab:SMC-parameters}
\end{center}
\vspace{-7mm}
\end{table}
Due to the fact that the power-optimal weights are found by computing them through the assumed ``true" channel $\bm{h}$, this optimized value would correspond closely to the ``true" reflection coefficients of the walls if the other channel parameters were known sufficiently well. 
The optimized reflection coefficient $\widetilde{\refcoeff}_4 = \SI{-4.8}{\dB}$ may therefore be a result of mismodeling, since we neglect diffraction effects in our visibility model and most of the visible part of the respective image source is located at the edge of the wall it is mirrored across.
From Table\,\ref{tab:SMC-parameters} it is evident that component $k=2$ is particularly weak, i.e., $PG_2$ is small compared with the other beams. 
Reflection coefficients of reflections at specular surfaces depend on the incidence angle of the waves and the electromagnetic parameters of the materials they are made of, particularly the relative permittivity of the concrete walls in our scenario.
It is worth noting that the transmit and receive antennas of our \gls{sa} measurement testbed have polarization vectors $\bm{\rho}$ aligned in parallel with the $z$-axis, which are thus parallel to the normal vector of the floor.
The reflection coefficient for such arrangements is described by~\cite[eq.\,(4-125)]{balanis}, which yields low values for typical relative permittivities of concrete (e.g., $\epsilon_\mrm{r} \approx 6$~\cite{Olkkonen13ConcretePermittivity}) and is generally unfavorable when compared to waves polarized orthogonal to the normal vector of specular surfaces.
Polarization-dependent reflection coefficients, however, are not regarded in our channel model, thus the optimized value of $\widetilde{\refcoeff}_2$ corresponds to the unfavorable polarization plane of our antennas w.r.t. the floor.
\begin{figure}[tb]
    \setlength{\figurewidth}{0.88\linewidth}
    \setlength{\figureheight}{0.8\linewidth}
    \hspace{-3mm}\input{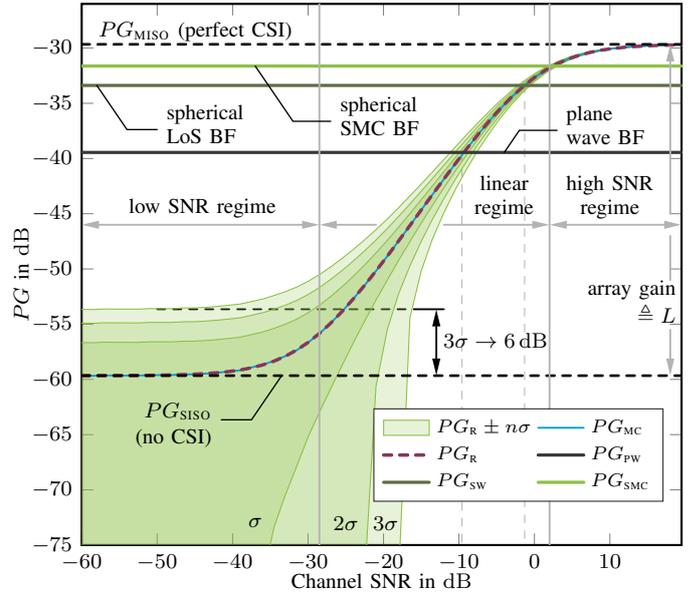} 
     \caption{Comparison of the \glspl{bf} introduced in Section~\ref{sec:beamforming-strategies} when applied on a measured channel vector $\bm{h}$ at $\bm{p}_\mrm{EN}^{(1)}$.
     The reciprocity-based \gls{bf} is evaluated by means of an MC analysis (i.e., $PG_\mrm{MC}$) and the analytical expression in~\eqref{eq:PG-reciprocity-analytical} (i.e., $PG_\textsc{r}$) with the respective \gls{snr} regimes in~\eqref{eq:PG-reciprocity-regimes}.}
     \label{fig:BF_comparison} \vspace{-5mm}
 \end{figure}

\setlength{\figurewidth}{0.4\linewidth}
\setlength{\figureheight}{0.4\linewidth}
\setlength{\hspaceing}{-1.31cm}
\begin{figure*}[bt]	
     \hspace{-4mm}
     \begin{subfigure}[t]{0.29\textwidth}
        \vskip 0pt	
         \centering
	    \ifdefined\ExcludeFigures   
	                \ExcludeFigures
        \else
%
%
\definecolor{mycolor1}{rgb}{0.21961,0.21961,0.21961}%

\pgfplotsset{every axis/.append style={
  label style={font=\footnotesize},
  legend style={font=\footnotesize},
  tick label style={font=\footnotesize},
}}

\begin{tikzpicture}

\begin{axis}[%
width=\figurewidth,
height=\figureheight,
at={(0\figurewidth,0\figureheight)},
axis line style = thick,	
scale only axis,
point meta min=-60,
point meta max=-30,
axis on top,
xmin=3.37290652572368,
xmax=3.58,
xlabel style={yshift=1.5mm},
xlabel={$x$ in \SI{}{\metre}},
ymin=-2.139,
ymax=-1.93190652572368,
ylabel style={yshift=-6pt},
ylabel={$y$ in \SI{}{\metre}},
xmajorgrids,
ymajorgrids,
grid style=dotted,
axis background/.style={fill=white},
colormap={mymap}{[1pt] rgb(0pt)=(0.995023,0.995023,0.995023); rgb(73pt)=(0.567253,0.659744,0.995023); rgb(74pt)=(0.561393,0.655151,0.995023); rgb(75pt)=(0.555533,0.650558,0.995023); rgb(76pt)=(0.549674,0.645965,0.995023); rgb(77pt)=(0.543814,0.641373,0.995023); rgb(78pt)=(0.537954,0.63678,0.995023); rgb(79pt)=(0.532094,0.632187,0.995023); rgb(80pt)=(0.526234,0.627594,0.995023); rgb(81pt)=(0.520374,0.623001,0.995023); rgb(82pt)=(0.514514,0.618408,0.995023); rgb(83pt)=(0.508655,0.613815,0.995023); rgb(84pt)=(0.502795,0.609223,0.995023); rgb(85pt)=(0.496935,0.60463,0.995023); rgb(86pt)=(0.491075,0.600037,0.995023); rgb(87pt)=(0.485215,0.595444,0.995023); rgb(88pt)=(0.479355,0.590851,0.995023); rgb(89pt)=(0.473495,0.586258,0.995023); rgb(90pt)=(0.467635,0.581665,0.995023); rgb(91pt)=(0.461776,0.577072,0.995023); rgb(92pt)=(0.455916,0.57248,0.995023); rgb(93pt)=(0.450056,0.567887,0.995023); rgb(94pt)=(0.444196,0.563294,0.995023); rgb(95pt)=(0.438336,0.558701,0.995023); rgb(96pt)=(0.432476,0.554108,0.995024); rgb(97pt)=(0.426616,0.549515,0.995024); rgb(98pt)=(0.420757,0.544922,0.995023); rgb(99pt)=(0.414897,0.540329,0.995023); rgb(100pt)=(0.409037,0.535737,0.995023); rgb(101pt)=(0.403177,0.531144,0.995023); rgb(102pt)=(0.397316,0.526551,0.995025); rgb(103pt)=(0.391455,0.521958,0.995025); rgb(104pt)=(0.385597,0.517365,0.995024); rgb(105pt)=(0.379741,0.512772,0.99502); rgb(106pt)=(0.373885,0.508178,0.995017); rgb(107pt)=(0.368022,0.503586,0.995019); rgb(108pt)=(0.362148,0.498995,0.995033); rgb(109pt)=(0.356273,0.494405,0.995048); rgb(110pt)=(0.350418,0.489811,0.995042); rgb(111pt)=(0.344605,0.48521,0.994998); rgb(112pt)=(0.338806,0.480607,0.994939); rgb(113pt)=(0.332944,0.476014,0.994941); rgb(114pt)=(0.326938,0.471447,0.995081); rgb(115pt)=(0.32084,0.466896,0.99531); rgb(116pt)=(0.31493,0.462312,0.995359); rgb(117pt)=(0.309513,0.457641,0.994933); rgb(118pt)=(0.304551,0.45289,0.994071); rgb(119pt)=(0.299066,0.448231,0.993711); rgb(120pt)=(0.29192,0.443865,0.994946); rgb(121pt)=(0.282789,0.439848,0.998086); rgb(122pt)=(0.274935,0.435607,1); rgb(123pt)=(0.272609,0.430392,0.996609); rgb(124pt)=(0.279306,0.423587,0.984555); rgb(125pt)=(0.293138,0.415525,0.965651); rgb(126pt)=(0.309867,0.406953,0.943967); rgb(127pt)=(0.325491,0.398575,0.923344); rgb(128pt)=(0.339058,0.39056,0.904696); rgb(129pt)=(0.351668,0.382714,0.886966); rgb(130pt)=(0.364431,0.37484,0.869089); rgb(131pt)=(0.377772,0.366865,0.850657); rgb(132pt)=(0.391418,0.358836,0.831932); rgb(133pt)=(0.405066,0.350807,0.813205); rgb(134pt)=(0.418557,0.342805,0.79463); rgb(135pt)=(0.431954,0.33482,0.776145); rgb(136pt)=(0.445337,0.326837,0.757672); rgb(137pt)=(0.458761,0.318847,0.739161); rgb(138pt)=(0.472215,0.310852,0.720621); rgb(139pt)=(0.485675,0.302856,0.702075); rgb(140pt)=(0.499125,0.294862,0.683539); rgb(141pt)=(0.512566,0.286869,0.665011); rgb(142pt)=(0.526005,0.278876,0.646485); rgb(143pt)=(0.539446,0.270883,0.627958); rgb(144pt)=(0.552889,0.26289,0.609428); rgb(145pt)=(0.566334,0.254896,0.590897); rgb(146pt)=(0.579778,0.246903,0.572366); rgb(147pt)=(0.593221,0.23891,0.553836); rgb(148pt)=(0.606664,0.230917,0.535307); rgb(149pt)=(0.620107,0.222923,0.516777); rgb(150pt)=(0.63355,0.21493,0.498247); rgb(151pt)=(0.646993,0.206937,0.479717); rgb(152pt)=(0.660437,0.198944,0.461187); rgb(153pt)=(0.67388,0.190951,0.442658); rgb(154pt)=(0.687323,0.182958,0.424128); rgb(155pt)=(0.700766,0.174963,0.405598); rgb(156pt)=(0.71421,0.166966,0.387067); rgb(157pt)=(0.727654,0.158972,0.368536); rgb(158pt)=(0.741096,0.150985,0.350008); rgb(159pt)=(0.754536,0.143004,0.331483); rgb(160pt)=(0.767977,0.135017,0.312955); rgb(161pt)=(0.781425,0.127007,0.29442); rgb(162pt)=(0.794879,0.118969,0.275874); rgb(163pt)=(0.808331,0.110944,0.257333); rgb(164pt)=(0.821763,0.102994,0.238819); rgb(165pt)=(0.835166,0.0951591,0.220344); rgb(166pt)=(0.848574,0.0873055,0.201863); rgb(167pt)=(0.862043,0.0792099,0.183297); rgb(168pt)=(0.875621,0.0706838,0.164581); rgb(169pt)=(0.889214,0.0621028,0.145846); rgb(170pt)=(0.902611,0.0542922,0.12738); rgb(171pt)=(0.915613,0.0480374,0.109458); rgb(172pt)=(0.928457,0.0424061,0.091754); rgb(173pt)=(0.941904,0.0343962,0.0732183); rgb(174pt)=(0.956736,0.0209305,0.052775); rgb(175pt)=(0.9725,0.00378572,0.0310453); rgb(176pt)=(0.986483,0.00633329,0.0117721); rgb(177pt)=(0.995725,0.0022327,0.000967649); rgb(178pt)=(0.998618,0.0358306,0.00495484); rgb(179pt)=(0.997427,0.0855268,0.00331317); rgb(180pt)=(0.995111,0.139657,0.000121229); rgb(181pt)=(0.99409,0.188685,0.00128695); rgb(182pt)=(0.994288,0.232904,0.00101351); rgb(183pt)=(0.994914,0.275439,0.000151196); rgb(184pt)=(0.995257,0.319085,0.000322596); rgb(185pt)=(0.995242,0.364147,0.000301319); rgb(186pt)=(0.995078,0.409793,7.57419e-05); rgb(187pt)=(0.994967,0.455231,7.71666e-05); rgb(188pt)=(0.99496,0.500262,8.74378e-05); rgb(189pt)=(0.995001,0.5451,3.04168e-05); rgb(190pt)=(0.995036,0.589964,1.73228e-05); rgb(191pt)=(0.995042,0.634943,2.51464e-05); rgb(192pt)=(0.995032,0.679984,1.1303e-05); rgb(193pt)=(0.995021,0.725028,3.71976e-06); rgb(194pt)=(0.995017,0.770042,8.33655e-06); rgb(195pt)=(0.99502,0.815035,5.36005e-06); rgb(196pt)=(0.995024,0.860017,1.0858e-06); rgb(197pt)=(0.995028,0.905002,6.87705e-06); rgb(198pt)=(0.995029,0.95,7.88977e-06); rgb(199pt)=(0.995023,0.995023,0)},
]
\addplot [forget plot] graphics [xmin=3.58020750849126, xmax=3.37269901723243, ymin=-2.13920750849126, ymax=-1.93169901723243] {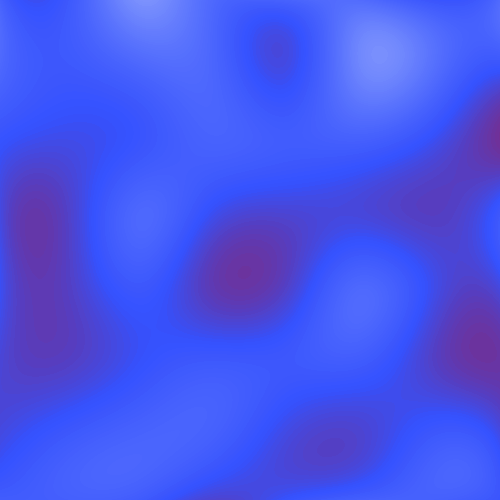};
\addplot[only marks, mark=*, mark options={}, mark size=1.5000pt, color=black, fill=RDlightgreen, forget plot] table[row sep=crcr]{%
x	y\\
3.46166087184211	-2.05024565388158\\
};
\draw[line width = 0.7pt, {Stealth[inset=0pt, scale=1.05, angle'=25]-},shorten < = 1mm] (3.462,-2.05) -- ++(0mm,14mm) node[anchor=south,font={\footnotesize},fill=white,opacity=0.75,inner sep=0.75mm] {\SI{-39.45}{\dB}};
\end{axis}
\end{tikzpicture}
        \fi 
        \vspace{-2.1mm}
         \caption{\Gls{pw} \gls{los} \gls{bf}}
         \label{fig:PW_pos1_switched0}
     \end{subfigure}%
     \hspace{-0.78cm}%
     \begin{subfigure}[t]{0.29\textwidth}
        \vskip 0pt	
         \centering
	    \ifdefined\ExcludeFigures   
            \ExcludeFigures
        \else
%
%
\definecolor{mycolor1}{rgb}{0.00000,0.44700,0.74100}%

\pgfplotsset{every axis/.append style={
  label style={font=\footnotesize},
  legend style={font=\footnotesize},
  tick label style={font=\footnotesize},
}}

\begin{tikzpicture}

\begin{axis}[%
width=\figurewidth,
height=\figureheight,
at={(0\figurewidth,0\figureheight)},
axis line style = thick,	
scale only axis,
point meta min=-60,
point meta max=-30,
axis on top,
xmin=3.37290652572368,
xmax=3.58,
xlabel style={yshift=1.5mm},
xlabel={$x$ in \SI{}{\metre}},
ymin=-2.139,
ymax=-1.93190652572368,
yticklabels=\empty,
xmajorgrids,
ymajorgrids,
grid style=dotted,
axis background/.style={fill=white},
colormap={mymap}{[1pt] rgb(0pt)=(0.995023,0.995023,0.995023); rgb(73pt)=(0.567253,0.659744,0.995023); rgb(74pt)=(0.561393,0.655151,0.995023); rgb(75pt)=(0.555533,0.650558,0.995023); rgb(76pt)=(0.549674,0.645965,0.995023); rgb(77pt)=(0.543814,0.641373,0.995023); rgb(78pt)=(0.537954,0.63678,0.995023); rgb(79pt)=(0.532094,0.632187,0.995023); rgb(80pt)=(0.526234,0.627594,0.995023); rgb(81pt)=(0.520374,0.623001,0.995023); rgb(82pt)=(0.514514,0.618408,0.995023); rgb(83pt)=(0.508655,0.613815,0.995023); rgb(84pt)=(0.502795,0.609223,0.995023); rgb(85pt)=(0.496935,0.60463,0.995023); rgb(86pt)=(0.491075,0.600037,0.995023); rgb(87pt)=(0.485215,0.595444,0.995023); rgb(88pt)=(0.479355,0.590851,0.995023); rgb(89pt)=(0.473495,0.586258,0.995023); rgb(90pt)=(0.467635,0.581665,0.995023); rgb(91pt)=(0.461776,0.577072,0.995023); rgb(92pt)=(0.455916,0.57248,0.995023); rgb(93pt)=(0.450056,0.567887,0.995023); rgb(94pt)=(0.444196,0.563294,0.995023); rgb(95pt)=(0.438336,0.558701,0.995023); rgb(96pt)=(0.432476,0.554108,0.995024); rgb(97pt)=(0.426616,0.549515,0.995024); rgb(98pt)=(0.420757,0.544922,0.995023); rgb(99pt)=(0.414897,0.540329,0.995023); rgb(100pt)=(0.409037,0.535737,0.995023); rgb(101pt)=(0.403177,0.531144,0.995023); rgb(102pt)=(0.397316,0.526551,0.995025); rgb(103pt)=(0.391455,0.521958,0.995025); rgb(104pt)=(0.385597,0.517365,0.995024); rgb(105pt)=(0.379741,0.512772,0.99502); rgb(106pt)=(0.373885,0.508178,0.995017); rgb(107pt)=(0.368022,0.503586,0.995019); rgb(108pt)=(0.362148,0.498995,0.995033); rgb(109pt)=(0.356273,0.494405,0.995048); rgb(110pt)=(0.350418,0.489811,0.995042); rgb(111pt)=(0.344605,0.48521,0.994998); rgb(112pt)=(0.338806,0.480607,0.994939); rgb(113pt)=(0.332944,0.476014,0.994941); rgb(114pt)=(0.326938,0.471447,0.995081); rgb(115pt)=(0.32084,0.466896,0.99531); rgb(116pt)=(0.31493,0.462312,0.995359); rgb(117pt)=(0.309513,0.457641,0.994933); rgb(118pt)=(0.304551,0.45289,0.994071); rgb(119pt)=(0.299066,0.448231,0.993711); rgb(120pt)=(0.29192,0.443865,0.994946); rgb(121pt)=(0.282789,0.439848,0.998086); rgb(122pt)=(0.274935,0.435607,1); rgb(123pt)=(0.272609,0.430392,0.996609); rgb(124pt)=(0.279306,0.423587,0.984555); rgb(125pt)=(0.293138,0.415525,0.965651); rgb(126pt)=(0.309867,0.406953,0.943967); rgb(127pt)=(0.325491,0.398575,0.923344); rgb(128pt)=(0.339058,0.39056,0.904696); rgb(129pt)=(0.351668,0.382714,0.886966); rgb(130pt)=(0.364431,0.37484,0.869089); rgb(131pt)=(0.377772,0.366865,0.850657); rgb(132pt)=(0.391418,0.358836,0.831932); rgb(133pt)=(0.405066,0.350807,0.813205); rgb(134pt)=(0.418557,0.342805,0.79463); rgb(135pt)=(0.431954,0.33482,0.776145); rgb(136pt)=(0.445337,0.326837,0.757672); rgb(137pt)=(0.458761,0.318847,0.739161); rgb(138pt)=(0.472215,0.310852,0.720621); rgb(139pt)=(0.485675,0.302856,0.702075); rgb(140pt)=(0.499125,0.294862,0.683539); rgb(141pt)=(0.512566,0.286869,0.665011); rgb(142pt)=(0.526005,0.278876,0.646485); rgb(143pt)=(0.539446,0.270883,0.627958); rgb(144pt)=(0.552889,0.26289,0.609428); rgb(145pt)=(0.566334,0.254896,0.590897); rgb(146pt)=(0.579778,0.246903,0.572366); rgb(147pt)=(0.593221,0.23891,0.553836); rgb(148pt)=(0.606664,0.230917,0.535307); rgb(149pt)=(0.620107,0.222923,0.516777); rgb(150pt)=(0.63355,0.21493,0.498247); rgb(151pt)=(0.646993,0.206937,0.479717); rgb(152pt)=(0.660437,0.198944,0.461187); rgb(153pt)=(0.67388,0.190951,0.442658); rgb(154pt)=(0.687323,0.182958,0.424128); rgb(155pt)=(0.700766,0.174963,0.405598); rgb(156pt)=(0.71421,0.166966,0.387067); rgb(157pt)=(0.727654,0.158972,0.368536); rgb(158pt)=(0.741096,0.150985,0.350008); rgb(159pt)=(0.754536,0.143004,0.331483); rgb(160pt)=(0.767977,0.135017,0.312955); rgb(161pt)=(0.781425,0.127007,0.29442); rgb(162pt)=(0.794879,0.118969,0.275874); rgb(163pt)=(0.808331,0.110944,0.257333); rgb(164pt)=(0.821763,0.102994,0.238819); rgb(165pt)=(0.835166,0.0951591,0.220344); rgb(166pt)=(0.848574,0.0873055,0.201863); rgb(167pt)=(0.862043,0.0792099,0.183297); rgb(168pt)=(0.875621,0.0706838,0.164581); rgb(169pt)=(0.889214,0.0621028,0.145846); rgb(170pt)=(0.902611,0.0542922,0.12738); rgb(171pt)=(0.915613,0.0480374,0.109458); rgb(172pt)=(0.928457,0.0424061,0.091754); rgb(173pt)=(0.941904,0.0343962,0.0732183); rgb(174pt)=(0.956736,0.0209305,0.052775); rgb(175pt)=(0.9725,0.00378572,0.0310453); rgb(176pt)=(0.986483,0.00633329,0.0117721); rgb(177pt)=(0.995725,0.0022327,0.000967649); rgb(178pt)=(0.998618,0.0358306,0.00495484); rgb(179pt)=(0.997427,0.0855268,0.00331317); rgb(180pt)=(0.995111,0.139657,0.000121229); rgb(181pt)=(0.99409,0.188685,0.00128695); rgb(182pt)=(0.994288,0.232904,0.00101351); rgb(183pt)=(0.994914,0.275439,0.000151196); rgb(184pt)=(0.995257,0.319085,0.000322596); rgb(185pt)=(0.995242,0.364147,0.000301319); rgb(186pt)=(0.995078,0.409793,7.57419e-05); rgb(187pt)=(0.994967,0.455231,7.71666e-05); rgb(188pt)=(0.99496,0.500262,8.74378e-05); rgb(189pt)=(0.995001,0.5451,3.04168e-05); rgb(190pt)=(0.995036,0.589964,1.73228e-05); rgb(191pt)=(0.995042,0.634943,2.51464e-05); rgb(192pt)=(0.995032,0.679984,1.1303e-05); rgb(193pt)=(0.995021,0.725028,3.71976e-06); rgb(194pt)=(0.995017,0.770042,8.33655e-06); rgb(195pt)=(0.99502,0.815035,5.36005e-06); rgb(196pt)=(0.995024,0.860017,1.0858e-06); rgb(197pt)=(0.995028,0.905002,6.87705e-06); rgb(198pt)=(0.995029,0.95,7.88977e-06); rgb(199pt)=(0.995023,0.995023,0)},
]
\addplot [forget plot] graphics [xmin=3.58020750849126, xmax=3.37269901723243, ymin=-2.13920750849126, ymax=-1.93169901723243] {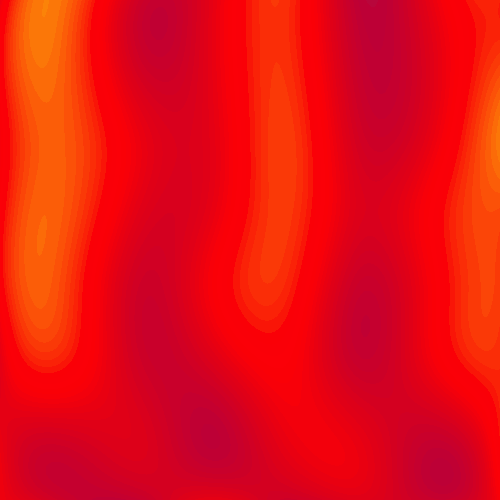};
\addplot[only marks, mark=*, mark options={}, mark size=1.5000pt, color=black, fill=RDlightgreen, forget plot] table[row sep=crcr]{%
x	y\\
3.46166087184211	-2.05024565388158\\
};

\draw[line width = 0.7pt, {Stealth[inset=0pt, scale=1.05, angle'=25]-},shorten < = 1mm] (3.462,-2.05) -- ++(0mm,14mm) node[anchor=south,font={\footnotesize},fill=white,opacity=0.75,inner sep=0.75mm] {\SI{-33.38}{\dB}};

\end{axis}
\end{tikzpicture}%
        \fi 
        \vspace{-2.1mm}
         \caption{\Gls{sw} \gls{los} \gls{bf}}
         \label{fig:SW_LoS_pos1_switched0}
     \end{subfigure}%
     \hspace{\hspaceing}%
     \begin{subfigure}[t]{0.29\textwidth}
         \vskip 0pt	
         \centering
	    \ifdefined\ExcludeFigures   
            \ExcludeFigures
        \else
%
%
\definecolor{mycolor1}{rgb}{0.36471,0.42745,0.26667}%

\pgfplotsset{every axis/.append style={
  label style={font=\footnotesize},
  legend style={font=\footnotesize},
  tick label style={font=\footnotesize},
}}

\begin{tikzpicture}

\begin{axis}[%
width=\figurewidth,
height=\figureheight,
at={(0\figurewidth,0\figureheight)},
axis line style = thick,	
scale only axis,
point meta min=-60,
point meta max=-30,
axis on top,
xmin=3.37290652572368,
xmax=3.58,
xlabel style={yshift=1.5mm},
xlabel={$x$ in \SI{}{\metre}},
ymin=-2.139,
ymax=-1.93190652572368,
yticklabels=\empty,
xmajorgrids,
ymajorgrids,
grid style=dotted,
axis background/.style={fill=white},
colormap={mymap}{[1pt] rgb(0pt)=(0.995023,0.995023,0.995023); rgb(73pt)=(0.567253,0.659744,0.995023); rgb(74pt)=(0.561393,0.655151,0.995023); rgb(75pt)=(0.555533,0.650558,0.995023); rgb(76pt)=(0.549674,0.645965,0.995023); rgb(77pt)=(0.543814,0.641373,0.995023); rgb(78pt)=(0.537954,0.63678,0.995023); rgb(79pt)=(0.532094,0.632187,0.995023); rgb(80pt)=(0.526234,0.627594,0.995023); rgb(81pt)=(0.520374,0.623001,0.995023); rgb(82pt)=(0.514514,0.618408,0.995023); rgb(83pt)=(0.508655,0.613815,0.995023); rgb(84pt)=(0.502795,0.609223,0.995023); rgb(85pt)=(0.496935,0.60463,0.995023); rgb(86pt)=(0.491075,0.600037,0.995023); rgb(87pt)=(0.485215,0.595444,0.995023); rgb(88pt)=(0.479355,0.590851,0.995023); rgb(89pt)=(0.473495,0.586258,0.995023); rgb(90pt)=(0.467635,0.581665,0.995023); rgb(91pt)=(0.461776,0.577072,0.995023); rgb(92pt)=(0.455916,0.57248,0.995023); rgb(93pt)=(0.450056,0.567887,0.995023); rgb(94pt)=(0.444196,0.563294,0.995023); rgb(95pt)=(0.438336,0.558701,0.995023); rgb(96pt)=(0.432476,0.554108,0.995024); rgb(97pt)=(0.426616,0.549515,0.995024); rgb(98pt)=(0.420757,0.544922,0.995023); rgb(99pt)=(0.414897,0.540329,0.995023); rgb(100pt)=(0.409037,0.535737,0.995023); rgb(101pt)=(0.403177,0.531144,0.995023); rgb(102pt)=(0.397316,0.526551,0.995025); rgb(103pt)=(0.391455,0.521958,0.995025); rgb(104pt)=(0.385597,0.517365,0.995024); rgb(105pt)=(0.379741,0.512772,0.99502); rgb(106pt)=(0.373885,0.508178,0.995017); rgb(107pt)=(0.368022,0.503586,0.995019); rgb(108pt)=(0.362148,0.498995,0.995033); rgb(109pt)=(0.356273,0.494405,0.995048); rgb(110pt)=(0.350418,0.489811,0.995042); rgb(111pt)=(0.344605,0.48521,0.994998); rgb(112pt)=(0.338806,0.480607,0.994939); rgb(113pt)=(0.332944,0.476014,0.994941); rgb(114pt)=(0.326938,0.471447,0.995081); rgb(115pt)=(0.32084,0.466896,0.99531); rgb(116pt)=(0.31493,0.462312,0.995359); rgb(117pt)=(0.309513,0.457641,0.994933); rgb(118pt)=(0.304551,0.45289,0.994071); rgb(119pt)=(0.299066,0.448231,0.993711); rgb(120pt)=(0.29192,0.443865,0.994946); rgb(121pt)=(0.282789,0.439848,0.998086); rgb(122pt)=(0.274935,0.435607,1); rgb(123pt)=(0.272609,0.430392,0.996609); rgb(124pt)=(0.279306,0.423587,0.984555); rgb(125pt)=(0.293138,0.415525,0.965651); rgb(126pt)=(0.309867,0.406953,0.943967); rgb(127pt)=(0.325491,0.398575,0.923344); rgb(128pt)=(0.339058,0.39056,0.904696); rgb(129pt)=(0.351668,0.382714,0.886966); rgb(130pt)=(0.364431,0.37484,0.869089); rgb(131pt)=(0.377772,0.366865,0.850657); rgb(132pt)=(0.391418,0.358836,0.831932); rgb(133pt)=(0.405066,0.350807,0.813205); rgb(134pt)=(0.418557,0.342805,0.79463); rgb(135pt)=(0.431954,0.33482,0.776145); rgb(136pt)=(0.445337,0.326837,0.757672); rgb(137pt)=(0.458761,0.318847,0.739161); rgb(138pt)=(0.472215,0.310852,0.720621); rgb(139pt)=(0.485675,0.302856,0.702075); rgb(140pt)=(0.499125,0.294862,0.683539); rgb(141pt)=(0.512566,0.286869,0.665011); rgb(142pt)=(0.526005,0.278876,0.646485); rgb(143pt)=(0.539446,0.270883,0.627958); rgb(144pt)=(0.552889,0.26289,0.609428); rgb(145pt)=(0.566334,0.254896,0.590897); rgb(146pt)=(0.579778,0.246903,0.572366); rgb(147pt)=(0.593221,0.23891,0.553836); rgb(148pt)=(0.606664,0.230917,0.535307); rgb(149pt)=(0.620107,0.222923,0.516777); rgb(150pt)=(0.63355,0.21493,0.498247); rgb(151pt)=(0.646993,0.206937,0.479717); rgb(152pt)=(0.660437,0.198944,0.461187); rgb(153pt)=(0.67388,0.190951,0.442658); rgb(154pt)=(0.687323,0.182958,0.424128); rgb(155pt)=(0.700766,0.174963,0.405598); rgb(156pt)=(0.71421,0.166966,0.387067); rgb(157pt)=(0.727654,0.158972,0.368536); rgb(158pt)=(0.741096,0.150985,0.350008); rgb(159pt)=(0.754536,0.143004,0.331483); rgb(160pt)=(0.767977,0.135017,0.312955); rgb(161pt)=(0.781425,0.127007,0.29442); rgb(162pt)=(0.794879,0.118969,0.275874); rgb(163pt)=(0.808331,0.110944,0.257333); rgb(164pt)=(0.821763,0.102994,0.238819); rgb(165pt)=(0.835166,0.0951591,0.220344); rgb(166pt)=(0.848574,0.0873055,0.201863); rgb(167pt)=(0.862043,0.0792099,0.183297); rgb(168pt)=(0.875621,0.0706838,0.164581); rgb(169pt)=(0.889214,0.0621028,0.145846); rgb(170pt)=(0.902611,0.0542922,0.12738); rgb(171pt)=(0.915613,0.0480374,0.109458); rgb(172pt)=(0.928457,0.0424061,0.091754); rgb(173pt)=(0.941904,0.0343962,0.0732183); rgb(174pt)=(0.956736,0.0209305,0.052775); rgb(175pt)=(0.9725,0.00378572,0.0310453); rgb(176pt)=(0.986483,0.00633329,0.0117721); rgb(177pt)=(0.995725,0.0022327,0.000967649); rgb(178pt)=(0.998618,0.0358306,0.00495484); rgb(179pt)=(0.997427,0.0855268,0.00331317); rgb(180pt)=(0.995111,0.139657,0.000121229); rgb(181pt)=(0.99409,0.188685,0.00128695); rgb(182pt)=(0.994288,0.232904,0.00101351); rgb(183pt)=(0.994914,0.275439,0.000151196); rgb(184pt)=(0.995257,0.319085,0.000322596); rgb(185pt)=(0.995242,0.364147,0.000301319); rgb(186pt)=(0.995078,0.409793,7.57419e-05); rgb(187pt)=(0.994967,0.455231,7.71666e-05); rgb(188pt)=(0.99496,0.500262,8.74378e-05); rgb(189pt)=(0.995001,0.5451,3.04168e-05); rgb(190pt)=(0.995036,0.589964,1.73228e-05); rgb(191pt)=(0.995042,0.634943,2.51464e-05); rgb(192pt)=(0.995032,0.679984,1.1303e-05); rgb(193pt)=(0.995021,0.725028,3.71976e-06); rgb(194pt)=(0.995017,0.770042,8.33655e-06); rgb(195pt)=(0.99502,0.815035,5.36005e-06); rgb(196pt)=(0.995024,0.860017,1.0858e-06); rgb(197pt)=(0.995028,0.905002,6.87705e-06); rgb(198pt)=(0.995029,0.95,7.88977e-06); rgb(199pt)=(0.995023,0.995023,0)},
]
\addplot [forget plot] graphics [xmin=3.58020750849126, xmax=3.37269901723243, ymin=-2.13920750849126, ymax=-1.93169901723243] {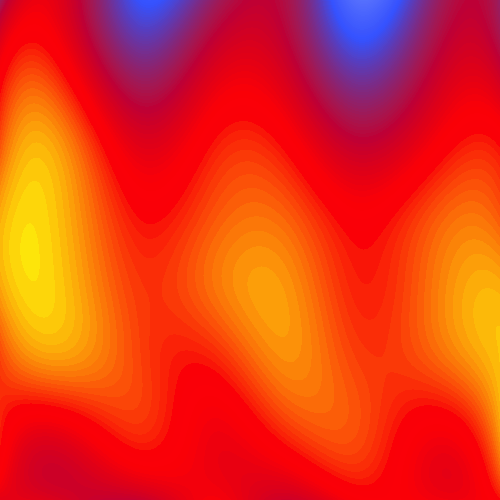};
\addplot[only marks, mark=*, mark options={}, mark size=1.5000pt, color=black, fill=RDlightgreen, forget plot] table[row sep=crcr]{%
x	y\\
3.46166087184211	-2.05024565388158\\
};

\draw[line width = 0.7pt, {Stealth[inset=0pt, scale=1.05, angle'=25]-},shorten < = 1mm] (3.462,-2.05) -- ++(0mm,14mm) node[anchor=south,font={\footnotesize},fill=white,opacity=0.75,inner sep=0.75mm] {\SI{-31.63}{\dB}};

\end{axis}
\end{tikzpicture}%
        \fi 
        \vspace{-2.1mm}
         \caption{\gls{sw} \gls{smc} \gls{bf}}
         \label{fig:SW_SMC_pos1_switched0}
     \end{subfigure}%
     \hspace{\hspaceing}%
     \begin{subfigure}[t]{0.29\textwidth}
         \vskip 0pt	
         \centering
	    \ifdefined\ExcludeFigures   
            \ExcludeFigures
        \else
%
%
\definecolor{mycolor1}{rgb}{0.21961,0.21961,0.21961}%

\pgfplotsset{every axis/.append style={
  label style={font=\footnotesize},
  legend style={font=\footnotesize},
  tick label style={font=\footnotesize},
}}

\begin{tikzpicture}

\begin{axis}[%
width=\figurewidth,
height=\figureheight,
at={(0\figurewidth,0\figureheight)},
axis line style = thick,	
scale only axis,
point meta min=-60,
point meta max=-30,
axis on top,
xmin=3.37290652572368,
xmax=3.58,
xlabel style={yshift=1.5mm},
xlabel={$x$ in \SI{}{\metre}},
ymin=-2.139,
ymax=-1.93190652572368,
yticklabels=\empty,
xmajorgrids,
ymajorgrids,
grid style=dotted,
axis background/.style={fill=white},
colormap={mymap}{[1pt] rgb(0pt)=(0.995023,0.995023,0.995023); rgb(73pt)=(0.567253,0.659744,0.995023); rgb(74pt)=(0.561393,0.655151,0.995023); rgb(75pt)=(0.555533,0.650558,0.995023); rgb(76pt)=(0.549674,0.645965,0.995023); rgb(77pt)=(0.543814,0.641373,0.995023); rgb(78pt)=(0.537954,0.63678,0.995023); rgb(79pt)=(0.532094,0.632187,0.995023); rgb(80pt)=(0.526234,0.627594,0.995023); rgb(81pt)=(0.520374,0.623001,0.995023); rgb(82pt)=(0.514514,0.618408,0.995023); rgb(83pt)=(0.508655,0.613815,0.995023); rgb(84pt)=(0.502795,0.609223,0.995023); rgb(85pt)=(0.496935,0.60463,0.995023); rgb(86pt)=(0.491075,0.600037,0.995023); rgb(87pt)=(0.485215,0.595444,0.995023); rgb(88pt)=(0.479355,0.590851,0.995023); rgb(89pt)=(0.473495,0.586258,0.995023); rgb(90pt)=(0.467635,0.581665,0.995023); rgb(91pt)=(0.461776,0.577072,0.995023); rgb(92pt)=(0.455916,0.57248,0.995023); rgb(93pt)=(0.450056,0.567887,0.995023); rgb(94pt)=(0.444196,0.563294,0.995023); rgb(95pt)=(0.438336,0.558701,0.995023); rgb(96pt)=(0.432476,0.554108,0.995024); rgb(97pt)=(0.426616,0.549515,0.995024); rgb(98pt)=(0.420757,0.544922,0.995023); rgb(99pt)=(0.414897,0.540329,0.995023); rgb(100pt)=(0.409037,0.535737,0.995023); rgb(101pt)=(0.403177,0.531144,0.995023); rgb(102pt)=(0.397316,0.526551,0.995025); rgb(103pt)=(0.391455,0.521958,0.995025); rgb(104pt)=(0.385597,0.517365,0.995024); rgb(105pt)=(0.379741,0.512772,0.99502); rgb(106pt)=(0.373885,0.508178,0.995017); rgb(107pt)=(0.368022,0.503586,0.995019); rgb(108pt)=(0.362148,0.498995,0.995033); rgb(109pt)=(0.356273,0.494405,0.995048); rgb(110pt)=(0.350418,0.489811,0.995042); rgb(111pt)=(0.344605,0.48521,0.994998); rgb(112pt)=(0.338806,0.480607,0.994939); rgb(113pt)=(0.332944,0.476014,0.994941); rgb(114pt)=(0.326938,0.471447,0.995081); rgb(115pt)=(0.32084,0.466896,0.99531); rgb(116pt)=(0.31493,0.462312,0.995359); rgb(117pt)=(0.309513,0.457641,0.994933); rgb(118pt)=(0.304551,0.45289,0.994071); rgb(119pt)=(0.299066,0.448231,0.993711); rgb(120pt)=(0.29192,0.443865,0.994946); rgb(121pt)=(0.282789,0.439848,0.998086); rgb(122pt)=(0.274935,0.435607,1); rgb(123pt)=(0.272609,0.430392,0.996609); rgb(124pt)=(0.279306,0.423587,0.984555); rgb(125pt)=(0.293138,0.415525,0.965651); rgb(126pt)=(0.309867,0.406953,0.943967); rgb(127pt)=(0.325491,0.398575,0.923344); rgb(128pt)=(0.339058,0.39056,0.904696); rgb(129pt)=(0.351668,0.382714,0.886966); rgb(130pt)=(0.364431,0.37484,0.869089); rgb(131pt)=(0.377772,0.366865,0.850657); rgb(132pt)=(0.391418,0.358836,0.831932); rgb(133pt)=(0.405066,0.350807,0.813205); rgb(134pt)=(0.418557,0.342805,0.79463); rgb(135pt)=(0.431954,0.33482,0.776145); rgb(136pt)=(0.445337,0.326837,0.757672); rgb(137pt)=(0.458761,0.318847,0.739161); rgb(138pt)=(0.472215,0.310852,0.720621); rgb(139pt)=(0.485675,0.302856,0.702075); rgb(140pt)=(0.499125,0.294862,0.683539); rgb(141pt)=(0.512566,0.286869,0.665011); rgb(142pt)=(0.526005,0.278876,0.646485); rgb(143pt)=(0.539446,0.270883,0.627958); rgb(144pt)=(0.552889,0.26289,0.609428); rgb(145pt)=(0.566334,0.254896,0.590897); rgb(146pt)=(0.579778,0.246903,0.572366); rgb(147pt)=(0.593221,0.23891,0.553836); rgb(148pt)=(0.606664,0.230917,0.535307); rgb(149pt)=(0.620107,0.222923,0.516777); rgb(150pt)=(0.63355,0.21493,0.498247); rgb(151pt)=(0.646993,0.206937,0.479717); rgb(152pt)=(0.660437,0.198944,0.461187); rgb(153pt)=(0.67388,0.190951,0.442658); rgb(154pt)=(0.687323,0.182958,0.424128); rgb(155pt)=(0.700766,0.174963,0.405598); rgb(156pt)=(0.71421,0.166966,0.387067); rgb(157pt)=(0.727654,0.158972,0.368536); rgb(158pt)=(0.741096,0.150985,0.350008); rgb(159pt)=(0.754536,0.143004,0.331483); rgb(160pt)=(0.767977,0.135017,0.312955); rgb(161pt)=(0.781425,0.127007,0.29442); rgb(162pt)=(0.794879,0.118969,0.275874); rgb(163pt)=(0.808331,0.110944,0.257333); rgb(164pt)=(0.821763,0.102994,0.238819); rgb(165pt)=(0.835166,0.0951591,0.220344); rgb(166pt)=(0.848574,0.0873055,0.201863); rgb(167pt)=(0.862043,0.0792099,0.183297); rgb(168pt)=(0.875621,0.0706838,0.164581); rgb(169pt)=(0.889214,0.0621028,0.145846); rgb(170pt)=(0.902611,0.0542922,0.12738); rgb(171pt)=(0.915613,0.0480374,0.109458); rgb(172pt)=(0.928457,0.0424061,0.091754); rgb(173pt)=(0.941904,0.0343962,0.0732183); rgb(174pt)=(0.956736,0.0209305,0.052775); rgb(175pt)=(0.9725,0.00378572,0.0310453); rgb(176pt)=(0.986483,0.00633329,0.0117721); rgb(177pt)=(0.995725,0.0022327,0.000967649); rgb(178pt)=(0.998618,0.0358306,0.00495484); rgb(179pt)=(0.997427,0.0855268,0.00331317); rgb(180pt)=(0.995111,0.139657,0.000121229); rgb(181pt)=(0.99409,0.188685,0.00128695); rgb(182pt)=(0.994288,0.232904,0.00101351); rgb(183pt)=(0.994914,0.275439,0.000151196); rgb(184pt)=(0.995257,0.319085,0.000322596); rgb(185pt)=(0.995242,0.364147,0.000301319); rgb(186pt)=(0.995078,0.409793,7.57419e-05); rgb(187pt)=(0.994967,0.455231,7.71666e-05); rgb(188pt)=(0.99496,0.500262,8.74378e-05); rgb(189pt)=(0.995001,0.5451,3.04168e-05); rgb(190pt)=(0.995036,0.589964,1.73228e-05); rgb(191pt)=(0.995042,0.634943,2.51464e-05); rgb(192pt)=(0.995032,0.679984,1.1303e-05); rgb(193pt)=(0.995021,0.725028,3.71976e-06); rgb(194pt)=(0.995017,0.770042,8.33655e-06); rgb(195pt)=(0.99502,0.815035,5.36005e-06); rgb(196pt)=(0.995024,0.860017,1.0858e-06); rgb(197pt)=(0.995028,0.905002,6.87705e-06); rgb(198pt)=(0.995029,0.95,7.88977e-06); rgb(199pt)=(0.995023,0.995023,0)},
]
\addplot [forget plot] graphics [xmin=3.58020750849126, xmax=3.37269901723243, ymin=-2.13920750849126, ymax=-1.93169901723243] {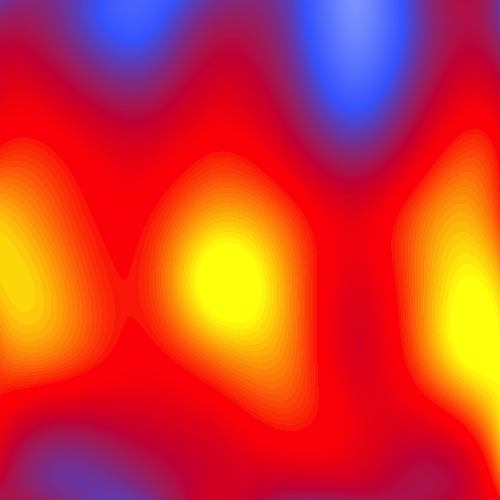};
\addplot[only marks, mark=*, mark options={}, mark size=1.5000pt, color=black, fill=RDlightgreen, forget plot] table[row sep=crcr]{%
x	y\\
3.46166087184211	-2.05024565388158\\
};

\draw[line width = 0.7pt, {Stealth[inset=0pt, scale=1.05, angle'=25]-},shorten < = 1mm] (3.462,-2.05) -- ++(0mm,14mm) node[anchor=south,font={\footnotesize},fill=white,opacity=0.75,inner sep=0.75mm] {\SI{-29.66}{\dB}};

\end{axis}
\end{tikzpicture}%
        \fi 
        \vspace{-2.1mm}
         \caption{Reciprocity-based \gls{bf} (full \gls{csi})}
         \label{fig:CSI_pos1_switched0}
     \end{subfigure}%
     \hspace{-0.65cm}%
     \begin{subfigure}[t]{0.058\textwidth}
        \vspace{-2.2mm} %
         \vskip 0pt	
         \centering
                  \setlength{\plotWidth}{1\textwidth}	
%
%

\pgfplotsset{every axis/.append style={
  label style={font=\footnotesize},
  legend style={font=\footnotesize},
  tick label style={font=\footnotesize}
}}

\begin{tikzpicture}

\begin{axis}[%
width=0.225\plotWidth,
height=3.38\plotWidth,
at={(0\plotWidth,0\plotWidth)},
scale only axis,
xmin=0,
xmax=1.02,
ymin=-1,
ymax=1,
zmin=-60,
zmax=-30,
zlabel={$PG$ in \SI{}{\dB}},
axis line style = thick,	
view={0}{0},
axis background/.style={fill=white},
axis z line*=right,			
xticklabel=\empty,			
tick align=center,		    
xtick=\empty,               
y axis line style= { draw opacity=0 }, 
z axis line style= { draw opacity=0 }, 
ztick distance={5},				
yticklabel pos=right,
xmajorgrids,
ymajorgrids,
zmajorgrids
]

\addplot3[%
surf,
shader=flat, z buffer=sort, colormap={mymap}{[1pt] rgb(0pt)=(1,1,1); rgb(26pt)=(0.51634,0.620915,1); rgb(27pt)=(0.497738,0.606335,1); rgb(34pt)=(0.367521,0.504274,1); rgb(35pt)=(0.348919,0.489693,1); rgb(39pt)=(0.27451,0.431373,1); rgb(40pt)=(0.317186,0.405998,0.941176); rgb(41pt)=(0.359862,0.380623,0.882353); rgb(43pt)=(0.445213,0.329873,0.764706); rgb(44pt)=(0.487889,0.304498,0.705882); rgb(45pt)=(0.530565,0.279123,0.647059); rgb(46pt)=(0.573241,0.253749,0.588235); rgb(47pt)=(0.615917,0.228374,0.529412); rgb(48pt)=(0.658593,0.202999,0.470588); rgb(49pt)=(0.701269,0.177624,0.411765); rgb(50pt)=(0.743945,0.152249,0.352941); rgb(51pt)=(0.786621,0.126874,0.294118); rgb(52pt)=(0.829296,0.101499,0.235294); rgb(53pt)=(0.871972,0.0761246,0.176471); rgb(54pt)=(0.914648,0.0507497,0.117647); rgb(55pt)=(0.957324,0.0253749,0.0588235); rgb(56pt)=(1,0,0); rgb(58pt)=(1,0.285714,0); rgb(59pt)=(1,0.428571,0); rgb(60pt)=(1,0.571429,0); rgb(63pt)=(1,1,0)}, mesh/rows=2]
table[row sep=crcr, point meta=\thisrow{c}] {%
x	y	z	c\\
0	-0.0001	-30.00	-30.00\\ 
0	-0.0001	-30.30	-30.30\\ 
0	-0.0001	-30.61	-30.61\\ 
0	-0.0001	-30.91	-30.91\\ 
0	-0.0001	-31.21	-31.21\\ 
0	-0.0001	-31.52	-31.52\\ 
0	-0.0001	-31.82	-31.82\\ 
0	-0.0001	-32.12	-32.12\\ 
0	-0.0001	-32.42	-32.42\\ 
0	-0.0001	-32.73	-32.73\\ 
0	-0.0001	-33.03	-33.03\\ 
0	-0.0001	-33.33	-33.33\\ 
0	-0.0001	-33.64	-33.64\\ 
0	-0.0001	-33.94	-33.94\\ 
0	-0.0001	-34.24	-34.24\\ 
0	-0.0001	-34.55	-34.55\\ 
0	-0.0001	-34.85	-34.85\\ 
0	-0.0001	-35.15	-35.15\\ 
0	-0.0001	-35.45	-35.45\\ 
0	-0.0001	-35.76	-35.76\\ 
0	-0.0001	-36.06	-36.06\\ 
0	-0.0001	-36.36	-36.36\\ 
0	-0.0001	-36.67	-36.67\\ 
0	-0.0001	-36.97	-36.97\\ 
0	-0.0001	-37.27	-37.27\\ 
0	-0.0001	-37.58	-37.58\\ 
0	-0.0001	-37.88	-37.88\\ 
0	-0.0001	-38.18	-38.18\\ 
0	-0.0001	-38.48	-38.48\\ 
0	-0.0001	-38.79	-38.79\\ 
0	-0.0001	-39.09	-39.09\\ 
0	-0.0001	-39.39	-39.39\\ 
0	-0.0001	-39.70	-39.70\\ 
0	-0.0001	-40.00	-40.00\\ 
0	-0.0001	-40.30	-40.30\\ 
0	-0.0001	-40.61	-40.61\\ 
0	-0.0001	-40.91	-40.91\\ 
0	-0.0001	-41.21	-41.21\\ 
0	-0.0001	-41.52	-41.52\\ 
0	-0.0001	-41.82	-41.82\\ 
0	-0.0001	-42.12	-42.12\\ 
0	-0.0001	-42.42	-42.42\\ 
0	-0.0001	-42.73	-42.73\\ 
0	-0.0001	-43.03	-43.03\\ 
0	-0.0001	-43.33	-43.33\\ 
0	-0.0001	-43.64	-43.64\\ 
0	-0.0001	-43.94	-43.94\\ 
0	-0.0001	-44.24	-44.24\\ 
0	-0.0001	-44.55	-44.55\\ 
0	-0.0001	-44.85	-44.85\\ 
0	-0.0001	-45.15	-45.15\\ 
0	-0.0001	-45.45	-45.45\\ 
0	-0.0001	-45.76	-45.76\\ 
0	-0.0001	-46.06	-46.06\\ 
0	-0.0001	-46.36	-46.36\\ 
0	-0.0001	-46.67	-46.67\\ 
0	-0.0001	-46.97	-46.97\\ 
0	-0.0001	-47.27	-47.27\\ 
0	-0.0001	-47.58	-47.58\\ 
0	-0.0001	-47.88	-47.88\\ 
0	-0.0001	-48.18	-48.18\\ 
0	-0.0001	-48.48	-48.48\\ 
0	-0.0001	-48.79	-48.79\\ 
0	-0.0001	-49.09	-49.09\\ 
0	-0.0001	-49.39	-49.39\\ 
0	-0.0001	-49.70	-49.70\\ 
0	-0.0001	-50.00	-50.00\\ 
0	-0.0001	-50.30	-50.30\\ 
0	-0.0001	-50.61	-50.61\\ 
0	-0.0001	-50.91	-50.91\\ 
0	-0.0001	-51.21	-51.21\\ 
0	-0.0001	-51.52	-51.52\\ 
0	-0.0001	-51.82	-51.82\\ 
0	-0.0001	-52.12	-52.12\\ 
0	-0.0001	-52.42	-52.42\\ 
0	-0.0001	-52.73	-52.73\\ 
0	-0.0001	-53.03	-53.03\\ 
0	-0.0001	-53.33	-53.33\\ 
0	-0.0001	-53.64	-53.64\\ 
0	-0.0001	-53.94	-53.94\\ 
0	-0.0001	-54.24	-54.24\\ 
0	-0.0001	-54.55	-54.55\\ 
0	-0.0001	-54.85	-54.85\\ 
0	-0.0001	-55.15	-55.15\\ 
0	-0.0001	-55.45	-55.45\\ 
0	-0.0001	-55.76	-55.76\\ 
0	-0.0001	-56.06	-56.06\\ 
0	-0.0001	-56.36	-56.36\\ 
0	-0.0001	-56.67	-56.67\\ 
0	-0.0001	-56.97	-56.97\\ 
0	-0.0001	-57.27	-57.27\\ 
0	-0.0001	-57.58	-57.58\\ 
0	-0.0001	-57.88	-57.88\\ 
0	-0.0001	-58.18	-58.18\\ 
0	-0.0001	-58.48	-58.48\\ 
0	-0.0001	-58.79	-58.79\\ 
0	-0.0001	-59.09	-59.09\\ 
0	-0.0001	-59.39	-59.39\\ 
0	-0.0001	-59.70	-59.70\\ 
0	-0.0001	-60.00	-60.00\\ 
1	-0.0001	-30.00	-30.00\\ 
1	-0.0001	-30.30	-30.30\\ 
1	-0.0001	-30.61	-30.61\\ 
1	-0.0001	-30.91	-30.91\\ 
1	-0.0001	-31.21	-31.21\\ 
1	-0.0001	-31.52	-31.52\\ 
1	-0.0001	-31.82	-31.82\\ 
1	-0.0001	-32.12	-32.12\\ 
1	-0.0001	-32.42	-32.42\\ 
1	-0.0001	-32.73	-32.73\\ 
1	-0.0001	-33.03	-33.03\\ 
1	-0.0001	-33.33	-33.33\\ 
1	-0.0001	-33.64	-33.64\\ 
1	-0.0001	-33.94	-33.94\\ 
1	-0.0001	-34.24	-34.24\\ 
1	-0.0001	-34.55	-34.55\\ 
1	-0.0001	-34.85	-34.85\\ 
1	-0.0001	-35.15	-35.15\\ 
1	-0.0001	-35.45	-35.45\\ 
1	-0.0001	-35.76	-35.76\\ 
1	-0.0001	-36.06	-36.06\\ 
1	-0.0001	-36.36	-36.36\\ 
1	-0.0001	-36.67	-36.67\\ 
1	-0.0001	-36.97	-36.97\\ 
1	-0.0001	-37.27	-37.27\\ 
1	-0.0001	-37.58	-37.58\\ 
1	-0.0001	-37.88	-37.88\\ 
1	-0.0001	-38.18	-38.18\\ 
1	-0.0001	-38.48	-38.48\\ 
1	-0.0001	-38.79	-38.79\\ 
1	-0.0001	-39.09	-39.09\\ 
1	-0.0001	-39.39	-39.39\\ 
1	-0.0001	-39.70	-39.70\\ 
1	-0.0001	-40.00	-40.00\\ 
1	-0.0001	-40.30	-40.30\\ 
1	-0.0001	-40.61	-40.61\\ 
1	-0.0001	-40.91	-40.91\\ 
1	-0.0001	-41.21	-41.21\\ 
1	-0.0001	-41.52	-41.52\\ 
1	-0.0001	-41.82	-41.82\\ 
1	-0.0001	-42.12	-42.12\\ 
1	-0.0001	-42.42	-42.42\\ 
1	-0.0001	-42.73	-42.73\\ 
1	-0.0001	-43.03	-43.03\\ 
1	-0.0001	-43.33	-43.33\\ 
1	-0.0001	-43.64	-43.64\\ 
1	-0.0001	-43.94	-43.94\\ 
1	-0.0001	-44.24	-44.24\\ 
1	-0.0001	-44.55	-44.55\\ 
1	-0.0001	-44.85	-44.85\\ 
1	-0.0001	-45.15	-45.15\\ 
1	-0.0001	-45.45	-45.45\\ 
1	-0.0001	-45.76	-45.76\\ 
1	-0.0001	-46.06	-46.06\\ 
1	-0.0001	-46.36	-46.36\\ 
1	-0.0001	-46.67	-46.67\\ 
1	-0.0001	-46.97	-46.97\\ 
1	-0.0001	-47.27	-47.27\\ 
1	-0.0001	-47.58	-47.58\\ 
1	-0.0001	-47.88	-47.88\\ 
1	-0.0001	-48.18	-48.18\\ 
1	-0.0001	-48.48	-48.48\\ 
1	-0.0001	-48.79	-48.79\\ 
1	-0.0001	-49.09	-49.09\\ 
1	-0.0001	-49.39	-49.39\\ 
1	-0.0001	-49.70	-49.70\\ 
1	-0.0001	-50.00	-50.00\\ 
1	-0.0001	-50.30	-50.30\\ 
1	-0.0001	-50.61	-50.61\\ 
1	-0.0001	-50.91	-50.91\\ 
1	-0.0001	-51.21	-51.21\\ 
1	-0.0001	-51.52	-51.52\\ 
1	-0.0001	-51.82	-51.82\\ 
1	-0.0001	-52.12	-52.12\\ 
1	-0.0001	-52.42	-52.42\\ 
1	-0.0001	-52.73	-52.73\\ 
1	-0.0001	-53.03	-53.03\\ 
1	-0.0001	-53.33	-53.33\\ 
1	-0.0001	-53.64	-53.64\\ 
1	-0.0001	-53.94	-53.94\\ 
1	-0.0001	-54.24	-54.24\\ 
1	-0.0001	-54.55	-54.55\\ 
1	-0.0001	-54.85	-54.85\\ 
1	-0.0001	-55.15	-55.15\\ 
1	-0.0001	-55.45	-55.45\\ 
1	-0.0001	-55.76	-55.76\\ 
1	-0.0001	-56.06	-56.06\\ 
1	-0.0001	-56.36	-56.36\\ 
1	-0.0001	-56.67	-56.67\\ 
1	-0.0001	-56.97	-56.97\\ 
1	-0.0001	-57.27	-57.27\\ 
1	-0.0001	-57.58	-57.58\\ 
1	-0.0001	-57.88	-57.88\\ 
1	-0.0001	-58.18	-58.18\\ 
1	-0.0001	-58.48	-58.48\\ 
1	-0.0001	-58.79	-58.79\\ 
1	-0.0001	-59.09	-59.09\\ 
1	-0.0001	-59.39	-59.39\\ 
1	-0.0001	-59.70	-59.70\\ 
1	-0.0001	-60.00	-60.00\\ 
};
\draw [draw=black, thick] (0.015,0.98,-60) rectangle (1,1,-30);    
\end{axis}
\end{tikzpicture}%
     \end{subfigure}
    	\vspace{-2.1mm}\caption{
    	Beamformers applied at $\bm{p}_\mrm{EN}^{(1)}$, $PG$ distribution evaluated and interpolated around  $\bm{p}_\mrm{EN}^{(1)}$.
    	}\label{fig:pos1_domains}
    	\vspace{-3mm}
\end{figure*}	

\begin{figure*}[bt]	
     \hspace{-4mm}
     \begin{subfigure}[t]{0.29\textwidth}
        \vskip 0pt	
         \centering
	    \ifdefined\ExcludeFigures   
	                \ExcludeFigures
        \else
%
%
\definecolor{mycolor1}{rgb}{0.21961,0.21961,0.21961}%

\pgfplotsset{every axis/.append style={
  label style={font=\footnotesize},
  legend style={font=\footnotesize},
  tick label style={font=\footnotesize},
}}

\begin{tikzpicture}

\begin{axis}[%
width=\figurewidth,
height=\figureheight,
at={(0\figurewidth,0\figureheight)},
scale only axis,
point meta min=-60,
point meta max=-30,
axis on top,
xmin=4.61190652572368,
xmax=4.819,
xlabel style={yshift=1.5mm},
xlabel={$x$ in \SI{}{\metre}},
ymin=-1.254,
ymax=-1.04690652572368,
ylabel style={yshift=-6pt},
ylabel={$y$ in \SI{}{\metre}},
ytick={-1.25,-1.2,-1.15,-1.1},
xmajorgrids,
ymajorgrids,
grid style=dotted,
axis line style = thick,	
axis background/.style={fill=white},
colormap={mymap}{[1pt] rgb(0pt)=(0.995023,0.995023,0.995023); rgb(73pt)=(0.567253,0.659744,0.995023); rgb(74pt)=(0.561393,0.655151,0.995023); rgb(75pt)=(0.555533,0.650558,0.995023); rgb(76pt)=(0.549674,0.645965,0.995023); rgb(77pt)=(0.543814,0.641373,0.995023); rgb(78pt)=(0.537954,0.63678,0.995023); rgb(79pt)=(0.532094,0.632187,0.995023); rgb(80pt)=(0.526234,0.627594,0.995023); rgb(81pt)=(0.520374,0.623001,0.995023); rgb(82pt)=(0.514514,0.618408,0.995023); rgb(83pt)=(0.508655,0.613815,0.995023); rgb(84pt)=(0.502795,0.609223,0.995023); rgb(85pt)=(0.496935,0.60463,0.995023); rgb(86pt)=(0.491075,0.600037,0.995023); rgb(87pt)=(0.485215,0.595444,0.995023); rgb(88pt)=(0.479355,0.590851,0.995023); rgb(89pt)=(0.473495,0.586258,0.995023); rgb(90pt)=(0.467635,0.581665,0.995023); rgb(91pt)=(0.461776,0.577072,0.995023); rgb(92pt)=(0.455916,0.57248,0.995023); rgb(93pt)=(0.450056,0.567887,0.995023); rgb(94pt)=(0.444196,0.563294,0.995023); rgb(95pt)=(0.438336,0.558701,0.995023); rgb(96pt)=(0.432476,0.554108,0.995024); rgb(97pt)=(0.426616,0.549515,0.995024); rgb(98pt)=(0.420757,0.544922,0.995023); rgb(99pt)=(0.414897,0.540329,0.995023); rgb(100pt)=(0.409037,0.535737,0.995023); rgb(101pt)=(0.403177,0.531144,0.995023); rgb(102pt)=(0.397316,0.526551,0.995025); rgb(103pt)=(0.391455,0.521958,0.995025); rgb(104pt)=(0.385597,0.517365,0.995024); rgb(105pt)=(0.379741,0.512772,0.99502); rgb(106pt)=(0.373885,0.508178,0.995017); rgb(107pt)=(0.368022,0.503586,0.995019); rgb(108pt)=(0.362148,0.498995,0.995033); rgb(109pt)=(0.356273,0.494405,0.995048); rgb(110pt)=(0.350418,0.489811,0.995042); rgb(111pt)=(0.344605,0.48521,0.994998); rgb(112pt)=(0.338806,0.480607,0.994939); rgb(113pt)=(0.332944,0.476014,0.994941); rgb(114pt)=(0.326938,0.471447,0.995081); rgb(115pt)=(0.32084,0.466896,0.99531); rgb(116pt)=(0.31493,0.462312,0.995359); rgb(117pt)=(0.309513,0.457641,0.994933); rgb(118pt)=(0.304551,0.45289,0.994071); rgb(119pt)=(0.299066,0.448231,0.993711); rgb(120pt)=(0.29192,0.443865,0.994946); rgb(121pt)=(0.282789,0.439848,0.998086); rgb(122pt)=(0.274935,0.435607,1); rgb(123pt)=(0.272609,0.430392,0.996609); rgb(124pt)=(0.279306,0.423587,0.984555); rgb(125pt)=(0.293138,0.415525,0.965651); rgb(126pt)=(0.309867,0.406953,0.943967); rgb(127pt)=(0.325491,0.398575,0.923344); rgb(128pt)=(0.339058,0.39056,0.904696); rgb(129pt)=(0.351668,0.382714,0.886966); rgb(130pt)=(0.364431,0.37484,0.869089); rgb(131pt)=(0.377772,0.366865,0.850657); rgb(132pt)=(0.391418,0.358836,0.831932); rgb(133pt)=(0.405066,0.350807,0.813205); rgb(134pt)=(0.418557,0.342805,0.79463); rgb(135pt)=(0.431954,0.33482,0.776145); rgb(136pt)=(0.445337,0.326837,0.757672); rgb(137pt)=(0.458761,0.318847,0.739161); rgb(138pt)=(0.472215,0.310852,0.720621); rgb(139pt)=(0.485675,0.302856,0.702075); rgb(140pt)=(0.499125,0.294862,0.683539); rgb(141pt)=(0.512566,0.286869,0.665011); rgb(142pt)=(0.526005,0.278876,0.646485); rgb(143pt)=(0.539446,0.270883,0.627958); rgb(144pt)=(0.552889,0.26289,0.609428); rgb(145pt)=(0.566334,0.254896,0.590897); rgb(146pt)=(0.579778,0.246903,0.572366); rgb(147pt)=(0.593221,0.23891,0.553836); rgb(148pt)=(0.606664,0.230917,0.535307); rgb(149pt)=(0.620107,0.222923,0.516777); rgb(150pt)=(0.63355,0.21493,0.498247); rgb(151pt)=(0.646993,0.206937,0.479717); rgb(152pt)=(0.660437,0.198944,0.461187); rgb(153pt)=(0.67388,0.190951,0.442658); rgb(154pt)=(0.687323,0.182958,0.424128); rgb(155pt)=(0.700766,0.174963,0.405598); rgb(156pt)=(0.71421,0.166966,0.387067); rgb(157pt)=(0.727654,0.158972,0.368536); rgb(158pt)=(0.741096,0.150985,0.350008); rgb(159pt)=(0.754536,0.143004,0.331483); rgb(160pt)=(0.767977,0.135017,0.312955); rgb(161pt)=(0.781425,0.127007,0.29442); rgb(162pt)=(0.794879,0.118969,0.275874); rgb(163pt)=(0.808331,0.110944,0.257333); rgb(164pt)=(0.821763,0.102994,0.238819); rgb(165pt)=(0.835166,0.0951591,0.220344); rgb(166pt)=(0.848574,0.0873055,0.201863); rgb(167pt)=(0.862043,0.0792099,0.183297); rgb(168pt)=(0.875621,0.0706838,0.164581); rgb(169pt)=(0.889214,0.0621028,0.145846); rgb(170pt)=(0.902611,0.0542922,0.12738); rgb(171pt)=(0.915613,0.0480374,0.109458); rgb(172pt)=(0.928457,0.0424061,0.091754); rgb(173pt)=(0.941904,0.0343962,0.0732183); rgb(174pt)=(0.956736,0.0209305,0.052775); rgb(175pt)=(0.9725,0.00378572,0.0310453); rgb(176pt)=(0.986483,0.00633329,0.0117721); rgb(177pt)=(0.995725,0.0022327,0.000967649); rgb(178pt)=(0.998618,0.0358306,0.00495484); rgb(179pt)=(0.997427,0.0855268,0.00331317); rgb(180pt)=(0.995111,0.139657,0.000121229); rgb(181pt)=(0.99409,0.188685,0.00128695); rgb(182pt)=(0.994288,0.232904,0.00101351); rgb(183pt)=(0.994914,0.275439,0.000151196); rgb(184pt)=(0.995257,0.319085,0.000322596); rgb(185pt)=(0.995242,0.364147,0.000301319); rgb(186pt)=(0.995078,0.409793,7.57419e-05); rgb(187pt)=(0.994967,0.455231,7.71666e-05); rgb(188pt)=(0.99496,0.500262,8.74378e-05); rgb(189pt)=(0.995001,0.5451,3.04168e-05); rgb(190pt)=(0.995036,0.589964,1.73228e-05); rgb(191pt)=(0.995042,0.634943,2.51464e-05); rgb(192pt)=(0.995032,0.679984,1.1303e-05); rgb(193pt)=(0.995021,0.725028,3.71976e-06); rgb(194pt)=(0.995017,0.770042,8.33655e-06); rgb(195pt)=(0.99502,0.815035,5.36005e-06); rgb(196pt)=(0.995024,0.860017,1.0858e-06); rgb(197pt)=(0.995028,0.905002,6.87705e-06); rgb(198pt)=(0.995029,0.95,7.88977e-06); rgb(199pt)=(0.995023,0.995023,0)},
]
\addplot [forget plot] graphics [xmin=4.81920750849126, xmax=4.61169901723243, ymin=-1.25420750849126, ymax=-1.04669901723243] {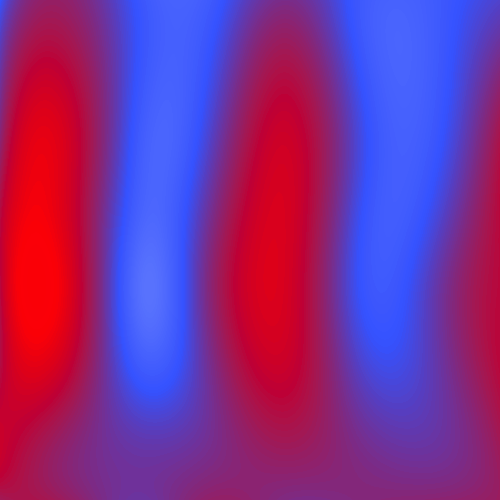};
\addplot[only marks, mark=*, mark options={}, mark size=1.5000pt, color=black, fill=RDlightgreen, forget plot] table[row sep=crcr]{%
x	y\\
4.70066087184211	-1.16524565388158\\
};

\draw[line width = 0.7pt, {Stealth[inset=0pt, scale=1.05, angle'=25]-},shorten < = 1mm] (4.701,-1.165) -- ++(0mm,14mm) node[anchor=south,font={\footnotesize},fill=white,opacity=0.75,inner sep=0.75mm] {\SI{-37.30}{\dB}};

\end{axis}
\end{tikzpicture}%
        \fi 
        \vspace{-2.1mm}
         \caption{\Gls{pw} \gls{los} \gls{bf}}
         \label{fig:PW_pos2_switched0}
     \end{subfigure}%
     \hspace{-0.78cm}%
     \begin{subfigure}[t]{0.29\textwidth}
        \vskip 0pt	
         \centering
	    \ifdefined\ExcludeFigures   
            \ExcludeFigures
        \else
            \vspace{-0.5mm}
%
%
\definecolor{mycolor1}{rgb}{0.00000,0.44700,0.74100}%
\pgfplotsset{every axis/.append style={
  label style={font=\footnotesize},
  legend style={font=\footnotesize},
  tick label style={font=\footnotesize},
}}

\begin{tikzpicture}

\begin{axis}[%
width=\figurewidth,
height=\figureheight,
at={(0\figurewidth,0\figureheight)},
scale only axis,
point meta min=-60,
point meta max=-30,
axis on top,
xmin=4.61190652572368,
xmax=4.819,
xlabel style={yshift=1.5mm},
xlabel={$x$ in \SI{}{\metre}},
ymin=-1.254,
ymax=-1.04690652572368,
yticklabels=\empty,
xmajorgrids,
ymajorgrids,
grid style=dotted,
axis line style = thick,	
axis background/.style={fill=white},
colormap={mymap}{[1pt] rgb(0pt)=(0.995023,0.995023,0.995023); rgb(73pt)=(0.567253,0.659744,0.995023); rgb(74pt)=(0.561393,0.655151,0.995023); rgb(75pt)=(0.555533,0.650558,0.995023); rgb(76pt)=(0.549674,0.645965,0.995023); rgb(77pt)=(0.543814,0.641373,0.995023); rgb(78pt)=(0.537954,0.63678,0.995023); rgb(79pt)=(0.532094,0.632187,0.995023); rgb(80pt)=(0.526234,0.627594,0.995023); rgb(81pt)=(0.520374,0.623001,0.995023); rgb(82pt)=(0.514514,0.618408,0.995023); rgb(83pt)=(0.508655,0.613815,0.995023); rgb(84pt)=(0.502795,0.609223,0.995023); rgb(85pt)=(0.496935,0.60463,0.995023); rgb(86pt)=(0.491075,0.600037,0.995023); rgb(87pt)=(0.485215,0.595444,0.995023); rgb(88pt)=(0.479355,0.590851,0.995023); rgb(89pt)=(0.473495,0.586258,0.995023); rgb(90pt)=(0.467635,0.581665,0.995023); rgb(91pt)=(0.461776,0.577072,0.995023); rgb(92pt)=(0.455916,0.57248,0.995023); rgb(93pt)=(0.450056,0.567887,0.995023); rgb(94pt)=(0.444196,0.563294,0.995023); rgb(95pt)=(0.438336,0.558701,0.995023); rgb(96pt)=(0.432476,0.554108,0.995024); rgb(97pt)=(0.426616,0.549515,0.995024); rgb(98pt)=(0.420757,0.544922,0.995023); rgb(99pt)=(0.414897,0.540329,0.995023); rgb(100pt)=(0.409037,0.535737,0.995023); rgb(101pt)=(0.403177,0.531144,0.995023); rgb(102pt)=(0.397316,0.526551,0.995025); rgb(103pt)=(0.391455,0.521958,0.995025); rgb(104pt)=(0.385597,0.517365,0.995024); rgb(105pt)=(0.379741,0.512772,0.99502); rgb(106pt)=(0.373885,0.508178,0.995017); rgb(107pt)=(0.368022,0.503586,0.995019); rgb(108pt)=(0.362148,0.498995,0.995033); rgb(109pt)=(0.356273,0.494405,0.995048); rgb(110pt)=(0.350418,0.489811,0.995042); rgb(111pt)=(0.344605,0.48521,0.994998); rgb(112pt)=(0.338806,0.480607,0.994939); rgb(113pt)=(0.332944,0.476014,0.994941); rgb(114pt)=(0.326938,0.471447,0.995081); rgb(115pt)=(0.32084,0.466896,0.99531); rgb(116pt)=(0.31493,0.462312,0.995359); rgb(117pt)=(0.309513,0.457641,0.994933); rgb(118pt)=(0.304551,0.45289,0.994071); rgb(119pt)=(0.299066,0.448231,0.993711); rgb(120pt)=(0.29192,0.443865,0.994946); rgb(121pt)=(0.282789,0.439848,0.998086); rgb(122pt)=(0.274935,0.435607,1); rgb(123pt)=(0.272609,0.430392,0.996609); rgb(124pt)=(0.279306,0.423587,0.984555); rgb(125pt)=(0.293138,0.415525,0.965651); rgb(126pt)=(0.309867,0.406953,0.943967); rgb(127pt)=(0.325491,0.398575,0.923344); rgb(128pt)=(0.339058,0.39056,0.904696); rgb(129pt)=(0.351668,0.382714,0.886966); rgb(130pt)=(0.364431,0.37484,0.869089); rgb(131pt)=(0.377772,0.366865,0.850657); rgb(132pt)=(0.391418,0.358836,0.831932); rgb(133pt)=(0.405066,0.350807,0.813205); rgb(134pt)=(0.418557,0.342805,0.79463); rgb(135pt)=(0.431954,0.33482,0.776145); rgb(136pt)=(0.445337,0.326837,0.757672); rgb(137pt)=(0.458761,0.318847,0.739161); rgb(138pt)=(0.472215,0.310852,0.720621); rgb(139pt)=(0.485675,0.302856,0.702075); rgb(140pt)=(0.499125,0.294862,0.683539); rgb(141pt)=(0.512566,0.286869,0.665011); rgb(142pt)=(0.526005,0.278876,0.646485); rgb(143pt)=(0.539446,0.270883,0.627958); rgb(144pt)=(0.552889,0.26289,0.609428); rgb(145pt)=(0.566334,0.254896,0.590897); rgb(146pt)=(0.579778,0.246903,0.572366); rgb(147pt)=(0.593221,0.23891,0.553836); rgb(148pt)=(0.606664,0.230917,0.535307); rgb(149pt)=(0.620107,0.222923,0.516777); rgb(150pt)=(0.63355,0.21493,0.498247); rgb(151pt)=(0.646993,0.206937,0.479717); rgb(152pt)=(0.660437,0.198944,0.461187); rgb(153pt)=(0.67388,0.190951,0.442658); rgb(154pt)=(0.687323,0.182958,0.424128); rgb(155pt)=(0.700766,0.174963,0.405598); rgb(156pt)=(0.71421,0.166966,0.387067); rgb(157pt)=(0.727654,0.158972,0.368536); rgb(158pt)=(0.741096,0.150985,0.350008); rgb(159pt)=(0.754536,0.143004,0.331483); rgb(160pt)=(0.767977,0.135017,0.312955); rgb(161pt)=(0.781425,0.127007,0.29442); rgb(162pt)=(0.794879,0.118969,0.275874); rgb(163pt)=(0.808331,0.110944,0.257333); rgb(164pt)=(0.821763,0.102994,0.238819); rgb(165pt)=(0.835166,0.0951591,0.220344); rgb(166pt)=(0.848574,0.0873055,0.201863); rgb(167pt)=(0.862043,0.0792099,0.183297); rgb(168pt)=(0.875621,0.0706838,0.164581); rgb(169pt)=(0.889214,0.0621028,0.145846); rgb(170pt)=(0.902611,0.0542922,0.12738); rgb(171pt)=(0.915613,0.0480374,0.109458); rgb(172pt)=(0.928457,0.0424061,0.091754); rgb(173pt)=(0.941904,0.0343962,0.0732183); rgb(174pt)=(0.956736,0.0209305,0.052775); rgb(175pt)=(0.9725,0.00378572,0.0310453); rgb(176pt)=(0.986483,0.00633329,0.0117721); rgb(177pt)=(0.995725,0.0022327,0.000967649); rgb(178pt)=(0.998618,0.0358306,0.00495484); rgb(179pt)=(0.997427,0.0855268,0.00331317); rgb(180pt)=(0.995111,0.139657,0.000121229); rgb(181pt)=(0.99409,0.188685,0.00128695); rgb(182pt)=(0.994288,0.232904,0.00101351); rgb(183pt)=(0.994914,0.275439,0.000151196); rgb(184pt)=(0.995257,0.319085,0.000322596); rgb(185pt)=(0.995242,0.364147,0.000301319); rgb(186pt)=(0.995078,0.409793,7.57419e-05); rgb(187pt)=(0.994967,0.455231,7.71666e-05); rgb(188pt)=(0.99496,0.500262,8.74378e-05); rgb(189pt)=(0.995001,0.5451,3.04168e-05); rgb(190pt)=(0.995036,0.589964,1.73228e-05); rgb(191pt)=(0.995042,0.634943,2.51464e-05); rgb(192pt)=(0.995032,0.679984,1.1303e-05); rgb(193pt)=(0.995021,0.725028,3.71976e-06); rgb(194pt)=(0.995017,0.770042,8.33655e-06); rgb(195pt)=(0.99502,0.815035,5.36005e-06); rgb(196pt)=(0.995024,0.860017,1.0858e-06); rgb(197pt)=(0.995028,0.905002,6.87705e-06); rgb(198pt)=(0.995029,0.95,7.88977e-06); rgb(199pt)=(0.995023,0.995023,0)},
]
\addplot [forget plot] graphics [xmin=4.81920750849126, xmax=4.61169901723243, ymin=-1.25420750849126, ymax=-1.04669901723243] {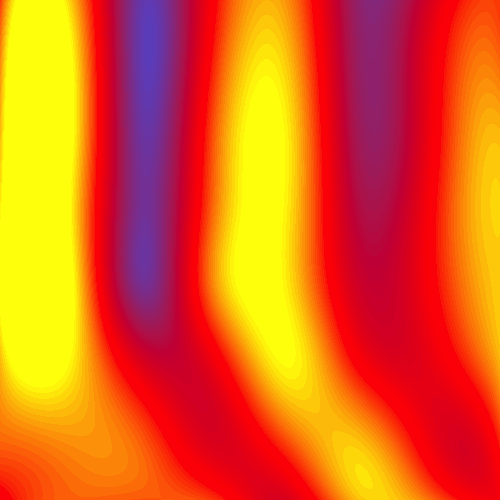};
\addplot[only marks, mark=*, mark options={}, mark size=1.5000pt, color=black, fill=RDlightgreen, forget plot] table[row sep=crcr]{%
x	y\\
4.70066087184211	-1.16524565388158\\
};

\draw[line width = 0.7pt, {Stealth[inset=0pt, scale=1.05, angle'=25]-},shorten < = 1mm] (4.701,-1.165) -- ++(0mm,14mm) node[anchor=south,font={\footnotesize},fill=white,opacity=0.75,inner sep=0.75mm] {\SI{-31.33}{\dB}};

\end{axis}
\end{tikzpicture}%
        \fi 
        \vspace{-2.1mm}
         \caption{\Gls{sw} \gls{los} \gls{bf}}
         \label{fig:SW_LoS_pos2_switched0}
     \end{subfigure}%
     \hspace{\hspaceing}%
     \begin{subfigure}[t]{0.29\textwidth}
         \vskip 0pt	
         \centering
	    \ifdefined\ExcludeFigures   
            \ExcludeFigures
        \else
            \vspace{-0.5mm}
%
%
\definecolor{mycolor1}{rgb}{0.36471,0.42745,0.26667}%
\pgfplotsset{every axis/.append style={
  label style={font=\footnotesize},
  legend style={font=\footnotesize},
  tick label style={font=\footnotesize},
}}

\begin{tikzpicture}

\begin{axis}[%
width=\figurewidth,
height=\figureheight,
at={(0\figurewidth,0\figureheight)},
scale only axis,
point meta min=-60,
point meta max=-30,
axis on top,
xmin=4.61190652572368,
xmax=4.819,
xlabel style={yshift=1.5mm},
xlabel={$x$ in \SI{}{\metre}},
ymin=-1.254,
ymax=-1.04690652572368,
yticklabels=\empty,
xmajorgrids,
ymajorgrids,
grid style=dotted,
axis line style = thick,	
axis background/.style={fill=white},
colormap={mymap}{[1pt] rgb(0pt)=(0.995023,0.995023,0.995023); rgb(73pt)=(0.567253,0.659744,0.995023); rgb(74pt)=(0.561393,0.655151,0.995023); rgb(75pt)=(0.555533,0.650558,0.995023); rgb(76pt)=(0.549674,0.645965,0.995023); rgb(77pt)=(0.543814,0.641373,0.995023); rgb(78pt)=(0.537954,0.63678,0.995023); rgb(79pt)=(0.532094,0.632187,0.995023); rgb(80pt)=(0.526234,0.627594,0.995023); rgb(81pt)=(0.520374,0.623001,0.995023); rgb(82pt)=(0.514514,0.618408,0.995023); rgb(83pt)=(0.508655,0.613815,0.995023); rgb(84pt)=(0.502795,0.609223,0.995023); rgb(85pt)=(0.496935,0.60463,0.995023); rgb(86pt)=(0.491075,0.600037,0.995023); rgb(87pt)=(0.485215,0.595444,0.995023); rgb(88pt)=(0.479355,0.590851,0.995023); rgb(89pt)=(0.473495,0.586258,0.995023); rgb(90pt)=(0.467635,0.581665,0.995023); rgb(91pt)=(0.461776,0.577072,0.995023); rgb(92pt)=(0.455916,0.57248,0.995023); rgb(93pt)=(0.450056,0.567887,0.995023); rgb(94pt)=(0.444196,0.563294,0.995023); rgb(95pt)=(0.438336,0.558701,0.995023); rgb(96pt)=(0.432476,0.554108,0.995024); rgb(97pt)=(0.426616,0.549515,0.995024); rgb(98pt)=(0.420757,0.544922,0.995023); rgb(99pt)=(0.414897,0.540329,0.995023); rgb(100pt)=(0.409037,0.535737,0.995023); rgb(101pt)=(0.403177,0.531144,0.995023); rgb(102pt)=(0.397316,0.526551,0.995025); rgb(103pt)=(0.391455,0.521958,0.995025); rgb(104pt)=(0.385597,0.517365,0.995024); rgb(105pt)=(0.379741,0.512772,0.99502); rgb(106pt)=(0.373885,0.508178,0.995017); rgb(107pt)=(0.368022,0.503586,0.995019); rgb(108pt)=(0.362148,0.498995,0.995033); rgb(109pt)=(0.356273,0.494405,0.995048); rgb(110pt)=(0.350418,0.489811,0.995042); rgb(111pt)=(0.344605,0.48521,0.994998); rgb(112pt)=(0.338806,0.480607,0.994939); rgb(113pt)=(0.332944,0.476014,0.994941); rgb(114pt)=(0.326938,0.471447,0.995081); rgb(115pt)=(0.32084,0.466896,0.99531); rgb(116pt)=(0.31493,0.462312,0.995359); rgb(117pt)=(0.309513,0.457641,0.994933); rgb(118pt)=(0.304551,0.45289,0.994071); rgb(119pt)=(0.299066,0.448231,0.993711); rgb(120pt)=(0.29192,0.443865,0.994946); rgb(121pt)=(0.282789,0.439848,0.998086); rgb(122pt)=(0.274935,0.435607,1); rgb(123pt)=(0.272609,0.430392,0.996609); rgb(124pt)=(0.279306,0.423587,0.984555); rgb(125pt)=(0.293138,0.415525,0.965651); rgb(126pt)=(0.309867,0.406953,0.943967); rgb(127pt)=(0.325491,0.398575,0.923344); rgb(128pt)=(0.339058,0.39056,0.904696); rgb(129pt)=(0.351668,0.382714,0.886966); rgb(130pt)=(0.364431,0.37484,0.869089); rgb(131pt)=(0.377772,0.366865,0.850657); rgb(132pt)=(0.391418,0.358836,0.831932); rgb(133pt)=(0.405066,0.350807,0.813205); rgb(134pt)=(0.418557,0.342805,0.79463); rgb(135pt)=(0.431954,0.33482,0.776145); rgb(136pt)=(0.445337,0.326837,0.757672); rgb(137pt)=(0.458761,0.318847,0.739161); rgb(138pt)=(0.472215,0.310852,0.720621); rgb(139pt)=(0.485675,0.302856,0.702075); rgb(140pt)=(0.499125,0.294862,0.683539); rgb(141pt)=(0.512566,0.286869,0.665011); rgb(142pt)=(0.526005,0.278876,0.646485); rgb(143pt)=(0.539446,0.270883,0.627958); rgb(144pt)=(0.552889,0.26289,0.609428); rgb(145pt)=(0.566334,0.254896,0.590897); rgb(146pt)=(0.579778,0.246903,0.572366); rgb(147pt)=(0.593221,0.23891,0.553836); rgb(148pt)=(0.606664,0.230917,0.535307); rgb(149pt)=(0.620107,0.222923,0.516777); rgb(150pt)=(0.63355,0.21493,0.498247); rgb(151pt)=(0.646993,0.206937,0.479717); rgb(152pt)=(0.660437,0.198944,0.461187); rgb(153pt)=(0.67388,0.190951,0.442658); rgb(154pt)=(0.687323,0.182958,0.424128); rgb(155pt)=(0.700766,0.174963,0.405598); rgb(156pt)=(0.71421,0.166966,0.387067); rgb(157pt)=(0.727654,0.158972,0.368536); rgb(158pt)=(0.741096,0.150985,0.350008); rgb(159pt)=(0.754536,0.143004,0.331483); rgb(160pt)=(0.767977,0.135017,0.312955); rgb(161pt)=(0.781425,0.127007,0.29442); rgb(162pt)=(0.794879,0.118969,0.275874); rgb(163pt)=(0.808331,0.110944,0.257333); rgb(164pt)=(0.821763,0.102994,0.238819); rgb(165pt)=(0.835166,0.0951591,0.220344); rgb(166pt)=(0.848574,0.0873055,0.201863); rgb(167pt)=(0.862043,0.0792099,0.183297); rgb(168pt)=(0.875621,0.0706838,0.164581); rgb(169pt)=(0.889214,0.0621028,0.145846); rgb(170pt)=(0.902611,0.0542922,0.12738); rgb(171pt)=(0.915613,0.0480374,0.109458); rgb(172pt)=(0.928457,0.0424061,0.091754); rgb(173pt)=(0.941904,0.0343962,0.0732183); rgb(174pt)=(0.956736,0.0209305,0.052775); rgb(175pt)=(0.9725,0.00378572,0.0310453); rgb(176pt)=(0.986483,0.00633329,0.0117721); rgb(177pt)=(0.995725,0.0022327,0.000967649); rgb(178pt)=(0.998618,0.0358306,0.00495484); rgb(179pt)=(0.997427,0.0855268,0.00331317); rgb(180pt)=(0.995111,0.139657,0.000121229); rgb(181pt)=(0.99409,0.188685,0.00128695); rgb(182pt)=(0.994288,0.232904,0.00101351); rgb(183pt)=(0.994914,0.275439,0.000151196); rgb(184pt)=(0.995257,0.319085,0.000322596); rgb(185pt)=(0.995242,0.364147,0.000301319); rgb(186pt)=(0.995078,0.409793,7.57419e-05); rgb(187pt)=(0.994967,0.455231,7.71666e-05); rgb(188pt)=(0.99496,0.500262,8.74378e-05); rgb(189pt)=(0.995001,0.5451,3.04168e-05); rgb(190pt)=(0.995036,0.589964,1.73228e-05); rgb(191pt)=(0.995042,0.634943,2.51464e-05); rgb(192pt)=(0.995032,0.679984,1.1303e-05); rgb(193pt)=(0.995021,0.725028,3.71976e-06); rgb(194pt)=(0.995017,0.770042,8.33655e-06); rgb(195pt)=(0.99502,0.815035,5.36005e-06); rgb(196pt)=(0.995024,0.860017,1.0858e-06); rgb(197pt)=(0.995028,0.905002,6.87705e-06); rgb(198pt)=(0.995029,0.95,7.88977e-06); rgb(199pt)=(0.995023,0.995023,0)},
]
\addplot [forget plot] graphics [xmin=4.81920750849126, xmax=4.61169901723243, ymin=-1.25420750849126, ymax=-1.04669901723243] {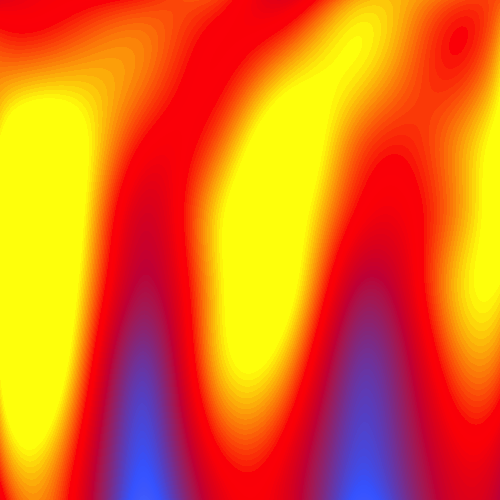};
\addplot[only marks, mark=*, mark options={}, mark size=1.5000pt, color=black, fill=RDlightgreen, forget plot] table[row sep=crcr]{%
x	y\\
4.70066087184211	-1.16524565388158\\
};

\draw[line width = 0.7pt, {Stealth[inset=0pt, scale=1.05, angle'=25]-},shorten < = 1mm] (4.701,-1.165) -- ++(0mm,14mm) node[anchor=south,font={\footnotesize},fill=white,opacity=0.75,inner sep=0.75mm] {\SI{-30.67}{\dB}};

\end{axis}
\end{tikzpicture}%
        \fi 
        \vspace{-2.1mm}
         \caption{\gls{sw} \gls{smc} \gls{bf}}
         \label{fig:SW_SMC_pos2_switched0}
     \end{subfigure}%
     \hspace{\hspaceing}%
     \begin{subfigure}[t]{0.29\textwidth}
         \vskip 0pt	
         \centering
	    \ifdefined\ExcludeFigures   
            \ExcludeFigures
        \else
            \vspace{-0.5mm}
%
%
\definecolor{mycolor1}{rgb}{0.21961,0.21961,0.21961}%

\pgfplotsset{every axis/.append style={
  label style={font=\footnotesize},
  legend style={font=\footnotesize},
  tick label style={font=\footnotesize},
}}

\begin{tikzpicture}

\begin{axis}[%
width=\figurewidth,
height=\figureheight,
at={(0\figurewidth,0\figureheight)},
scale only axis,
point meta min=-60,
point meta max=-30,
axis on top,
xmin=4.61190652572368,
xmax=4.819,
xlabel style={yshift=1.5mm},
xlabel={$x$ in \SI{}{\metre}},
ymin=-1.254,
ymax=-1.04690652572368,
yticklabels=\empty,
xmajorgrids,
ymajorgrids,
grid style=dotted,
axis line style = thick,	
axis background/.style={fill=white},
colormap={mymap}{[1pt] rgb(0pt)=(0.995023,0.995023,0.995023); rgb(73pt)=(0.567253,0.659744,0.995023); rgb(74pt)=(0.561393,0.655151,0.995023); rgb(75pt)=(0.555533,0.650558,0.995023); rgb(76pt)=(0.549674,0.645965,0.995023); rgb(77pt)=(0.543814,0.641373,0.995023); rgb(78pt)=(0.537954,0.63678,0.995023); rgb(79pt)=(0.532094,0.632187,0.995023); rgb(80pt)=(0.526234,0.627594,0.995023); rgb(81pt)=(0.520374,0.623001,0.995023); rgb(82pt)=(0.514514,0.618408,0.995023); rgb(83pt)=(0.508655,0.613815,0.995023); rgb(84pt)=(0.502795,0.609223,0.995023); rgb(85pt)=(0.496935,0.60463,0.995023); rgb(86pt)=(0.491075,0.600037,0.995023); rgb(87pt)=(0.485215,0.595444,0.995023); rgb(88pt)=(0.479355,0.590851,0.995023); rgb(89pt)=(0.473495,0.586258,0.995023); rgb(90pt)=(0.467635,0.581665,0.995023); rgb(91pt)=(0.461776,0.577072,0.995023); rgb(92pt)=(0.455916,0.57248,0.995023); rgb(93pt)=(0.450056,0.567887,0.995023); rgb(94pt)=(0.444196,0.563294,0.995023); rgb(95pt)=(0.438336,0.558701,0.995023); rgb(96pt)=(0.432476,0.554108,0.995024); rgb(97pt)=(0.426616,0.549515,0.995024); rgb(98pt)=(0.420757,0.544922,0.995023); rgb(99pt)=(0.414897,0.540329,0.995023); rgb(100pt)=(0.409037,0.535737,0.995023); rgb(101pt)=(0.403177,0.531144,0.995023); rgb(102pt)=(0.397316,0.526551,0.995025); rgb(103pt)=(0.391455,0.521958,0.995025); rgb(104pt)=(0.385597,0.517365,0.995024); rgb(105pt)=(0.379741,0.512772,0.99502); rgb(106pt)=(0.373885,0.508178,0.995017); rgb(107pt)=(0.368022,0.503586,0.995019); rgb(108pt)=(0.362148,0.498995,0.995033); rgb(109pt)=(0.356273,0.494405,0.995048); rgb(110pt)=(0.350418,0.489811,0.995042); rgb(111pt)=(0.344605,0.48521,0.994998); rgb(112pt)=(0.338806,0.480607,0.994939); rgb(113pt)=(0.332944,0.476014,0.994941); rgb(114pt)=(0.326938,0.471447,0.995081); rgb(115pt)=(0.32084,0.466896,0.99531); rgb(116pt)=(0.31493,0.462312,0.995359); rgb(117pt)=(0.309513,0.457641,0.994933); rgb(118pt)=(0.304551,0.45289,0.994071); rgb(119pt)=(0.299066,0.448231,0.993711); rgb(120pt)=(0.29192,0.443865,0.994946); rgb(121pt)=(0.282789,0.439848,0.998086); rgb(122pt)=(0.274935,0.435607,1); rgb(123pt)=(0.272609,0.430392,0.996609); rgb(124pt)=(0.279306,0.423587,0.984555); rgb(125pt)=(0.293138,0.415525,0.965651); rgb(126pt)=(0.309867,0.406953,0.943967); rgb(127pt)=(0.325491,0.398575,0.923344); rgb(128pt)=(0.339058,0.39056,0.904696); rgb(129pt)=(0.351668,0.382714,0.886966); rgb(130pt)=(0.364431,0.37484,0.869089); rgb(131pt)=(0.377772,0.366865,0.850657); rgb(132pt)=(0.391418,0.358836,0.831932); rgb(133pt)=(0.405066,0.350807,0.813205); rgb(134pt)=(0.418557,0.342805,0.79463); rgb(135pt)=(0.431954,0.33482,0.776145); rgb(136pt)=(0.445337,0.326837,0.757672); rgb(137pt)=(0.458761,0.318847,0.739161); rgb(138pt)=(0.472215,0.310852,0.720621); rgb(139pt)=(0.485675,0.302856,0.702075); rgb(140pt)=(0.499125,0.294862,0.683539); rgb(141pt)=(0.512566,0.286869,0.665011); rgb(142pt)=(0.526005,0.278876,0.646485); rgb(143pt)=(0.539446,0.270883,0.627958); rgb(144pt)=(0.552889,0.26289,0.609428); rgb(145pt)=(0.566334,0.254896,0.590897); rgb(146pt)=(0.579778,0.246903,0.572366); rgb(147pt)=(0.593221,0.23891,0.553836); rgb(148pt)=(0.606664,0.230917,0.535307); rgb(149pt)=(0.620107,0.222923,0.516777); rgb(150pt)=(0.63355,0.21493,0.498247); rgb(151pt)=(0.646993,0.206937,0.479717); rgb(152pt)=(0.660437,0.198944,0.461187); rgb(153pt)=(0.67388,0.190951,0.442658); rgb(154pt)=(0.687323,0.182958,0.424128); rgb(155pt)=(0.700766,0.174963,0.405598); rgb(156pt)=(0.71421,0.166966,0.387067); rgb(157pt)=(0.727654,0.158972,0.368536); rgb(158pt)=(0.741096,0.150985,0.350008); rgb(159pt)=(0.754536,0.143004,0.331483); rgb(160pt)=(0.767977,0.135017,0.312955); rgb(161pt)=(0.781425,0.127007,0.29442); rgb(162pt)=(0.794879,0.118969,0.275874); rgb(163pt)=(0.808331,0.110944,0.257333); rgb(164pt)=(0.821763,0.102994,0.238819); rgb(165pt)=(0.835166,0.0951591,0.220344); rgb(166pt)=(0.848574,0.0873055,0.201863); rgb(167pt)=(0.862043,0.0792099,0.183297); rgb(168pt)=(0.875621,0.0706838,0.164581); rgb(169pt)=(0.889214,0.0621028,0.145846); rgb(170pt)=(0.902611,0.0542922,0.12738); rgb(171pt)=(0.915613,0.0480374,0.109458); rgb(172pt)=(0.928457,0.0424061,0.091754); rgb(173pt)=(0.941904,0.0343962,0.0732183); rgb(174pt)=(0.956736,0.0209305,0.052775); rgb(175pt)=(0.9725,0.00378572,0.0310453); rgb(176pt)=(0.986483,0.00633329,0.0117721); rgb(177pt)=(0.995725,0.0022327,0.000967649); rgb(178pt)=(0.998618,0.0358306,0.00495484); rgb(179pt)=(0.997427,0.0855268,0.00331317); rgb(180pt)=(0.995111,0.139657,0.000121229); rgb(181pt)=(0.99409,0.188685,0.00128695); rgb(182pt)=(0.994288,0.232904,0.00101351); rgb(183pt)=(0.994914,0.275439,0.000151196); rgb(184pt)=(0.995257,0.319085,0.000322596); rgb(185pt)=(0.995242,0.364147,0.000301319); rgb(186pt)=(0.995078,0.409793,7.57419e-05); rgb(187pt)=(0.994967,0.455231,7.71666e-05); rgb(188pt)=(0.99496,0.500262,8.74378e-05); rgb(189pt)=(0.995001,0.5451,3.04168e-05); rgb(190pt)=(0.995036,0.589964,1.73228e-05); rgb(191pt)=(0.995042,0.634943,2.51464e-05); rgb(192pt)=(0.995032,0.679984,1.1303e-05); rgb(193pt)=(0.995021,0.725028,3.71976e-06); rgb(194pt)=(0.995017,0.770042,8.33655e-06); rgb(195pt)=(0.99502,0.815035,5.36005e-06); rgb(196pt)=(0.995024,0.860017,1.0858e-06); rgb(197pt)=(0.995028,0.905002,6.87705e-06); rgb(198pt)=(0.995029,0.95,7.88977e-06); rgb(199pt)=(0.995023,0.995023,0)},
]
\addplot [forget plot] graphics [xmin=4.81920750849126, xmax=4.61169901723243, ymin=-1.25420750849126, ymax=-1.04669901723243] {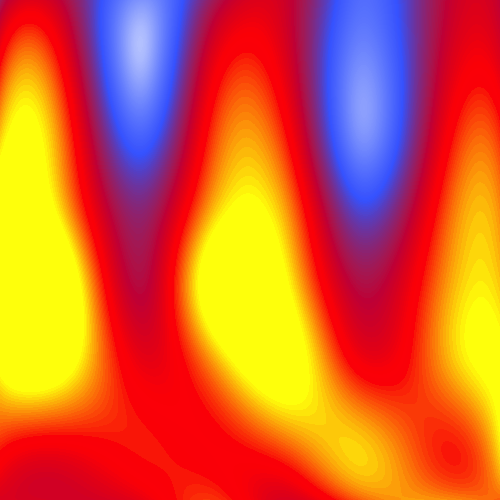};
\addplot[only marks, mark=*, mark options={}, mark size=1.5000pt, color=black, fill=RDlightgreen, forget plot] table[row sep=crcr]{%
x	y\\
4.70066087184211	-1.16524565388158\\
};

\draw[line width = 0.7pt, {Stealth[inset=0pt, scale=1.05, angle'=25]-},shorten < = 1mm] (4.701,-1.165) -- ++(0mm,14mm) node[anchor=south,font={\footnotesize},fill=white,opacity=0.75,inner sep=0.75mm] {\SI{-28.95}{\dB}};

\end{axis}
\end{tikzpicture}%
        \fi 
        \vspace{-2.1mm}
         \caption{Reciprocity-based \gls{bf} (full \gls{csi})}
         \label{fig:CSI_pos2_switched0}
     \end{subfigure}%
     \hspace{-0.65cm}%
     \begin{subfigure}[t]{0.058\textwidth}
        \vspace{-2.2mm} %
         \vskip 0pt	
         \centering
                  \setlength{\plotWidth}{1\textwidth}	
%
%

\pgfplotsset{every axis/.append style={
  label style={font=\footnotesize},
  legend style={font=\footnotesize},
  tick label style={font=\footnotesize}
}}

\begin{tikzpicture}

\begin{axis}[%
width=0.225\plotWidth,
height=3.38\plotWidth,
at={(0\plotWidth,0\plotWidth)},
scale only axis,
xmin=0,
xmax=1.02,
ymin=-1,
ymax=1,
zmin=-60,
zmax=-30,
zlabel={$PG$ in \SI{}{\dB}},
axis line style = thick,	
view={0}{0},
axis background/.style={fill=white},
axis z line*=right,			
xticklabel=\empty,			
tick align=center,		    
xtick=\empty,               
y axis line style= { draw opacity=0 }, 
z axis line style= { draw opacity=0 }, 
ztick distance={5},				
yticklabel pos=right,
xmajorgrids,
ymajorgrids,
zmajorgrids
]

\addplot3[%
surf,
shader=flat, z buffer=sort, colormap={mymap}{[1pt] rgb(0pt)=(1,1,1); rgb(26pt)=(0.51634,0.620915,1); rgb(27pt)=(0.497738,0.606335,1); rgb(34pt)=(0.367521,0.504274,1); rgb(35pt)=(0.348919,0.489693,1); rgb(39pt)=(0.27451,0.431373,1); rgb(40pt)=(0.317186,0.405998,0.941176); rgb(41pt)=(0.359862,0.380623,0.882353); rgb(43pt)=(0.445213,0.329873,0.764706); rgb(44pt)=(0.487889,0.304498,0.705882); rgb(45pt)=(0.530565,0.279123,0.647059); rgb(46pt)=(0.573241,0.253749,0.588235); rgb(47pt)=(0.615917,0.228374,0.529412); rgb(48pt)=(0.658593,0.202999,0.470588); rgb(49pt)=(0.701269,0.177624,0.411765); rgb(50pt)=(0.743945,0.152249,0.352941); rgb(51pt)=(0.786621,0.126874,0.294118); rgb(52pt)=(0.829296,0.101499,0.235294); rgb(53pt)=(0.871972,0.0761246,0.176471); rgb(54pt)=(0.914648,0.0507497,0.117647); rgb(55pt)=(0.957324,0.0253749,0.0588235); rgb(56pt)=(1,0,0); rgb(58pt)=(1,0.285714,0); rgb(59pt)=(1,0.428571,0); rgb(60pt)=(1,0.571429,0); rgb(63pt)=(1,1,0)}, mesh/rows=2]
table[row sep=crcr, point meta=\thisrow{c}] {%
x	y	z	c\\
0	-0.0001	-30.00	-30.00\\ 
0	-0.0001	-30.30	-30.30\\ 
0	-0.0001	-30.61	-30.61\\ 
0	-0.0001	-30.91	-30.91\\ 
0	-0.0001	-31.21	-31.21\\ 
0	-0.0001	-31.52	-31.52\\ 
0	-0.0001	-31.82	-31.82\\ 
0	-0.0001	-32.12	-32.12\\ 
0	-0.0001	-32.42	-32.42\\ 
0	-0.0001	-32.73	-32.73\\ 
0	-0.0001	-33.03	-33.03\\ 
0	-0.0001	-33.33	-33.33\\ 
0	-0.0001	-33.64	-33.64\\ 
0	-0.0001	-33.94	-33.94\\ 
0	-0.0001	-34.24	-34.24\\ 
0	-0.0001	-34.55	-34.55\\ 
0	-0.0001	-34.85	-34.85\\ 
0	-0.0001	-35.15	-35.15\\ 
0	-0.0001	-35.45	-35.45\\ 
0	-0.0001	-35.76	-35.76\\ 
0	-0.0001	-36.06	-36.06\\ 
0	-0.0001	-36.36	-36.36\\ 
0	-0.0001	-36.67	-36.67\\ 
0	-0.0001	-36.97	-36.97\\ 
0	-0.0001	-37.27	-37.27\\ 
0	-0.0001	-37.58	-37.58\\ 
0	-0.0001	-37.88	-37.88\\ 
0	-0.0001	-38.18	-38.18\\ 
0	-0.0001	-38.48	-38.48\\ 
0	-0.0001	-38.79	-38.79\\ 
0	-0.0001	-39.09	-39.09\\ 
0	-0.0001	-39.39	-39.39\\ 
0	-0.0001	-39.70	-39.70\\ 
0	-0.0001	-40.00	-40.00\\ 
0	-0.0001	-40.30	-40.30\\ 
0	-0.0001	-40.61	-40.61\\ 
0	-0.0001	-40.91	-40.91\\ 
0	-0.0001	-41.21	-41.21\\ 
0	-0.0001	-41.52	-41.52\\ 
0	-0.0001	-41.82	-41.82\\ 
0	-0.0001	-42.12	-42.12\\ 
0	-0.0001	-42.42	-42.42\\ 
0	-0.0001	-42.73	-42.73\\ 
0	-0.0001	-43.03	-43.03\\ 
0	-0.0001	-43.33	-43.33\\ 
0	-0.0001	-43.64	-43.64\\ 
0	-0.0001	-43.94	-43.94\\ 
0	-0.0001	-44.24	-44.24\\ 
0	-0.0001	-44.55	-44.55\\ 
0	-0.0001	-44.85	-44.85\\ 
0	-0.0001	-45.15	-45.15\\ 
0	-0.0001	-45.45	-45.45\\ 
0	-0.0001	-45.76	-45.76\\ 
0	-0.0001	-46.06	-46.06\\ 
0	-0.0001	-46.36	-46.36\\ 
0	-0.0001	-46.67	-46.67\\ 
0	-0.0001	-46.97	-46.97\\ 
0	-0.0001	-47.27	-47.27\\ 
0	-0.0001	-47.58	-47.58\\ 
0	-0.0001	-47.88	-47.88\\ 
0	-0.0001	-48.18	-48.18\\ 
0	-0.0001	-48.48	-48.48\\ 
0	-0.0001	-48.79	-48.79\\ 
0	-0.0001	-49.09	-49.09\\ 
0	-0.0001	-49.39	-49.39\\ 
0	-0.0001	-49.70	-49.70\\ 
0	-0.0001	-50.00	-50.00\\ 
0	-0.0001	-50.30	-50.30\\ 
0	-0.0001	-50.61	-50.61\\ 
0	-0.0001	-50.91	-50.91\\ 
0	-0.0001	-51.21	-51.21\\ 
0	-0.0001	-51.52	-51.52\\ 
0	-0.0001	-51.82	-51.82\\ 
0	-0.0001	-52.12	-52.12\\ 
0	-0.0001	-52.42	-52.42\\ 
0	-0.0001	-52.73	-52.73\\ 
0	-0.0001	-53.03	-53.03\\ 
0	-0.0001	-53.33	-53.33\\ 
0	-0.0001	-53.64	-53.64\\ 
0	-0.0001	-53.94	-53.94\\ 
0	-0.0001	-54.24	-54.24\\ 
0	-0.0001	-54.55	-54.55\\ 
0	-0.0001	-54.85	-54.85\\ 
0	-0.0001	-55.15	-55.15\\ 
0	-0.0001	-55.45	-55.45\\ 
0	-0.0001	-55.76	-55.76\\ 
0	-0.0001	-56.06	-56.06\\ 
0	-0.0001	-56.36	-56.36\\ 
0	-0.0001	-56.67	-56.67\\ 
0	-0.0001	-56.97	-56.97\\ 
0	-0.0001	-57.27	-57.27\\ 
0	-0.0001	-57.58	-57.58\\ 
0	-0.0001	-57.88	-57.88\\ 
0	-0.0001	-58.18	-58.18\\ 
0	-0.0001	-58.48	-58.48\\ 
0	-0.0001	-58.79	-58.79\\ 
0	-0.0001	-59.09	-59.09\\ 
0	-0.0001	-59.39	-59.39\\ 
0	-0.0001	-59.70	-59.70\\ 
0	-0.0001	-60.00	-60.00\\ 
1	-0.0001	-30.00	-30.00\\ 
1	-0.0001	-30.30	-30.30\\ 
1	-0.0001	-30.61	-30.61\\ 
1	-0.0001	-30.91	-30.91\\ 
1	-0.0001	-31.21	-31.21\\ 
1	-0.0001	-31.52	-31.52\\ 
1	-0.0001	-31.82	-31.82\\ 
1	-0.0001	-32.12	-32.12\\ 
1	-0.0001	-32.42	-32.42\\ 
1	-0.0001	-32.73	-32.73\\ 
1	-0.0001	-33.03	-33.03\\ 
1	-0.0001	-33.33	-33.33\\ 
1	-0.0001	-33.64	-33.64\\ 
1	-0.0001	-33.94	-33.94\\ 
1	-0.0001	-34.24	-34.24\\ 
1	-0.0001	-34.55	-34.55\\ 
1	-0.0001	-34.85	-34.85\\ 
1	-0.0001	-35.15	-35.15\\ 
1	-0.0001	-35.45	-35.45\\ 
1	-0.0001	-35.76	-35.76\\ 
1	-0.0001	-36.06	-36.06\\ 
1	-0.0001	-36.36	-36.36\\ 
1	-0.0001	-36.67	-36.67\\ 
1	-0.0001	-36.97	-36.97\\ 
1	-0.0001	-37.27	-37.27\\ 
1	-0.0001	-37.58	-37.58\\ 
1	-0.0001	-37.88	-37.88\\ 
1	-0.0001	-38.18	-38.18\\ 
1	-0.0001	-38.48	-38.48\\ 
1	-0.0001	-38.79	-38.79\\ 
1	-0.0001	-39.09	-39.09\\ 
1	-0.0001	-39.39	-39.39\\ 
1	-0.0001	-39.70	-39.70\\ 
1	-0.0001	-40.00	-40.00\\ 
1	-0.0001	-40.30	-40.30\\ 
1	-0.0001	-40.61	-40.61\\ 
1	-0.0001	-40.91	-40.91\\ 
1	-0.0001	-41.21	-41.21\\ 
1	-0.0001	-41.52	-41.52\\ 
1	-0.0001	-41.82	-41.82\\ 
1	-0.0001	-42.12	-42.12\\ 
1	-0.0001	-42.42	-42.42\\ 
1	-0.0001	-42.73	-42.73\\ 
1	-0.0001	-43.03	-43.03\\ 
1	-0.0001	-43.33	-43.33\\ 
1	-0.0001	-43.64	-43.64\\ 
1	-0.0001	-43.94	-43.94\\ 
1	-0.0001	-44.24	-44.24\\ 
1	-0.0001	-44.55	-44.55\\ 
1	-0.0001	-44.85	-44.85\\ 
1	-0.0001	-45.15	-45.15\\ 
1	-0.0001	-45.45	-45.45\\ 
1	-0.0001	-45.76	-45.76\\ 
1	-0.0001	-46.06	-46.06\\ 
1	-0.0001	-46.36	-46.36\\ 
1	-0.0001	-46.67	-46.67\\ 
1	-0.0001	-46.97	-46.97\\ 
1	-0.0001	-47.27	-47.27\\ 
1	-0.0001	-47.58	-47.58\\ 
1	-0.0001	-47.88	-47.88\\ 
1	-0.0001	-48.18	-48.18\\ 
1	-0.0001	-48.48	-48.48\\ 
1	-0.0001	-48.79	-48.79\\ 
1	-0.0001	-49.09	-49.09\\ 
1	-0.0001	-49.39	-49.39\\ 
1	-0.0001	-49.70	-49.70\\ 
1	-0.0001	-50.00	-50.00\\ 
1	-0.0001	-50.30	-50.30\\ 
1	-0.0001	-50.61	-50.61\\ 
1	-0.0001	-50.91	-50.91\\ 
1	-0.0001	-51.21	-51.21\\ 
1	-0.0001	-51.52	-51.52\\ 
1	-0.0001	-51.82	-51.82\\ 
1	-0.0001	-52.12	-52.12\\ 
1	-0.0001	-52.42	-52.42\\ 
1	-0.0001	-52.73	-52.73\\ 
1	-0.0001	-53.03	-53.03\\ 
1	-0.0001	-53.33	-53.33\\ 
1	-0.0001	-53.64	-53.64\\ 
1	-0.0001	-53.94	-53.94\\ 
1	-0.0001	-54.24	-54.24\\ 
1	-0.0001	-54.55	-54.55\\ 
1	-0.0001	-54.85	-54.85\\ 
1	-0.0001	-55.15	-55.15\\ 
1	-0.0001	-55.45	-55.45\\ 
1	-0.0001	-55.76	-55.76\\ 
1	-0.0001	-56.06	-56.06\\ 
1	-0.0001	-56.36	-56.36\\ 
1	-0.0001	-56.67	-56.67\\ 
1	-0.0001	-56.97	-56.97\\ 
1	-0.0001	-57.27	-57.27\\ 
1	-0.0001	-57.58	-57.58\\ 
1	-0.0001	-57.88	-57.88\\ 
1	-0.0001	-58.18	-58.18\\ 
1	-0.0001	-58.48	-58.48\\ 
1	-0.0001	-58.79	-58.79\\ 
1	-0.0001	-59.09	-59.09\\ 
1	-0.0001	-59.39	-59.39\\ 
1	-0.0001	-59.70	-59.70\\ 
1	-0.0001	-60.00	-60.00\\ 
};
\draw [draw=black, thick] (0.015,0.98,-60) rectangle (1,1,-30);    
\end{axis}
\end{tikzpicture}%
     \end{subfigure}
    	\vspace{-1mm}\caption{
    	Beamformers applied at $\bm{p}_\mrm{EN}^{(2)}$, $PG$ distribution evaluated and interpolated around  $\bm{p}_\mrm{EN}^{(2)}$.
    	}\label{fig:pos2_domains}
    	\vspace{-3mm}
\end{figure*}	

\ifdefined\dontreduceSize  
    ~
\else
    \begin{figure*}[bt]	
         \hspace{-4mm}
         \begin{subfigure}[t]{0.29\textwidth}
            \vskip 0pt	
             \centering
    	    \ifdefined\ExcludeFigures
    	                \ExcludeFigures
            \else
%
%
\definecolor{mycolor1}{rgb}{0.21961,0.21961,0.21961}%
\pgfplotsset{every axis/.append style={
  label style={font=\footnotesize},
  legend style={font=\footnotesize},
  tick label style={font=\footnotesize},
}}

\begin{tikzpicture}

\begin{axis}[%
width=\figurewidth,
height=\figureheight,
at={(0\figurewidth,0\figureheight)},
scale only axis,
point meta min=-60,
point meta max=-30,
axis on top,
xmin=4.61190652572368,
xmax=4.819,
xlabel style={yshift=1.5mm},
xlabel={$x$ in \SI{}{\metre}},
ymin=-1.254,
ymax=-1.04690652572368,
ylabel style={yshift=-6pt},
ylabel={$y$ in \SI{}{\metre}},
ytick={-1.25,-1.2,-1.15,-1.1},
axis line style = thick,	
axis background/.style={fill=white},
colormap={mymap}{[1pt] rgb(0pt)=(0.995023,0.995023,0.995023); rgb(73pt)=(0.567253,0.659744,0.995023); rgb(74pt)=(0.561393,0.655151,0.995023); rgb(75pt)=(0.555533,0.650558,0.995023); rgb(76pt)=(0.549674,0.645965,0.995023); rgb(77pt)=(0.543814,0.641373,0.995023); rgb(78pt)=(0.537954,0.63678,0.995023); rgb(79pt)=(0.532094,0.632187,0.995023); rgb(80pt)=(0.526234,0.627594,0.995023); rgb(81pt)=(0.520374,0.623001,0.995023); rgb(82pt)=(0.514514,0.618408,0.995023); rgb(83pt)=(0.508655,0.613815,0.995023); rgb(84pt)=(0.502795,0.609223,0.995023); rgb(85pt)=(0.496935,0.60463,0.995023); rgb(86pt)=(0.491075,0.600037,0.995023); rgb(87pt)=(0.485215,0.595444,0.995023); rgb(88pt)=(0.479355,0.590851,0.995023); rgb(89pt)=(0.473495,0.586258,0.995023); rgb(90pt)=(0.467635,0.581665,0.995023); rgb(91pt)=(0.461776,0.577072,0.995023); rgb(92pt)=(0.455916,0.57248,0.995023); rgb(93pt)=(0.450056,0.567887,0.995023); rgb(94pt)=(0.444196,0.563294,0.995023); rgb(95pt)=(0.438336,0.558701,0.995023); rgb(96pt)=(0.432476,0.554108,0.995024); rgb(97pt)=(0.426616,0.549515,0.995024); rgb(98pt)=(0.420757,0.544922,0.995023); rgb(99pt)=(0.414897,0.540329,0.995023); rgb(100pt)=(0.409037,0.535737,0.995023); rgb(101pt)=(0.403177,0.531144,0.995023); rgb(102pt)=(0.397316,0.526551,0.995025); rgb(103pt)=(0.391455,0.521958,0.995025); rgb(104pt)=(0.385597,0.517365,0.995024); rgb(105pt)=(0.379741,0.512772,0.99502); rgb(106pt)=(0.373885,0.508178,0.995017); rgb(107pt)=(0.368022,0.503586,0.995019); rgb(108pt)=(0.362148,0.498995,0.995033); rgb(109pt)=(0.356273,0.494405,0.995048); rgb(110pt)=(0.350418,0.489811,0.995042); rgb(111pt)=(0.344605,0.48521,0.994998); rgb(112pt)=(0.338806,0.480607,0.994939); rgb(113pt)=(0.332944,0.476014,0.994941); rgb(114pt)=(0.326938,0.471447,0.995081); rgb(115pt)=(0.32084,0.466896,0.99531); rgb(116pt)=(0.31493,0.462312,0.995359); rgb(117pt)=(0.309513,0.457641,0.994933); rgb(118pt)=(0.304551,0.45289,0.994071); rgb(119pt)=(0.299066,0.448231,0.993711); rgb(120pt)=(0.29192,0.443865,0.994946); rgb(121pt)=(0.282789,0.439848,0.998086); rgb(122pt)=(0.274935,0.435607,1); rgb(123pt)=(0.272609,0.430392,0.996609); rgb(124pt)=(0.279306,0.423587,0.984555); rgb(125pt)=(0.293138,0.415525,0.965651); rgb(126pt)=(0.309867,0.406953,0.943967); rgb(127pt)=(0.325491,0.398575,0.923344); rgb(128pt)=(0.339058,0.39056,0.904696); rgb(129pt)=(0.351668,0.382714,0.886966); rgb(130pt)=(0.364431,0.37484,0.869089); rgb(131pt)=(0.377772,0.366865,0.850657); rgb(132pt)=(0.391418,0.358836,0.831932); rgb(133pt)=(0.405066,0.350807,0.813205); rgb(134pt)=(0.418557,0.342805,0.79463); rgb(135pt)=(0.431954,0.33482,0.776145); rgb(136pt)=(0.445337,0.326837,0.757672); rgb(137pt)=(0.458761,0.318847,0.739161); rgb(138pt)=(0.472215,0.310852,0.720621); rgb(139pt)=(0.485675,0.302856,0.702075); rgb(140pt)=(0.499125,0.294862,0.683539); rgb(141pt)=(0.512566,0.286869,0.665011); rgb(142pt)=(0.526005,0.278876,0.646485); rgb(143pt)=(0.539446,0.270883,0.627958); rgb(144pt)=(0.552889,0.26289,0.609428); rgb(145pt)=(0.566334,0.254896,0.590897); rgb(146pt)=(0.579778,0.246903,0.572366); rgb(147pt)=(0.593221,0.23891,0.553836); rgb(148pt)=(0.606664,0.230917,0.535307); rgb(149pt)=(0.620107,0.222923,0.516777); rgb(150pt)=(0.63355,0.21493,0.498247); rgb(151pt)=(0.646993,0.206937,0.479717); rgb(152pt)=(0.660437,0.198944,0.461187); rgb(153pt)=(0.67388,0.190951,0.442658); rgb(154pt)=(0.687323,0.182958,0.424128); rgb(155pt)=(0.700766,0.174963,0.405598); rgb(156pt)=(0.71421,0.166966,0.387067); rgb(157pt)=(0.727654,0.158972,0.368536); rgb(158pt)=(0.741096,0.150985,0.350008); rgb(159pt)=(0.754536,0.143004,0.331483); rgb(160pt)=(0.767977,0.135017,0.312955); rgb(161pt)=(0.781425,0.127007,0.29442); rgb(162pt)=(0.794879,0.118969,0.275874); rgb(163pt)=(0.808331,0.110944,0.257333); rgb(164pt)=(0.821763,0.102994,0.238819); rgb(165pt)=(0.835166,0.0951591,0.220344); rgb(166pt)=(0.848574,0.0873055,0.201863); rgb(167pt)=(0.862043,0.0792099,0.183297); rgb(168pt)=(0.875621,0.0706838,0.164581); rgb(169pt)=(0.889214,0.0621028,0.145846); rgb(170pt)=(0.902611,0.0542922,0.12738); rgb(171pt)=(0.915613,0.0480374,0.109458); rgb(172pt)=(0.928457,0.0424061,0.091754); rgb(173pt)=(0.941904,0.0343962,0.0732183); rgb(174pt)=(0.956736,0.0209305,0.052775); rgb(175pt)=(0.9725,0.00378572,0.0310453); rgb(176pt)=(0.986483,0.00633329,0.0117721); rgb(177pt)=(0.995725,0.0022327,0.000967649); rgb(178pt)=(0.998618,0.0358306,0.00495484); rgb(179pt)=(0.997427,0.0855268,0.00331317); rgb(180pt)=(0.995111,0.139657,0.000121229); rgb(181pt)=(0.99409,0.188685,0.00128695); rgb(182pt)=(0.994288,0.232904,0.00101351); rgb(183pt)=(0.994914,0.275439,0.000151196); rgb(184pt)=(0.995257,0.319085,0.000322596); rgb(185pt)=(0.995242,0.364147,0.000301319); rgb(186pt)=(0.995078,0.409793,7.57419e-05); rgb(187pt)=(0.994967,0.455231,7.71666e-05); rgb(188pt)=(0.99496,0.500262,8.74378e-05); rgb(189pt)=(0.995001,0.5451,3.04168e-05); rgb(190pt)=(0.995036,0.589964,1.73228e-05); rgb(191pt)=(0.995042,0.634943,2.51464e-05); rgb(192pt)=(0.995032,0.679984,1.1303e-05); rgb(193pt)=(0.995021,0.725028,3.71976e-06); rgb(194pt)=(0.995017,0.770042,8.33655e-06); rgb(195pt)=(0.99502,0.815035,5.36005e-06); rgb(196pt)=(0.995024,0.860017,1.0858e-06); rgb(197pt)=(0.995028,0.905002,6.87705e-06); rgb(198pt)=(0.995029,0.95,7.88977e-06); rgb(199pt)=(0.995023,0.995023,0)},
]
\addplot [forget plot] graphics [xmin=4.81920750849126, xmax=4.61169901723243, ymin=-1.25420750849126, ymax=-1.04669901723243] {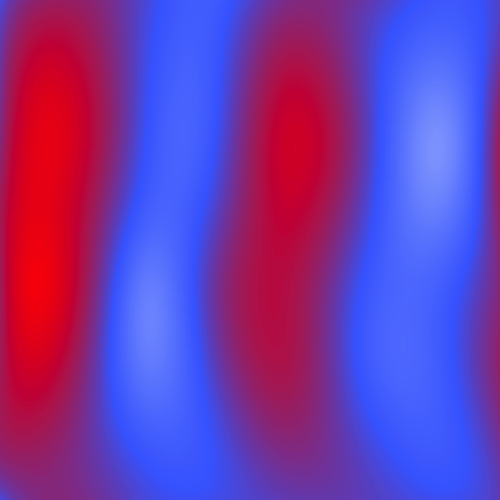};
\addplot[only marks, mark=*, mark options={}, mark size=1.7678pt, color=mycolor1, fill=mycolor1, forget plot] table[row sep=crcr]{%
x	y\\
3.46166087184211	-2.05024565388158\\
};
\end{axis}
\end{tikzpicture}%
            \fi 
            \vspace{-2.1mm}
             \caption{\Gls{pw} \gls{los} \gls{bf}}
             \label{fig:PW_pos1_switched1}
         \end{subfigure}%
         \hspace{-0.78cm}%
         \begin{subfigure}[t]{0.29\textwidth}
            \vskip 0pt	
             \centering
    	    \ifdefined\ExcludeFigures   
                \ExcludeFigures
            \else
                \vspace{-0.5mm}
%
%
\definecolor{mycolor1}{rgb}{0.00000,0.44700,0.74100}%
\pgfplotsset{every axis/.append style={
  label style={font=\footnotesize},
  legend style={font=\footnotesize},
  tick label style={font=\footnotesize},
}}

\begin{tikzpicture}

\begin{axis}[%
width=\figurewidth,
height=\figureheight,
at={(0\figurewidth,0\figureheight)},
scale only axis,
point meta min=-60,
point meta max=-30,
axis on top,
xmin=4.61190652572368,
xmax=4.819,
xlabel style={yshift=1.5mm},
xlabel={$x$ in \SI{}{\metre}},
ymin=-1.254,
ymax=-1.04690652572368,
yticklabels=\empty,
axis line style = thick,	
axis background/.style={fill=white},
colormap={mymap}{[1pt] rgb(0pt)=(0.995023,0.995023,0.995023); rgb(73pt)=(0.567253,0.659744,0.995023); rgb(74pt)=(0.561393,0.655151,0.995023); rgb(75pt)=(0.555533,0.650558,0.995023); rgb(76pt)=(0.549674,0.645965,0.995023); rgb(77pt)=(0.543814,0.641373,0.995023); rgb(78pt)=(0.537954,0.63678,0.995023); rgb(79pt)=(0.532094,0.632187,0.995023); rgb(80pt)=(0.526234,0.627594,0.995023); rgb(81pt)=(0.520374,0.623001,0.995023); rgb(82pt)=(0.514514,0.618408,0.995023); rgb(83pt)=(0.508655,0.613815,0.995023); rgb(84pt)=(0.502795,0.609223,0.995023); rgb(85pt)=(0.496935,0.60463,0.995023); rgb(86pt)=(0.491075,0.600037,0.995023); rgb(87pt)=(0.485215,0.595444,0.995023); rgb(88pt)=(0.479355,0.590851,0.995023); rgb(89pt)=(0.473495,0.586258,0.995023); rgb(90pt)=(0.467635,0.581665,0.995023); rgb(91pt)=(0.461776,0.577072,0.995023); rgb(92pt)=(0.455916,0.57248,0.995023); rgb(93pt)=(0.450056,0.567887,0.995023); rgb(94pt)=(0.444196,0.563294,0.995023); rgb(95pt)=(0.438336,0.558701,0.995023); rgb(96pt)=(0.432476,0.554108,0.995024); rgb(97pt)=(0.426616,0.549515,0.995024); rgb(98pt)=(0.420757,0.544922,0.995023); rgb(99pt)=(0.414897,0.540329,0.995023); rgb(100pt)=(0.409037,0.535737,0.995023); rgb(101pt)=(0.403177,0.531144,0.995023); rgb(102pt)=(0.397316,0.526551,0.995025); rgb(103pt)=(0.391455,0.521958,0.995025); rgb(104pt)=(0.385597,0.517365,0.995024); rgb(105pt)=(0.379741,0.512772,0.99502); rgb(106pt)=(0.373885,0.508178,0.995017); rgb(107pt)=(0.368022,0.503586,0.995019); rgb(108pt)=(0.362148,0.498995,0.995033); rgb(109pt)=(0.356273,0.494405,0.995048); rgb(110pt)=(0.350418,0.489811,0.995042); rgb(111pt)=(0.344605,0.48521,0.994998); rgb(112pt)=(0.338806,0.480607,0.994939); rgb(113pt)=(0.332944,0.476014,0.994941); rgb(114pt)=(0.326938,0.471447,0.995081); rgb(115pt)=(0.32084,0.466896,0.99531); rgb(116pt)=(0.31493,0.462312,0.995359); rgb(117pt)=(0.309513,0.457641,0.994933); rgb(118pt)=(0.304551,0.45289,0.994071); rgb(119pt)=(0.299066,0.448231,0.993711); rgb(120pt)=(0.29192,0.443865,0.994946); rgb(121pt)=(0.282789,0.439848,0.998086); rgb(122pt)=(0.274935,0.435607,1); rgb(123pt)=(0.272609,0.430392,0.996609); rgb(124pt)=(0.279306,0.423587,0.984555); rgb(125pt)=(0.293138,0.415525,0.965651); rgb(126pt)=(0.309867,0.406953,0.943967); rgb(127pt)=(0.325491,0.398575,0.923344); rgb(128pt)=(0.339058,0.39056,0.904696); rgb(129pt)=(0.351668,0.382714,0.886966); rgb(130pt)=(0.364431,0.37484,0.869089); rgb(131pt)=(0.377772,0.366865,0.850657); rgb(132pt)=(0.391418,0.358836,0.831932); rgb(133pt)=(0.405066,0.350807,0.813205); rgb(134pt)=(0.418557,0.342805,0.79463); rgb(135pt)=(0.431954,0.33482,0.776145); rgb(136pt)=(0.445337,0.326837,0.757672); rgb(137pt)=(0.458761,0.318847,0.739161); rgb(138pt)=(0.472215,0.310852,0.720621); rgb(139pt)=(0.485675,0.302856,0.702075); rgb(140pt)=(0.499125,0.294862,0.683539); rgb(141pt)=(0.512566,0.286869,0.665011); rgb(142pt)=(0.526005,0.278876,0.646485); rgb(143pt)=(0.539446,0.270883,0.627958); rgb(144pt)=(0.552889,0.26289,0.609428); rgb(145pt)=(0.566334,0.254896,0.590897); rgb(146pt)=(0.579778,0.246903,0.572366); rgb(147pt)=(0.593221,0.23891,0.553836); rgb(148pt)=(0.606664,0.230917,0.535307); rgb(149pt)=(0.620107,0.222923,0.516777); rgb(150pt)=(0.63355,0.21493,0.498247); rgb(151pt)=(0.646993,0.206937,0.479717); rgb(152pt)=(0.660437,0.198944,0.461187); rgb(153pt)=(0.67388,0.190951,0.442658); rgb(154pt)=(0.687323,0.182958,0.424128); rgb(155pt)=(0.700766,0.174963,0.405598); rgb(156pt)=(0.71421,0.166966,0.387067); rgb(157pt)=(0.727654,0.158972,0.368536); rgb(158pt)=(0.741096,0.150985,0.350008); rgb(159pt)=(0.754536,0.143004,0.331483); rgb(160pt)=(0.767977,0.135017,0.312955); rgb(161pt)=(0.781425,0.127007,0.29442); rgb(162pt)=(0.794879,0.118969,0.275874); rgb(163pt)=(0.808331,0.110944,0.257333); rgb(164pt)=(0.821763,0.102994,0.238819); rgb(165pt)=(0.835166,0.0951591,0.220344); rgb(166pt)=(0.848574,0.0873055,0.201863); rgb(167pt)=(0.862043,0.0792099,0.183297); rgb(168pt)=(0.875621,0.0706838,0.164581); rgb(169pt)=(0.889214,0.0621028,0.145846); rgb(170pt)=(0.902611,0.0542922,0.12738); rgb(171pt)=(0.915613,0.0480374,0.109458); rgb(172pt)=(0.928457,0.0424061,0.091754); rgb(173pt)=(0.941904,0.0343962,0.0732183); rgb(174pt)=(0.956736,0.0209305,0.052775); rgb(175pt)=(0.9725,0.00378572,0.0310453); rgb(176pt)=(0.986483,0.00633329,0.0117721); rgb(177pt)=(0.995725,0.0022327,0.000967649); rgb(178pt)=(0.998618,0.0358306,0.00495484); rgb(179pt)=(0.997427,0.0855268,0.00331317); rgb(180pt)=(0.995111,0.139657,0.000121229); rgb(181pt)=(0.99409,0.188685,0.00128695); rgb(182pt)=(0.994288,0.232904,0.00101351); rgb(183pt)=(0.994914,0.275439,0.000151196); rgb(184pt)=(0.995257,0.319085,0.000322596); rgb(185pt)=(0.995242,0.364147,0.000301319); rgb(186pt)=(0.995078,0.409793,7.57419e-05); rgb(187pt)=(0.994967,0.455231,7.71666e-05); rgb(188pt)=(0.99496,0.500262,8.74378e-05); rgb(189pt)=(0.995001,0.5451,3.04168e-05); rgb(190pt)=(0.995036,0.589964,1.73228e-05); rgb(191pt)=(0.995042,0.634943,2.51464e-05); rgb(192pt)=(0.995032,0.679984,1.1303e-05); rgb(193pt)=(0.995021,0.725028,3.71976e-06); rgb(194pt)=(0.995017,0.770042,8.33655e-06); rgb(195pt)=(0.99502,0.815035,5.36005e-06); rgb(196pt)=(0.995024,0.860017,1.0858e-06); rgb(197pt)=(0.995028,0.905002,6.87705e-06); rgb(198pt)=(0.995029,0.95,7.88977e-06); rgb(199pt)=(0.995023,0.995023,0)},
]
\addplot [forget plot] graphics [xmin=4.81920750849126, xmax=4.61169901723243, ymin=-1.25420750849126, ymax=-1.04669901723243] {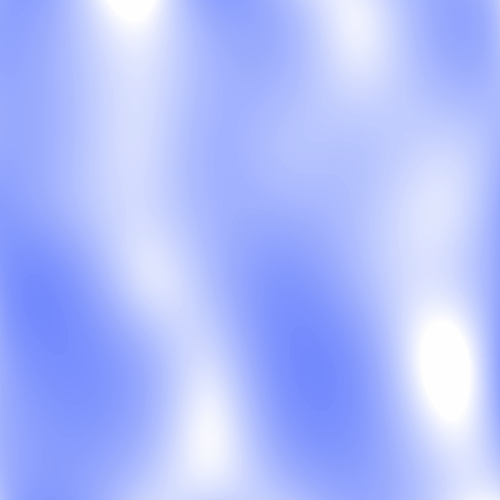};
\addplot[only marks, mark=*, mark options={}, mark size=1.5000pt, color=mycolor1, fill=mycolor1, forget plot] table[row sep=crcr]{%
x	y\\
3.46166087184211	-2.05024565388158\\
};
\end{axis}
\end{tikzpicture}%
            \fi 
             \vspace{-2.1mm}
             \caption{\Gls{sw} \gls{los} \gls{bf}}
             \label{fig:SW_LoS_pos1_switched1}
         \end{subfigure}%
         \hspace{\hspaceing}%
         \begin{subfigure}[t]{0.29\textwidth}
             \vskip 0pt	
             \centering
    	    \ifdefined\ExcludeFigures   
                \ExcludeFigures
            \else
                \vspace{-0.5mm}
%
%
\definecolor{mycolor1}{rgb}{0.36471,0.42745,0.26667}%

\pgfplotsset{every axis/.append style={
  label style={font=\footnotesize},
  legend style={font=\footnotesize},
  tick label style={font=\footnotesize},
}}

\begin{tikzpicture}

\begin{axis}[%
width=\figurewidth,
height=\figureheight,
at={(0\figurewidth,0\figureheight)},
scale only axis,
point meta min=-60,
point meta max=-30,
axis on top,
xmin=4.61190652572368,
xmax=4.819,
xlabel style={yshift=1.5mm},
xlabel={$x$ in \SI{}{\metre}},
ymin=-1.254,
ymax=-1.04690652572368,
yticklabels=\empty,
axis line style = thick,	
axis background/.style={fill=white},
colormap={mymap}{[1pt] rgb(0pt)=(0.995023,0.995023,0.995023); rgb(73pt)=(0.567253,0.659744,0.995023); rgb(74pt)=(0.561393,0.655151,0.995023); rgb(75pt)=(0.555533,0.650558,0.995023); rgb(76pt)=(0.549674,0.645965,0.995023); rgb(77pt)=(0.543814,0.641373,0.995023); rgb(78pt)=(0.537954,0.63678,0.995023); rgb(79pt)=(0.532094,0.632187,0.995023); rgb(80pt)=(0.526234,0.627594,0.995023); rgb(81pt)=(0.520374,0.623001,0.995023); rgb(82pt)=(0.514514,0.618408,0.995023); rgb(83pt)=(0.508655,0.613815,0.995023); rgb(84pt)=(0.502795,0.609223,0.995023); rgb(85pt)=(0.496935,0.60463,0.995023); rgb(86pt)=(0.491075,0.600037,0.995023); rgb(87pt)=(0.485215,0.595444,0.995023); rgb(88pt)=(0.479355,0.590851,0.995023); rgb(89pt)=(0.473495,0.586258,0.995023); rgb(90pt)=(0.467635,0.581665,0.995023); rgb(91pt)=(0.461776,0.577072,0.995023); rgb(92pt)=(0.455916,0.57248,0.995023); rgb(93pt)=(0.450056,0.567887,0.995023); rgb(94pt)=(0.444196,0.563294,0.995023); rgb(95pt)=(0.438336,0.558701,0.995023); rgb(96pt)=(0.432476,0.554108,0.995024); rgb(97pt)=(0.426616,0.549515,0.995024); rgb(98pt)=(0.420757,0.544922,0.995023); rgb(99pt)=(0.414897,0.540329,0.995023); rgb(100pt)=(0.409037,0.535737,0.995023); rgb(101pt)=(0.403177,0.531144,0.995023); rgb(102pt)=(0.397316,0.526551,0.995025); rgb(103pt)=(0.391455,0.521958,0.995025); rgb(104pt)=(0.385597,0.517365,0.995024); rgb(105pt)=(0.379741,0.512772,0.99502); rgb(106pt)=(0.373885,0.508178,0.995017); rgb(107pt)=(0.368022,0.503586,0.995019); rgb(108pt)=(0.362148,0.498995,0.995033); rgb(109pt)=(0.356273,0.494405,0.995048); rgb(110pt)=(0.350418,0.489811,0.995042); rgb(111pt)=(0.344605,0.48521,0.994998); rgb(112pt)=(0.338806,0.480607,0.994939); rgb(113pt)=(0.332944,0.476014,0.994941); rgb(114pt)=(0.326938,0.471447,0.995081); rgb(115pt)=(0.32084,0.466896,0.99531); rgb(116pt)=(0.31493,0.462312,0.995359); rgb(117pt)=(0.309513,0.457641,0.994933); rgb(118pt)=(0.304551,0.45289,0.994071); rgb(119pt)=(0.299066,0.448231,0.993711); rgb(120pt)=(0.29192,0.443865,0.994946); rgb(121pt)=(0.282789,0.439848,0.998086); rgb(122pt)=(0.274935,0.435607,1); rgb(123pt)=(0.272609,0.430392,0.996609); rgb(124pt)=(0.279306,0.423587,0.984555); rgb(125pt)=(0.293138,0.415525,0.965651); rgb(126pt)=(0.309867,0.406953,0.943967); rgb(127pt)=(0.325491,0.398575,0.923344); rgb(128pt)=(0.339058,0.39056,0.904696); rgb(129pt)=(0.351668,0.382714,0.886966); rgb(130pt)=(0.364431,0.37484,0.869089); rgb(131pt)=(0.377772,0.366865,0.850657); rgb(132pt)=(0.391418,0.358836,0.831932); rgb(133pt)=(0.405066,0.350807,0.813205); rgb(134pt)=(0.418557,0.342805,0.79463); rgb(135pt)=(0.431954,0.33482,0.776145); rgb(136pt)=(0.445337,0.326837,0.757672); rgb(137pt)=(0.458761,0.318847,0.739161); rgb(138pt)=(0.472215,0.310852,0.720621); rgb(139pt)=(0.485675,0.302856,0.702075); rgb(140pt)=(0.499125,0.294862,0.683539); rgb(141pt)=(0.512566,0.286869,0.665011); rgb(142pt)=(0.526005,0.278876,0.646485); rgb(143pt)=(0.539446,0.270883,0.627958); rgb(144pt)=(0.552889,0.26289,0.609428); rgb(145pt)=(0.566334,0.254896,0.590897); rgb(146pt)=(0.579778,0.246903,0.572366); rgb(147pt)=(0.593221,0.23891,0.553836); rgb(148pt)=(0.606664,0.230917,0.535307); rgb(149pt)=(0.620107,0.222923,0.516777); rgb(150pt)=(0.63355,0.21493,0.498247); rgb(151pt)=(0.646993,0.206937,0.479717); rgb(152pt)=(0.660437,0.198944,0.461187); rgb(153pt)=(0.67388,0.190951,0.442658); rgb(154pt)=(0.687323,0.182958,0.424128); rgb(155pt)=(0.700766,0.174963,0.405598); rgb(156pt)=(0.71421,0.166966,0.387067); rgb(157pt)=(0.727654,0.158972,0.368536); rgb(158pt)=(0.741096,0.150985,0.350008); rgb(159pt)=(0.754536,0.143004,0.331483); rgb(160pt)=(0.767977,0.135017,0.312955); rgb(161pt)=(0.781425,0.127007,0.29442); rgb(162pt)=(0.794879,0.118969,0.275874); rgb(163pt)=(0.808331,0.110944,0.257333); rgb(164pt)=(0.821763,0.102994,0.238819); rgb(165pt)=(0.835166,0.0951591,0.220344); rgb(166pt)=(0.848574,0.0873055,0.201863); rgb(167pt)=(0.862043,0.0792099,0.183297); rgb(168pt)=(0.875621,0.0706838,0.164581); rgb(169pt)=(0.889214,0.0621028,0.145846); rgb(170pt)=(0.902611,0.0542922,0.12738); rgb(171pt)=(0.915613,0.0480374,0.109458); rgb(172pt)=(0.928457,0.0424061,0.091754); rgb(173pt)=(0.941904,0.0343962,0.0732183); rgb(174pt)=(0.956736,0.0209305,0.052775); rgb(175pt)=(0.9725,0.00378572,0.0310453); rgb(176pt)=(0.986483,0.00633329,0.0117721); rgb(177pt)=(0.995725,0.0022327,0.000967649); rgb(178pt)=(0.998618,0.0358306,0.00495484); rgb(179pt)=(0.997427,0.0855268,0.00331317); rgb(180pt)=(0.995111,0.139657,0.000121229); rgb(181pt)=(0.99409,0.188685,0.00128695); rgb(182pt)=(0.994288,0.232904,0.00101351); rgb(183pt)=(0.994914,0.275439,0.000151196); rgb(184pt)=(0.995257,0.319085,0.000322596); rgb(185pt)=(0.995242,0.364147,0.000301319); rgb(186pt)=(0.995078,0.409793,7.57419e-05); rgb(187pt)=(0.994967,0.455231,7.71666e-05); rgb(188pt)=(0.99496,0.500262,8.74378e-05); rgb(189pt)=(0.995001,0.5451,3.04168e-05); rgb(190pt)=(0.995036,0.589964,1.73228e-05); rgb(191pt)=(0.995042,0.634943,2.51464e-05); rgb(192pt)=(0.995032,0.679984,1.1303e-05); rgb(193pt)=(0.995021,0.725028,3.71976e-06); rgb(194pt)=(0.995017,0.770042,8.33655e-06); rgb(195pt)=(0.99502,0.815035,5.36005e-06); rgb(196pt)=(0.995024,0.860017,1.0858e-06); rgb(197pt)=(0.995028,0.905002,6.87705e-06); rgb(198pt)=(0.995029,0.95,7.88977e-06); rgb(199pt)=(0.995023,0.995023,0)},
]
\addplot [forget plot] graphics [xmin=4.81920750849126, xmax=4.61169901723243, ymin=-1.25420750849126, ymax=-1.04669901723243] {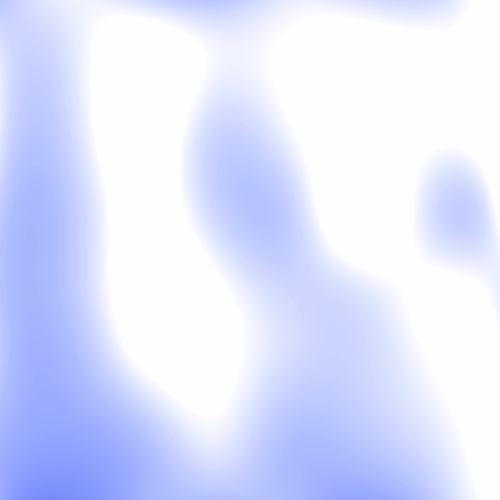};
\addplot[only marks, mark=*, mark options={}, mark size=1.7678pt, color=mycolor1, fill=mycolor1, forget plot] table[row sep=crcr]{%
x	y\\
3.46166087184211	-2.05024565388158\\
};
\end{axis}
\end{tikzpicture}%
            \fi 
             \vspace{-2.1mm}
             \caption{\gls{sw} \gls{smc} \gls{bf}}
             \label{fig:SW_SMC_pos1_switched1}
         \end{subfigure}%
         \hspace{\hspaceing}%
         \begin{subfigure}[t]{0.29\textwidth}
             \vskip 0pt	
             \centering
    	    \ifdefined\ExcludeFigures   
                \ExcludeFigures
            \else
                \vspace{-0.5mm}
%
%
\definecolor{mycolor1}{rgb}{0.21961,0.21961,0.21961}%
\pgfplotsset{every axis/.append style={
  label style={font=\footnotesize},
  legend style={font=\footnotesize},
  tick label style={font=\footnotesize},
}}

\begin{tikzpicture}

\begin{axis}[%
width=\figurewidth,
height=\figureheight,
at={(0\figurewidth,0\figureheight)},
scale only axis,
point meta min=-60,
point meta max=-30,
axis on top,
xmin=4.61190652572368,
xmax=4.819,
xlabel style={yshift=1.5mm},
xlabel={$x$ in \SI{}{\metre}},
ymin=-1.254,
ymax=-1.04690652572368,
yticklabels=\empty,
axis line style = thick,	
axis background/.style={fill=white},
colormap={mymap}{[1pt] rgb(0pt)=(0.995023,0.995023,0.995023); rgb(73pt)=(0.567253,0.659744,0.995023); rgb(74pt)=(0.561393,0.655151,0.995023); rgb(75pt)=(0.555533,0.650558,0.995023); rgb(76pt)=(0.549674,0.645965,0.995023); rgb(77pt)=(0.543814,0.641373,0.995023); rgb(78pt)=(0.537954,0.63678,0.995023); rgb(79pt)=(0.532094,0.632187,0.995023); rgb(80pt)=(0.526234,0.627594,0.995023); rgb(81pt)=(0.520374,0.623001,0.995023); rgb(82pt)=(0.514514,0.618408,0.995023); rgb(83pt)=(0.508655,0.613815,0.995023); rgb(84pt)=(0.502795,0.609223,0.995023); rgb(85pt)=(0.496935,0.60463,0.995023); rgb(86pt)=(0.491075,0.600037,0.995023); rgb(87pt)=(0.485215,0.595444,0.995023); rgb(88pt)=(0.479355,0.590851,0.995023); rgb(89pt)=(0.473495,0.586258,0.995023); rgb(90pt)=(0.467635,0.581665,0.995023); rgb(91pt)=(0.461776,0.577072,0.995023); rgb(92pt)=(0.455916,0.57248,0.995023); rgb(93pt)=(0.450056,0.567887,0.995023); rgb(94pt)=(0.444196,0.563294,0.995023); rgb(95pt)=(0.438336,0.558701,0.995023); rgb(96pt)=(0.432476,0.554108,0.995024); rgb(97pt)=(0.426616,0.549515,0.995024); rgb(98pt)=(0.420757,0.544922,0.995023); rgb(99pt)=(0.414897,0.540329,0.995023); rgb(100pt)=(0.409037,0.535737,0.995023); rgb(101pt)=(0.403177,0.531144,0.995023); rgb(102pt)=(0.397316,0.526551,0.995025); rgb(103pt)=(0.391455,0.521958,0.995025); rgb(104pt)=(0.385597,0.517365,0.995024); rgb(105pt)=(0.379741,0.512772,0.99502); rgb(106pt)=(0.373885,0.508178,0.995017); rgb(107pt)=(0.368022,0.503586,0.995019); rgb(108pt)=(0.362148,0.498995,0.995033); rgb(109pt)=(0.356273,0.494405,0.995048); rgb(110pt)=(0.350418,0.489811,0.995042); rgb(111pt)=(0.344605,0.48521,0.994998); rgb(112pt)=(0.338806,0.480607,0.994939); rgb(113pt)=(0.332944,0.476014,0.994941); rgb(114pt)=(0.326938,0.471447,0.995081); rgb(115pt)=(0.32084,0.466896,0.99531); rgb(116pt)=(0.31493,0.462312,0.995359); rgb(117pt)=(0.309513,0.457641,0.994933); rgb(118pt)=(0.304551,0.45289,0.994071); rgb(119pt)=(0.299066,0.448231,0.993711); rgb(120pt)=(0.29192,0.443865,0.994946); rgb(121pt)=(0.282789,0.439848,0.998086); rgb(122pt)=(0.274935,0.435607,1); rgb(123pt)=(0.272609,0.430392,0.996609); rgb(124pt)=(0.279306,0.423587,0.984555); rgb(125pt)=(0.293138,0.415525,0.965651); rgb(126pt)=(0.309867,0.406953,0.943967); rgb(127pt)=(0.325491,0.398575,0.923344); rgb(128pt)=(0.339058,0.39056,0.904696); rgb(129pt)=(0.351668,0.382714,0.886966); rgb(130pt)=(0.364431,0.37484,0.869089); rgb(131pt)=(0.377772,0.366865,0.850657); rgb(132pt)=(0.391418,0.358836,0.831932); rgb(133pt)=(0.405066,0.350807,0.813205); rgb(134pt)=(0.418557,0.342805,0.79463); rgb(135pt)=(0.431954,0.33482,0.776145); rgb(136pt)=(0.445337,0.326837,0.757672); rgb(137pt)=(0.458761,0.318847,0.739161); rgb(138pt)=(0.472215,0.310852,0.720621); rgb(139pt)=(0.485675,0.302856,0.702075); rgb(140pt)=(0.499125,0.294862,0.683539); rgb(141pt)=(0.512566,0.286869,0.665011); rgb(142pt)=(0.526005,0.278876,0.646485); rgb(143pt)=(0.539446,0.270883,0.627958); rgb(144pt)=(0.552889,0.26289,0.609428); rgb(145pt)=(0.566334,0.254896,0.590897); rgb(146pt)=(0.579778,0.246903,0.572366); rgb(147pt)=(0.593221,0.23891,0.553836); rgb(148pt)=(0.606664,0.230917,0.535307); rgb(149pt)=(0.620107,0.222923,0.516777); rgb(150pt)=(0.63355,0.21493,0.498247); rgb(151pt)=(0.646993,0.206937,0.479717); rgb(152pt)=(0.660437,0.198944,0.461187); rgb(153pt)=(0.67388,0.190951,0.442658); rgb(154pt)=(0.687323,0.182958,0.424128); rgb(155pt)=(0.700766,0.174963,0.405598); rgb(156pt)=(0.71421,0.166966,0.387067); rgb(157pt)=(0.727654,0.158972,0.368536); rgb(158pt)=(0.741096,0.150985,0.350008); rgb(159pt)=(0.754536,0.143004,0.331483); rgb(160pt)=(0.767977,0.135017,0.312955); rgb(161pt)=(0.781425,0.127007,0.29442); rgb(162pt)=(0.794879,0.118969,0.275874); rgb(163pt)=(0.808331,0.110944,0.257333); rgb(164pt)=(0.821763,0.102994,0.238819); rgb(165pt)=(0.835166,0.0951591,0.220344); rgb(166pt)=(0.848574,0.0873055,0.201863); rgb(167pt)=(0.862043,0.0792099,0.183297); rgb(168pt)=(0.875621,0.0706838,0.164581); rgb(169pt)=(0.889214,0.0621028,0.145846); rgb(170pt)=(0.902611,0.0542922,0.12738); rgb(171pt)=(0.915613,0.0480374,0.109458); rgb(172pt)=(0.928457,0.0424061,0.091754); rgb(173pt)=(0.941904,0.0343962,0.0732183); rgb(174pt)=(0.956736,0.0209305,0.052775); rgb(175pt)=(0.9725,0.00378572,0.0310453); rgb(176pt)=(0.986483,0.00633329,0.0117721); rgb(177pt)=(0.995725,0.0022327,0.000967649); rgb(178pt)=(0.998618,0.0358306,0.00495484); rgb(179pt)=(0.997427,0.0855268,0.00331317); rgb(180pt)=(0.995111,0.139657,0.000121229); rgb(181pt)=(0.99409,0.188685,0.00128695); rgb(182pt)=(0.994288,0.232904,0.00101351); rgb(183pt)=(0.994914,0.275439,0.000151196); rgb(184pt)=(0.995257,0.319085,0.000322596); rgb(185pt)=(0.995242,0.364147,0.000301319); rgb(186pt)=(0.995078,0.409793,7.57419e-05); rgb(187pt)=(0.994967,0.455231,7.71666e-05); rgb(188pt)=(0.99496,0.500262,8.74378e-05); rgb(189pt)=(0.995001,0.5451,3.04168e-05); rgb(190pt)=(0.995036,0.589964,1.73228e-05); rgb(191pt)=(0.995042,0.634943,2.51464e-05); rgb(192pt)=(0.995032,0.679984,1.1303e-05); rgb(193pt)=(0.995021,0.725028,3.71976e-06); rgb(194pt)=(0.995017,0.770042,8.33655e-06); rgb(195pt)=(0.99502,0.815035,5.36005e-06); rgb(196pt)=(0.995024,0.860017,1.0858e-06); rgb(197pt)=(0.995028,0.905002,6.87705e-06); rgb(198pt)=(0.995029,0.95,7.88977e-06); rgb(199pt)=(0.995023,0.995023,0)},
]
\addplot [forget plot] graphics [xmin=4.81920750849126, xmax=4.61169901723243, ymin=-1.25420750849126, ymax=-1.04669901723243] {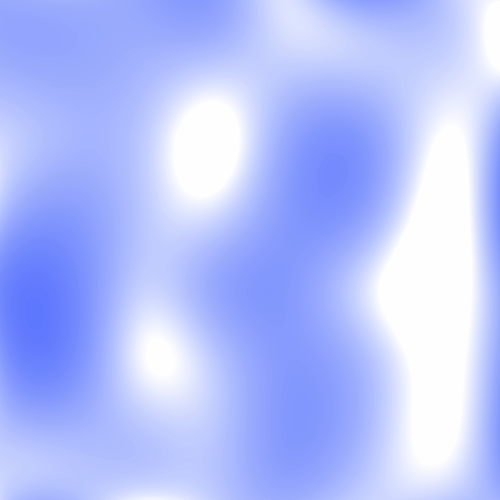};
\addplot[only marks, mark=*, mark options={}, mark size=1.7678pt, color=mycolor1, fill=mycolor1, forget plot] table[row sep=crcr]{%
x	y\\
3.46166087184211	-2.05024565388158\\
};
\end{axis}
\end{tikzpicture}%
            \fi 
            \vspace{-2.1mm}
             \caption{Reciprocity-based \gls{bf} (full \gls{csi})}
             \label{fig:CSI_pos1_switched1}
         \end{subfigure}%
         \hspace{-0.65cm}%
         \begin{subfigure}[t]{0.058\textwidth}
            \vspace{-2.2mm} %
             \vskip 0pt	
             \centering
                      \setlength{\plotWidth}{1\textwidth}	
%
%

\pgfplotsset{every axis/.append style={
  label style={font=\footnotesize},
  legend style={font=\footnotesize},
  tick label style={font=\footnotesize}
}}

\begin{tikzpicture}

\begin{axis}[%
width=0.225\plotWidth,
height=3.38\plotWidth,
at={(0\plotWidth,0\plotWidth)},
scale only axis,
xmin=0,
xmax=1.02,
ymin=-1,
ymax=1,
zmin=-60,
zmax=-30,
zlabel={$PG$ in \SI{}{\dB}},
axis line style = thick,	
view={0}{0},
axis background/.style={fill=white},
axis z line*=right,			
xticklabel=\empty,			
tick align=center,		    
xtick=\empty,               
y axis line style= { draw opacity=0 }, 
z axis line style= { draw opacity=0 }, 
ztick distance={5},				
yticklabel pos=right,
xmajorgrids,
ymajorgrids,
zmajorgrids
]

\addplot3[%
surf,
shader=flat, z buffer=sort, colormap={mymap}{[1pt] rgb(0pt)=(1,1,1); rgb(26pt)=(0.51634,0.620915,1); rgb(27pt)=(0.497738,0.606335,1); rgb(34pt)=(0.367521,0.504274,1); rgb(35pt)=(0.348919,0.489693,1); rgb(39pt)=(0.27451,0.431373,1); rgb(40pt)=(0.317186,0.405998,0.941176); rgb(41pt)=(0.359862,0.380623,0.882353); rgb(43pt)=(0.445213,0.329873,0.764706); rgb(44pt)=(0.487889,0.304498,0.705882); rgb(45pt)=(0.530565,0.279123,0.647059); rgb(46pt)=(0.573241,0.253749,0.588235); rgb(47pt)=(0.615917,0.228374,0.529412); rgb(48pt)=(0.658593,0.202999,0.470588); rgb(49pt)=(0.701269,0.177624,0.411765); rgb(50pt)=(0.743945,0.152249,0.352941); rgb(51pt)=(0.786621,0.126874,0.294118); rgb(52pt)=(0.829296,0.101499,0.235294); rgb(53pt)=(0.871972,0.0761246,0.176471); rgb(54pt)=(0.914648,0.0507497,0.117647); rgb(55pt)=(0.957324,0.0253749,0.0588235); rgb(56pt)=(1,0,0); rgb(58pt)=(1,0.285714,0); rgb(59pt)=(1,0.428571,0); rgb(60pt)=(1,0.571429,0); rgb(63pt)=(1,1,0)}, mesh/rows=2]
table[row sep=crcr, point meta=\thisrow{c}] {%
x	y	z	c\\
0	-0.0001	-30.00	-30.00\\ 
0	-0.0001	-30.30	-30.30\\ 
0	-0.0001	-30.61	-30.61\\ 
0	-0.0001	-30.91	-30.91\\ 
0	-0.0001	-31.21	-31.21\\ 
0	-0.0001	-31.52	-31.52\\ 
0	-0.0001	-31.82	-31.82\\ 
0	-0.0001	-32.12	-32.12\\ 
0	-0.0001	-32.42	-32.42\\ 
0	-0.0001	-32.73	-32.73\\ 
0	-0.0001	-33.03	-33.03\\ 
0	-0.0001	-33.33	-33.33\\ 
0	-0.0001	-33.64	-33.64\\ 
0	-0.0001	-33.94	-33.94\\ 
0	-0.0001	-34.24	-34.24\\ 
0	-0.0001	-34.55	-34.55\\ 
0	-0.0001	-34.85	-34.85\\ 
0	-0.0001	-35.15	-35.15\\ 
0	-0.0001	-35.45	-35.45\\ 
0	-0.0001	-35.76	-35.76\\ 
0	-0.0001	-36.06	-36.06\\ 
0	-0.0001	-36.36	-36.36\\ 
0	-0.0001	-36.67	-36.67\\ 
0	-0.0001	-36.97	-36.97\\ 
0	-0.0001	-37.27	-37.27\\ 
0	-0.0001	-37.58	-37.58\\ 
0	-0.0001	-37.88	-37.88\\ 
0	-0.0001	-38.18	-38.18\\ 
0	-0.0001	-38.48	-38.48\\ 
0	-0.0001	-38.79	-38.79\\ 
0	-0.0001	-39.09	-39.09\\ 
0	-0.0001	-39.39	-39.39\\ 
0	-0.0001	-39.70	-39.70\\ 
0	-0.0001	-40.00	-40.00\\ 
0	-0.0001	-40.30	-40.30\\ 
0	-0.0001	-40.61	-40.61\\ 
0	-0.0001	-40.91	-40.91\\ 
0	-0.0001	-41.21	-41.21\\ 
0	-0.0001	-41.52	-41.52\\ 
0	-0.0001	-41.82	-41.82\\ 
0	-0.0001	-42.12	-42.12\\ 
0	-0.0001	-42.42	-42.42\\ 
0	-0.0001	-42.73	-42.73\\ 
0	-0.0001	-43.03	-43.03\\ 
0	-0.0001	-43.33	-43.33\\ 
0	-0.0001	-43.64	-43.64\\ 
0	-0.0001	-43.94	-43.94\\ 
0	-0.0001	-44.24	-44.24\\ 
0	-0.0001	-44.55	-44.55\\ 
0	-0.0001	-44.85	-44.85\\ 
0	-0.0001	-45.15	-45.15\\ 
0	-0.0001	-45.45	-45.45\\ 
0	-0.0001	-45.76	-45.76\\ 
0	-0.0001	-46.06	-46.06\\ 
0	-0.0001	-46.36	-46.36\\ 
0	-0.0001	-46.67	-46.67\\ 
0	-0.0001	-46.97	-46.97\\ 
0	-0.0001	-47.27	-47.27\\ 
0	-0.0001	-47.58	-47.58\\ 
0	-0.0001	-47.88	-47.88\\ 
0	-0.0001	-48.18	-48.18\\ 
0	-0.0001	-48.48	-48.48\\ 
0	-0.0001	-48.79	-48.79\\ 
0	-0.0001	-49.09	-49.09\\ 
0	-0.0001	-49.39	-49.39\\ 
0	-0.0001	-49.70	-49.70\\ 
0	-0.0001	-50.00	-50.00\\ 
0	-0.0001	-50.30	-50.30\\ 
0	-0.0001	-50.61	-50.61\\ 
0	-0.0001	-50.91	-50.91\\ 
0	-0.0001	-51.21	-51.21\\ 
0	-0.0001	-51.52	-51.52\\ 
0	-0.0001	-51.82	-51.82\\ 
0	-0.0001	-52.12	-52.12\\ 
0	-0.0001	-52.42	-52.42\\ 
0	-0.0001	-52.73	-52.73\\ 
0	-0.0001	-53.03	-53.03\\ 
0	-0.0001	-53.33	-53.33\\ 
0	-0.0001	-53.64	-53.64\\ 
0	-0.0001	-53.94	-53.94\\ 
0	-0.0001	-54.24	-54.24\\ 
0	-0.0001	-54.55	-54.55\\ 
0	-0.0001	-54.85	-54.85\\ 
0	-0.0001	-55.15	-55.15\\ 
0	-0.0001	-55.45	-55.45\\ 
0	-0.0001	-55.76	-55.76\\ 
0	-0.0001	-56.06	-56.06\\ 
0	-0.0001	-56.36	-56.36\\ 
0	-0.0001	-56.67	-56.67\\ 
0	-0.0001	-56.97	-56.97\\ 
0	-0.0001	-57.27	-57.27\\ 
0	-0.0001	-57.58	-57.58\\ 
0	-0.0001	-57.88	-57.88\\ 
0	-0.0001	-58.18	-58.18\\ 
0	-0.0001	-58.48	-58.48\\ 
0	-0.0001	-58.79	-58.79\\ 
0	-0.0001	-59.09	-59.09\\ 
0	-0.0001	-59.39	-59.39\\ 
0	-0.0001	-59.70	-59.70\\ 
0	-0.0001	-60.00	-60.00\\ 
1	-0.0001	-30.00	-30.00\\ 
1	-0.0001	-30.30	-30.30\\ 
1	-0.0001	-30.61	-30.61\\ 
1	-0.0001	-30.91	-30.91\\ 
1	-0.0001	-31.21	-31.21\\ 
1	-0.0001	-31.52	-31.52\\ 
1	-0.0001	-31.82	-31.82\\ 
1	-0.0001	-32.12	-32.12\\ 
1	-0.0001	-32.42	-32.42\\ 
1	-0.0001	-32.73	-32.73\\ 
1	-0.0001	-33.03	-33.03\\ 
1	-0.0001	-33.33	-33.33\\ 
1	-0.0001	-33.64	-33.64\\ 
1	-0.0001	-33.94	-33.94\\ 
1	-0.0001	-34.24	-34.24\\ 
1	-0.0001	-34.55	-34.55\\ 
1	-0.0001	-34.85	-34.85\\ 
1	-0.0001	-35.15	-35.15\\ 
1	-0.0001	-35.45	-35.45\\ 
1	-0.0001	-35.76	-35.76\\ 
1	-0.0001	-36.06	-36.06\\ 
1	-0.0001	-36.36	-36.36\\ 
1	-0.0001	-36.67	-36.67\\ 
1	-0.0001	-36.97	-36.97\\ 
1	-0.0001	-37.27	-37.27\\ 
1	-0.0001	-37.58	-37.58\\ 
1	-0.0001	-37.88	-37.88\\ 
1	-0.0001	-38.18	-38.18\\ 
1	-0.0001	-38.48	-38.48\\ 
1	-0.0001	-38.79	-38.79\\ 
1	-0.0001	-39.09	-39.09\\ 
1	-0.0001	-39.39	-39.39\\ 
1	-0.0001	-39.70	-39.70\\ 
1	-0.0001	-40.00	-40.00\\ 
1	-0.0001	-40.30	-40.30\\ 
1	-0.0001	-40.61	-40.61\\ 
1	-0.0001	-40.91	-40.91\\ 
1	-0.0001	-41.21	-41.21\\ 
1	-0.0001	-41.52	-41.52\\ 
1	-0.0001	-41.82	-41.82\\ 
1	-0.0001	-42.12	-42.12\\ 
1	-0.0001	-42.42	-42.42\\ 
1	-0.0001	-42.73	-42.73\\ 
1	-0.0001	-43.03	-43.03\\ 
1	-0.0001	-43.33	-43.33\\ 
1	-0.0001	-43.64	-43.64\\ 
1	-0.0001	-43.94	-43.94\\ 
1	-0.0001	-44.24	-44.24\\ 
1	-0.0001	-44.55	-44.55\\ 
1	-0.0001	-44.85	-44.85\\ 
1	-0.0001	-45.15	-45.15\\ 
1	-0.0001	-45.45	-45.45\\ 
1	-0.0001	-45.76	-45.76\\ 
1	-0.0001	-46.06	-46.06\\ 
1	-0.0001	-46.36	-46.36\\ 
1	-0.0001	-46.67	-46.67\\ 
1	-0.0001	-46.97	-46.97\\ 
1	-0.0001	-47.27	-47.27\\ 
1	-0.0001	-47.58	-47.58\\ 
1	-0.0001	-47.88	-47.88\\ 
1	-0.0001	-48.18	-48.18\\ 
1	-0.0001	-48.48	-48.48\\ 
1	-0.0001	-48.79	-48.79\\ 
1	-0.0001	-49.09	-49.09\\ 
1	-0.0001	-49.39	-49.39\\ 
1	-0.0001	-49.70	-49.70\\ 
1	-0.0001	-50.00	-50.00\\ 
1	-0.0001	-50.30	-50.30\\ 
1	-0.0001	-50.61	-50.61\\ 
1	-0.0001	-50.91	-50.91\\ 
1	-0.0001	-51.21	-51.21\\ 
1	-0.0001	-51.52	-51.52\\ 
1	-0.0001	-51.82	-51.82\\ 
1	-0.0001	-52.12	-52.12\\ 
1	-0.0001	-52.42	-52.42\\ 
1	-0.0001	-52.73	-52.73\\ 
1	-0.0001	-53.03	-53.03\\ 
1	-0.0001	-53.33	-53.33\\ 
1	-0.0001	-53.64	-53.64\\ 
1	-0.0001	-53.94	-53.94\\ 
1	-0.0001	-54.24	-54.24\\ 
1	-0.0001	-54.55	-54.55\\ 
1	-0.0001	-54.85	-54.85\\ 
1	-0.0001	-55.15	-55.15\\ 
1	-0.0001	-55.45	-55.45\\ 
1	-0.0001	-55.76	-55.76\\ 
1	-0.0001	-56.06	-56.06\\ 
1	-0.0001	-56.36	-56.36\\ 
1	-0.0001	-56.67	-56.67\\ 
1	-0.0001	-56.97	-56.97\\ 
1	-0.0001	-57.27	-57.27\\ 
1	-0.0001	-57.58	-57.58\\ 
1	-0.0001	-57.88	-57.88\\ 
1	-0.0001	-58.18	-58.18\\ 
1	-0.0001	-58.48	-58.48\\ 
1	-0.0001	-58.79	-58.79\\ 
1	-0.0001	-59.09	-59.09\\ 
1	-0.0001	-59.39	-59.39\\ 
1	-0.0001	-59.70	-59.70\\ 
1	-0.0001	-60.00	-60.00\\ 
};
\draw [draw=black, thick] (0.015,0.98,-60) rectangle (1,1,-30);    
\end{axis}
\end{tikzpicture}%
         \end{subfigure}
        	\vspace{-1mm}\caption{
        	Beamformers applied at $\bm{p}_\mrm{EN}^{(1)}$, $PG$ distribution evaluated and interpolated around  $\bm{p}_\mrm{EN}^{(2)}$.
        	}\label{fig:pos1_domains_switched}
        	\vspace{-3mm}
    \end{figure*}
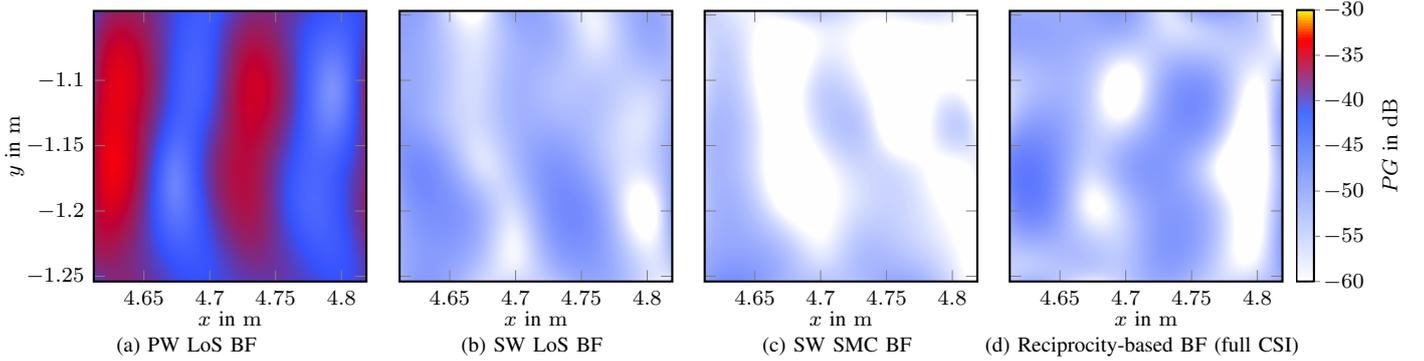

\fi

\subsection{Measurement-based Validation of Beamforming Strategies}\label{sec:validation}
We compare the \glspl{bf} introduced in Section~\ref{sec:beamforming-strategies} by means of their efficiencies, i.e., evaluating their achieved $PG$ in~\eqref{eq:PG} with geometry-based channel vectors $\widetilde{\bm{h}}$ generated with the ``true" \gls{en} device position $\bm{p}_\mrm{EN}$. 
We further investigate the efficiency of a reciprocity-based \gls{bf} as a function of the quality of its \gls{csi}. 
Mishra and Larsson~\cite{Mishra2019} have demonstrated that the $PG$ reduces with imperfect \gls{csi}. 
Through our definition of the channel \gls{snr} in \eqref{eq:channel-SNR}, we can find an approximate 
expression for the expected efficiency of a reciprocity-based \gls{bf} 
\begin{align}\label{eq:PG-reciprocity-analytical}
    PG_\mrm{R} &= \mathbb{E} \left\{ 
        \left|
            \frac{\hat{\bm{h}}^\herm\bm{h}}{\lVert\hat{\bm{h}}\rVert}
        \right|^2
    \right\}
    \approx 
    \frac{\SNR}{1+\SNR}\left(
    L\, P_\mrm{ch} + P_\mrm{n}
    \right) 
    \\
    &\approx \left\{
        \begin{array}{@{}lrl@{}}
        P_\mrm{ch},     &   {\scriptstyle  \SNR < 1/L} & {\footnotesize \text{ low-SNR regime}}\\
        L \,P_\mrm{ch} \, \SNR , & {\scriptstyle 1/L < \SNR < 1} & {\footnotesize \text{ linear regime}} \\
         L \, P_\mrm{ch}, & {\scriptstyle \SNR > 1} & {\footnotesize \text{ high-SNR regime}}
        \end{array}
    \right. \label{eq:PG-reciprocity-regimes}
\end{align}
which is derived in Appendix~\ref{app:CSI-loss}.
Fig.\,\ref{fig:BF_comparison} shows the performance comparison of all introduced \glspl{bf} evaluated on the measured channel at $\bm{p}_\mrm{EN}^{(1)}$, which is assumed to be the ``true" channel, i.e., perfect \gls{csi}. 
The analytical expression for the efficiency of the reciprocity-based \gls{bf} in~\eqref{eq:PG-reciprocity-analytical} is compared against a \gls{mc} analysis with $M=10^{5}$ realizations of~\eqref{eq:channel-estimate}, denoted $PG_\textsc{mc}$, where both curves show good correspondence.
Furthermore, the standard deviation $\sigma$ of the path gain distribution is evaluated by means of the \gls{mc} analysis.
The indicated \gls{snr} regimes are defined in~\eqref{eq:PG-reciprocity-regimes}.
In the low \gls{snr} regime, the expected efficiency of the reciprocity-based \gls{bf} is approximately equal to the path gain $PG_\textsc{siso}$ of an equivalent \gls{siso} system. 
However, through ``random beamforming" corresponding to $\hat{\bm{h}}=\bm{n}_h$, an efficiency improvement of up to \SI{6}{\dB} is achievable within the $3\sigma$-interval in Fig.\,\ref{fig:BF_comparison} (capturing approx. \SI{98}{\percent} of the path gain realizations\footnote{In the low \gls{snr} regime, the path gain is a chi-squared distributed \gls{rv} with the intervals $PG_{\textsc{r}}\pm n\sigma$ capturing $\{86.4, 95.0, 98.2\}\SI{}{\percent}$ of its distribution for $n \in \{1, 2, 3\}$. In the linear regime it transitions to a Gaussian \gls{rv} where the intervals cover $\{68.3, 95.4, 99.7\}\SI{}{\percent}$.}).
The \SI{6}{\dB} gain results from
$\sigma \approx P_\mrm{ch} \approx PG_\textsc{siso}$
in the low \gls{snr} regime (see Appendix~\ref{app:CSI-loss}), and thus $PG_\mrm{R} + 3 \sigma \approx 4 \, PG_\textsc{siso}$, which is achievable irrespective of $L$.
Targeting higher gains will be practically unreasonable for most applications due to the large number of weight realizations involved. 
The amplitude $|y|$ is Rayleigh-distributed in this regime.
As the \gls{snr} increases, $PG_\textsc{r}$ enters a regime of linear increase with the \gls{snr}, where both the amplitude $|y|$ and the path gain are 
normally distributed. 
In this transition region from a ``stochastic" \gls{bf} to a deterministic \gls{bf}, the relative uncertainty of the path gain  
decreases 
with increasing \gls{snr}. 
Entering the high \gls{snr} regime, $PG_\textsc{r}$ saturates at the \gls{miso} path gain $PG_\textsc{miso}$ 
requiring perfect \gls{csi} and leveraging the full array gain $L$.

\ifdefined\reduceSize  
    ~
\else
    \setlength{\figurewidth}{0.7\linewidth}
    \setlength{\figureheight}{0.7\linewidth}
    \begin{figure}[tb]
    \centering %
%
%
\definecolor{mycolor1}{rgb}{0.55294,0.75294,0.27059}%

\pgfplotsset{every axis/.append style={
  label style={font=\footnotesize},
  legend style={font=\footnotesize},
  tick label style={font=\footnotesize},
}}

\begin{tikzpicture}

\begin{axis}[%
width=0.904\figurewidth,
tick align=inside,		        
axis line style = thick,	    
line cap = round,               
height=0.889\figurewidth,
at={(0\figurewidth,0\figureheight)},
scale only axis,
point meta min=-79.8056051632163,
point meta max=-39.8056051632163,
axis on top,
x dir=reverse,
xmin=-45.9183673469388,
xmax=45.9183673469388,
xlabel={Azimuth angle $\varphi$ in \SI{}{\degree}},
xlabel style={yshift=0.1cm},
y dir=reverse,
ymin=44.0816326530612,
ymax=135.918367346939,
ylabel={Elevation angle $\theta$ in \SI{}{\degree}},
ylabel style={yshift=-0.15cm},
axis background/.style={fill=white},
colormap={mymap}{[1pt] rgb(0pt)=(0.995023,0.995023,0.995023); rgb(73pt)=(0.567253,0.659744,0.995023); rgb(74pt)=(0.561393,0.655151,0.995023); rgb(75pt)=(0.555533,0.650558,0.995023); rgb(76pt)=(0.549674,0.645965,0.995023); rgb(77pt)=(0.543814,0.641373,0.995023); rgb(78pt)=(0.537954,0.63678,0.995023); rgb(79pt)=(0.532094,0.632187,0.995023); rgb(80pt)=(0.526234,0.627594,0.995023); rgb(81pt)=(0.520374,0.623001,0.995023); rgb(82pt)=(0.514514,0.618408,0.995023); rgb(83pt)=(0.508655,0.613815,0.995023); rgb(84pt)=(0.502795,0.609223,0.995023); rgb(85pt)=(0.496935,0.60463,0.995023); rgb(86pt)=(0.491075,0.600037,0.995023); rgb(87pt)=(0.485215,0.595444,0.995023); rgb(88pt)=(0.479355,0.590851,0.995023); rgb(89pt)=(0.473495,0.586258,0.995023); rgb(90pt)=(0.467635,0.581665,0.995023); rgb(91pt)=(0.461776,0.577072,0.995023); rgb(92pt)=(0.455916,0.57248,0.995023); rgb(93pt)=(0.450056,0.567887,0.995023); rgb(94pt)=(0.444196,0.563294,0.995023); rgb(95pt)=(0.438336,0.558701,0.995023); rgb(96pt)=(0.432476,0.554108,0.995024); rgb(97pt)=(0.426616,0.549515,0.995024); rgb(98pt)=(0.420757,0.544922,0.995023); rgb(99pt)=(0.414897,0.540329,0.995023); rgb(100pt)=(0.409037,0.535737,0.995023); rgb(101pt)=(0.403177,0.531144,0.995023); rgb(102pt)=(0.397316,0.526551,0.995025); rgb(103pt)=(0.391455,0.521958,0.995025); rgb(104pt)=(0.385597,0.517365,0.995024); rgb(105pt)=(0.379741,0.512772,0.99502); rgb(106pt)=(0.373885,0.508178,0.995017); rgb(107pt)=(0.368022,0.503586,0.995019); rgb(108pt)=(0.362148,0.498995,0.995033); rgb(109pt)=(0.356273,0.494405,0.995048); rgb(110pt)=(0.350418,0.489811,0.995042); rgb(111pt)=(0.344605,0.48521,0.994998); rgb(112pt)=(0.338806,0.480607,0.994939); rgb(113pt)=(0.332944,0.476014,0.994941); rgb(114pt)=(0.326938,0.471447,0.995081); rgb(115pt)=(0.32084,0.466896,0.99531); rgb(116pt)=(0.31493,0.462312,0.995359); rgb(117pt)=(0.309513,0.457641,0.994933); rgb(118pt)=(0.304551,0.45289,0.994071); rgb(119pt)=(0.299066,0.448231,0.993711); rgb(120pt)=(0.29192,0.443865,0.994946); rgb(121pt)=(0.282789,0.439848,0.998086); rgb(122pt)=(0.274935,0.435607,1); rgb(123pt)=(0.272609,0.430392,0.996609); rgb(124pt)=(0.279306,0.423587,0.984555); rgb(125pt)=(0.293138,0.415525,0.965651); rgb(126pt)=(0.309867,0.406953,0.943967); rgb(127pt)=(0.325491,0.398575,0.923344); rgb(128pt)=(0.339058,0.39056,0.904696); rgb(129pt)=(0.351668,0.382714,0.886966); rgb(130pt)=(0.364431,0.37484,0.869089); rgb(131pt)=(0.377772,0.366865,0.850657); rgb(132pt)=(0.391418,0.358836,0.831932); rgb(133pt)=(0.405066,0.350807,0.813205); rgb(134pt)=(0.418557,0.342805,0.79463); rgb(135pt)=(0.431954,0.33482,0.776145); rgb(136pt)=(0.445337,0.326837,0.757672); rgb(137pt)=(0.458761,0.318847,0.739161); rgb(138pt)=(0.472215,0.310852,0.720621); rgb(139pt)=(0.485675,0.302856,0.702075); rgb(140pt)=(0.499125,0.294862,0.683539); rgb(141pt)=(0.512566,0.286869,0.665011); rgb(142pt)=(0.526005,0.278876,0.64648); rgb(143pt)=(0.539446,0.270883,0.627958); rgb(144pt)=(0.552889,0.26289,0.609428); rgb(145pt)=(0.566334,0.254896,0.590897); rgb(146pt)=(0.579778,0.246903,0.572366); rgb(147pt)=(0.593221,0.23891,0.553836); rgb(148pt)=(0.606664,0.230917,0.535307); rgb(149pt)=(0.620107,0.222923,0.516777); rgb(150pt)=(0.63355,0.21493,0.498247); rgb(151pt)=(0.646993,0.206937,0.479717); rgb(152pt)=(0.660437,0.198944,0.461187); rgb(153pt)=(0.67388,0.190951,0.442658); rgb(154pt)=(0.687323,0.182958,0.424128); rgb(155pt)=(0.700766,0.174963,0.405598); rgb(156pt)=(0.71421,0.166966,0.387067); rgb(157pt)=(0.727654,0.158972,0.368536); rgb(158pt)=(0.741096,0.150985,0.350008); rgb(159pt)=(0.754536,0.143004,0.331483); rgb(160pt)=(0.767977,0.135017,0.312955); rgb(161pt)=(0.781425,0.127007,0.29442); rgb(162pt)=(0.794879,0.118969,0.275874); rgb(163pt)=(0.808331,0.110944,0.257333); rgb(164pt)=(0.821763,0.102994,0.238819); rgb(165pt)=(0.835166,0.0951591,0.220344); rgb(166pt)=(0.848574,0.0873055,0.201863); rgb(167pt)=(0.862043,0.0792099,0.183297); rgb(168pt)=(0.875621,0.0706838,0.164581); rgb(169pt)=(0.889214,0.0621028,0.145846); rgb(170pt)=(0.902611,0.0542922,0.12738); rgb(171pt)=(0.915613,0.0480374,0.109458); rgb(172pt)=(0.928457,0.0424061,0.091754); rgb(173pt)=(0.941904,0.0343962,0.0732183); rgb(174pt)=(0.956736,0.0209305,0.052775); rgb(175pt)=(0.9725,0.00378572,0.0310453); rgb(176pt)=(0.986483,0.00633329,0.0117721); rgb(177pt)=(0.995725,0.0022327,0.000967649); rgb(178pt)=(0.998618,0.0358306,0.00495484); rgb(179pt)=(0.997427,0.0855268,0.00331317); rgb(180pt)=(0.995111,0.139657,0.000121229); rgb(181pt)=(0.99409,0.188685,0.00128695); rgb(182pt)=(0.994288,0.232904,0.00101351); rgb(183pt)=(0.994914,0.275439,0.000151196); rgb(184pt)=(0.995257,0.319085,0.000322596); rgb(185pt)=(0.995242,0.364147,0.000301319); rgb(186pt)=(0.995078,0.409793,7.57419e-05); rgb(187pt)=(0.994967,0.455231,7.71666e-05); rgb(188pt)=(0.99496,0.500262,8.74378e-05); rgb(189pt)=(0.995001,0.5451,3.04168e-05); rgb(190pt)=(0.995036,0.589964,1.73228e-05); rgb(191pt)=(0.995042,0.634943,2.51464e-05); rgb(192pt)=(0.995032,0.679984,1.1303e-05); rgb(193pt)=(0.995021,0.725028,3.71976e-06); rgb(194pt)=(0.995017,0.770042,8.33655e-06); rgb(195pt)=(0.99502,0.815035,5.36005e-06); rgb(196pt)=(0.995024,0.860017,1.0858e-06); rgb(197pt)=(0.995028,0.905002,6.87705e-06); rgb(198pt)=(0.995029,0.95,7.88977e-06); rgb(199pt)=(0.995023,0.995023,0)},
colorbar,
colorbar style={ylabel style={font=\color{white!15!black}},%
width=0.3cm,%
ytick pos=right, %
ytick distance={5}, %
tick align=center, %
axis line style = thick,	
xshift = -0.2cm, %
ylabel={\footnotesize $PG$ in \SI{}{\dB}} }
]
\addplot [forget plot] graphics [xmin=-45.9183673469388, xmax=45.9183673469388, ymin=44.0816326530612, ymax=135.918367346939] {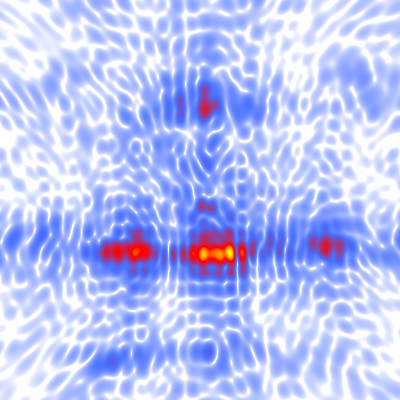};
\addplot [color=mycolor1, line width = 2.0, only marks, mark size=0.8pt, mark=*, mark options={solid, mycolor1}, forget plot]
  table[row sep=crcr]{%
-3.32510908549622	101.612187227894\\
};

\draw[color=black, line width=0.5pt]
(-3.32,101.6) -- %
(-11.32,93.6) -- %
(-16.32,93.6); %
\node[right, align=left ,font={\footnotesize},fill=white,opacity=0.75,inner sep=0.75mm] at (axis cs:-16.32,93.6) {\SI{-42.21}{\dB}};

\draw[color=black, line width=0.5pt]
(-7,102.5) -- %
(-11,106.5) -- %
(-16.32,106.5); %
\node[right, align=left ,font={\footnotesize},fill=white,opacity=0.75,inner sep=0.75mm] at (axis cs:-16.32,106.5)
{\SI{-39.45}{\dB}};

\end{axis}
\end{tikzpicture}%
\vspace{-3mm}
    \caption{\Gls{pw} \gls{bf} beamsweep at position $\bm{p}_\mrm{EN}^{(1)}$: Depicted is the path gain evaluated on a portion of the elevation-azimuth plane. 
    At the ``true" position of the \gls{en} device, the \gls{pw} \gls{bf} achieves $PG = \SI{-42.21}{\dB}$, while it achieves a maximum of $\SI{-39.45}{\dB}$ with constructive \gls{smc} interference.}
    \label{fig:PW_beamsweeping_array_pos1}
    \vspace{-5mm}
    \end{figure}
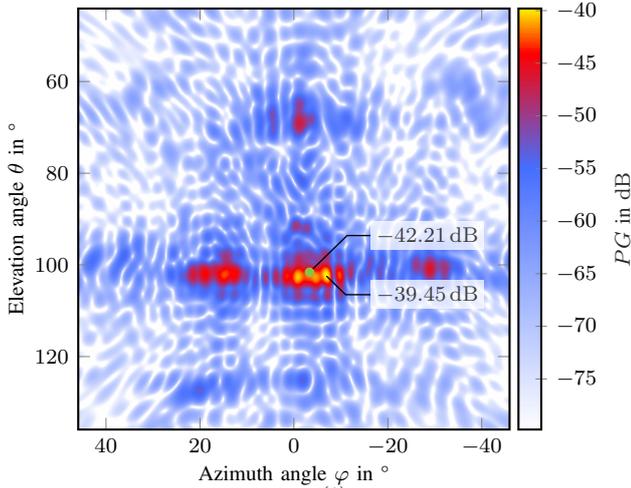	
    
    In Fig.\,\ref{fig:PW_beamsweeping_array_pos1}, we show the path gain of an exhaustive beam sweep with the \gls{pw} \gls{bf}, evaluated on a portion of the elevation-azimuth plane w.r.t the center of gravity of the \gls{rw}. 
    It achieves an efficiency of $PG = \SI{-42.21}{\dB}$ at the ``true" position of the \gls{en} device. 
    In its vicinity across the azimuth plane, there is a fading pattern visible which originates from constructive and destructive interferences with the \glspl{smc} possibly caused by walls.
    As is indicated in the figure, the beam sweep reaches its maximum path gain of $\SI{-39.45}{\dB}$ not at the ``true" position, but rather at an azimuth angle where constructive interference occurs. 
    Therefore it is no coincidence that $\bm{p}_\mrm{EN}^{(1)}$ is located at a ``peak'' of the path gain distribution in Fig.\,\ref{fig:PW_pos1_switched0}. 
    The shape of that distribution is consequently a result of the interference of \glspl{smc}.
\fi

We further compare the reciprocity-based \gls{bf} to our geometry-based \glspl{bf}, defined in Section~\ref{sec:beamforming-strategies}. 
The \gls{pw} \gls{bf} shows an efficiency loss of $\Delta PG \approx \SI{10}{\dB}$, when compared with $PG_\textsc{miso}$ and is thus comparably inefficient, however, it can strongly reduce the duration of a possible beam sweeping procedure due to its reduced, two-dimensional search space. The \gls{sw} \gls{los} \gls{bf} is approx. \SI{6}{\dB} more efficient than the \gls{pw} \gls{bf} and constitutes a good trade-off between model complexity and efficiency.
In the given scenario, the \gls{sw} \gls{smc} \gls{bf} exhibits another \SI{1.75}{\dB} increase in efficiency when compared with the \gls{sw} \gls{los} \gls{bf}. 
It is based on our most complete \gls{smc} channel model and is dependent on multiple parameters as well as a geometric environment model.  
Given perfect \gls{csi}, a reciprocity-based \gls{bf} could still attain \SI{2}{\dB} more efficiency than the \gls{sw} \gls{smc} \gls{bf}.
Fig.\,\ref{fig:pos1_domains} and Fig.\,\ref{fig:pos2_domains} depict the interpolated path gain distributions around $\bm{p}_\mrm{EN}^{(1)}$ and $\bm{p}_\mrm{EN}^{(2)}$, when applying the described \glspl{bf}.
It is observable that a rather homogeneous power distribution is achievable with the \gls{pw} and \gls{sw} \glspl{bf} around $\bm{p}_\mrm{EN}^{(1)}$ (see Fig.\,\ref{fig:pos1_domains}).
Exploiting \glspl{smc} virtually increases the array aperture and thus narrows its beam width, which is observable for the \gls{sw} \gls{smc} \gls{bf} and the reciprocity-based \gls{bf}.
The path gain distribution around $\bm{p}_\mrm{EN}^{(2)}$ (see Fig.\,\ref{fig:pos2_domains}) shows a strong standing wave pattern, possibly originating from the bottles within the shelf and the metal grid mounted on its backside. 
    Fig.\,\ref{fig:pos1_domains_switched} shows the path gain distribution around $\bm{p}_\mrm{EN}^{(2)}$ when beamforming to $\bm{p}_\mrm{EN}^{(1)}$. 
    It is clearly observable that the \gls{pw} \gls{bf} ``pollutes" its environment with radiated power even in unintended positions.
    For the \gls{sw} \gls{los} and \gls{smc} \glspl{bf}, the power density around $\bm{p}_\mrm{EN}^{(2)}$ is strongly reduced since no specular components are exploited in its vicinity.
    However, the power density around $\bm{p}_\mrm{EN}^{(2)}$ increases again for the reciprocity-based \gls{bf} that inherently exploits the complete channel \emph{including diffuse reflections}, possibly originating from the shelf, to increase its efficiency.

\section{Conclusion}\label{sec:conclusion}
We presented a geometry-based channel model that represents channel vectors through \gls{s-parameter} vectors of transmission coefficients and is thus suitable to model \gls{wpt} in a \gls{miso} system. 
We formulated \glspl{bf} for \gls{pw} and \gls{sw} propagation capable of exploiting \glspl{smc} to establish a simultaneous multibeam transmission. 
We derived an approximate 
expression for the expected efficiency of a reciprocity-based \gls{bf} depending on the quality of its \gls{csi}.
We acquired realistic measurement data with an \gls{xlmimo} testbed and evaluated the performance of our \glspl{bf} on that data. 
We showed that random beamforming may achieve a gain of around \SI{6}{\dB}, far behind the \gls{pw} \gls{bf} which constitutes a low-complexity solution for the initial access problem to \gls{en} devices. 
Our \gls{sw} \gls{los} \gls{bf} comes at higher complexity but still holds considerable gains over the \gls{pw} \gls{bf}.
Our \gls{sw} \gls{smc} \gls{bf} suffers a loss of only \SI{2}{\dB} when compared with perfect \gls{csi}.
We demonstrated both the benefit of increased efficiency as well as the drawback of increased parametric complexity when exploiting \glspl{smc}.

\begin{appendices}

\section{Reciprocity-based BF Efficiency}\label{app:CSI-loss}
Given the noisy channel estimate $\hat{\bm{h}}$ in~\eqref{eq:channel-estimate}, we derive the expected path gain as
\begin{align}
    PG_\textsc{r} &= 
    \frac{\mathbb{E} \left\{ 
        \left|
            y
        \right|^2
    \right\}}{P_\mrm{t}}
    =
    \mathbb{E} \left\{ 
        \left|
            \frac{\hat{\bm{h}}^\herm\bm{h}}{\lVert\hat{\bm{h}}\rVert}
        \right|^2
    \right\}
    \nonumber \\ 
    &= \mathbb{E} \left\{ 
        \frac{\left( 
            \bm{h}^\herm\bm{h} +
            \bm{h}^\herm\bm{n}
        \right)
        \left( 
            \bm{h}^\herm\bm{h} +
            \bm{h}^\herm\bm{n}
        \right)^\ast}{\lVert \bm{h} + \bm{n} \rVert^2}
    \right\} \nonumber \\ 
    &\approx 
    \frac{\lVert\bm{h}\rVert^4 + 2 \lVert\bm{h}\rVert^2 \mathbb{E}\left\{ 
             (\bm{h}^\herm\bm{n})
        \right\}
        + 
        \bm{h}^\herm \mathbb{E}\left\{ 
            \bm{n}\bm{n}^\herm
        \right\} \bm{h}
        }{\left(P_\mrm{ch}+P_\mrm{n}\right)L}
    \nonumber \\
    &= \frac{L^2\,P_\mrm{ch}^2 + 0 + L\,P_\mrm{ch}\,P_\mrm{n}}{\left(P_\mrm{ch}+P_\mrm{n}\right)L} 
    = \frac{L\,P_\mrm{ch}^2 + P_\mrm{ch}\,P_\mrm{n}}{P_\mrm{ch}+P_\mrm{n}} 
    \nonumber \\ 
    & =  \underbrace{\frac{\SNR}{1+\SNR} L\, P_\mrm{ch}}_{\approx \frac{\mathbb{E}\{|y|\}^2}{P_\mrm{t}}} 
    \hspace{0.3cm}
    + 
\underbrace{\frac{\SNR}{1+\SNR}  P_\mrm{n}}_{\substack{\approx \frac{\sqrt{\Var\{|y|^2\}}}{P_\mrm{t}} \approx \sqrt{\Var\{PG\}} = \sigma \\ \text{in the low-SNR regime}}}
    \,.
\end{align}

\ifdefined\reduceSize  
    ~
\else
    \section{Definition of Walls}
    To perform both the visibility analysis as well as the mirroring of an \gls{rw} across walls, we need to define a plane in a \gls{3d} space as 
    \begin{equation}
        \bm{w} = w_{\mathrm{off}} \, \bm{n}_{\mathrm{w}},
    \end{equation}
    where $\bm{n}_\mrm{w}$ represents the wall normal vector and $w_{\mathrm{off}}$ a scalar offset. 
    Further, the extent of a wall, i.e., the limits in each direction, need to be known.
    \section{Visibility Vector}\label{app:visibility}
    A ray $\bm{r}(l)$, cast from one antenna to another, is defined as
    \begin{equation}
        \bm{r}(l) = \bm{p}_0 + l \, \bm{e}_r,
    \end{equation}
    where $\bm{p}_0$ is the point of origin, $\bm{e}_{r}$ a unit-vector pointing from the antenna at the ray origin to the respective other antenna, and $l$ the length of the ray. 
    If the condition $\bm{e}_r^\trp \bm{n}_{\mathrm{w}} \neq 0$ holds, i.e., the ray is not orthogonal to the wall-normal, there exists a unique intersection point between the ray and the wall. 
    The parameter $l$ of such an intersection lies at
    \begin{equation}
        l_\mathrm{int} = \frac{w_{\mathrm{off}} - \bm{p}_0^\trp  \bm{n}_{\mathrm{w}}}{\bm{e}_r^\trp \bm{n}_{\mathrm{w}}} \,.
    \end{equation} 
    If $\bm{r}(l_\mathrm{int})$ then lies within the boundaries of the wall, the ray hits the wall along its path~\cite{ShirleyRaytracing2000}.
    
    To iteratively construct the visibility vector 
    \begin{align}
        [\bm{\vis}]_\ell = 
        \begin{cases}
            1, & \text{component visible} \\
            0, & \text{component not visible} \,, \\
        \end{cases} 
    \end{align}
    a ray is cast from each receiving antenna position to each transmitting antenna position and if there exists a valid intersection, either a zero 
    or one 
    is entered. 
    This leads to a vector that can be multiplied element-wise with the channel vector, i.e., $\bm{h}_{\bm{\vis}} = \bm{\vis} \odot \bm{h}$. 
    Note, that for the \gls{los} component $k=1$ the elements of the visibility vector are $[\bm{\vis}]_\ell=1$ only if they intersect with no walls and $[\bm{\vis}]_\ell=0$ else.

    \section{Householder Matrix}\label{app:householder}
    To perform the reflection of an antenna position $\bm{p}_\mrm{RW}^{(\ell)}$ across a given wall, using its normal vector $\bm{n}_{\mathrm{w}}$, the following linear transformation matrix
    \begin{equation}
        \bm{\house} = \mathrm{\bm{I}} - 2\bm{n}_{\mathrm{w}}\bm{n}_{\mathrm{w}}^\trp \, ,
    \end{equation}
    can be used, which is also known as the Householder matrix~\cite{Householder58}. 
    With the array layout defined by its center of gravity $\bm{p}_\mrm{RW}$ and antenna positions $\widetilde{\bm{P}}_\mrm{RW}$ relative to it, i.e., centered around $\bm{p} = \bm{0}$, these relative positions can be transformed separately with 
    \begin{equation}
        \widetilde{\bm{P}}_\mrm{RW,\bm{\house}} = \bm{\house} \widetilde{\bm{P}}_\mrm{RW} \ \in \realset{3}{L} \, ,
    \end{equation}
    while the center of gravity of the array needs to be shifted by twice its distance $d_{\mathrm{w}}$ to the wall in the opposite direction of $\bm{n}_{\mathrm{w}}$ through
    \begin{equation}
        \bm{p}_{\bm{\house}} = \bm{p}_\mrm{RW} - 2d_{\mathrm{w}}\bm{n}_{\mathrm{w}} \ \in \realset{3}{1} \, .
    \end{equation}
    The complete mirrored array layout is thus to be found by
    \begin{equation}
        \bm{P}_{\mrm{RW},\bm{\house}} = \bm{p}_{\bm{\house}} \, \bm{1}_{1\times L} + \widetilde{\bm{P}}_\mrm{RW,\bm{\house}} \ \in \realset{3}{L} \, .
    \end{equation}
\fi

\end{appendices}

\bibliographystyle{IEEEtran}
\bibliography{IEEEabrv,ICC_2023}
\balance


\end{document}